\documentclass[twocolumn]{aastex61} 
\usepackage{graphicx}				
\usepackage{amssymb}

\begin{document}

\title{Identifying AGNs in low-mass galaxies via long-term optical variability}
\author{Vivienne F. Baldassare}
\altaffiliation{Einstein Fellow}
\affiliation{Yale University \\
Department of Astronomy \\
52 Hillhouse Avenue \\
New Haven, CT 06511, USA}

\author{Marla Geha}
\affiliation{Yale University \\
Department of Astronomy \\
52 Hillhouse Avenue \\
New Haven, CT 06511, USA}

\author{Jenny Greene}
\affiliation{Princeton University \\
Department of Astrophysical Sciences\\
4 Ivy Lane\\
Princeton University, Princeton, NJ 08544 }

\correspondingauthor{Vivienne F. Baldassare} \email{vivienne.baldassare@yale.edu}

\received{28 August 2018}
\submitjournal{ApJ}

\begin{abstract}

We present an analysis of the nuclear variability of $\sim28,000$ nearby ($z<0.15$) galaxies with Sloan Digital Sky Survey (SDSS) spectroscopy in Stripe 82. We construct light curves using difference imaging of SDSS \textit{g-}band images, which allows us to detect subtle variations in the central light output. We select variable AGN by assessing whether detected variability is well-described by a damped random walk model. We find 135 galaxies with AGN-like nuclear variability. While most of the variability-selected AGNs have narrow emission lines consistent with the presence of an AGN, a small fraction have narrow emission lines dominated by star formation. The star-forming systems with nuclear AGN-like variability tend to be low-mass ($M_{\ast}<10^{10}~M_{\odot}$), and may be AGNs missed by other selection techniques due to star formation dilution or low-metallicities. We explore the AGN fraction as a function of stellar mass, and find that the fraction of variable AGN increases with stellar mass, even after taking into account the fact that lower mass systems are fainter. There are several possible explanations for an observed decline in the fraction of variable AGN with decreasing stellar mass, including a drop in the supermassive black hole occupation fraction, a decrease in the ratio of black hole mass to galaxy stellar mass, or a change in the variability properties of lower-mass AGNs. We demonstrate that optical photometric variability is a promising avenue for detecting AGNs in low-mass, star formation-dominated galaxies, which has implications for the upcoming Large Synoptic Survey Telescope. 

\end{abstract}

\section{Introduction}
Supermassive black holes (BHs; $M_{\rm BH}\gtrsim 10^{5}~M_{\odot}$) are ubiquitous in the centers of galaxies with stellar masses $\gtrsim10^{10}M_{\odot}$. Less is known about the population of BHs in the centers of low-mass galaxies (here defined as galaxies with $M_{\ast}\lesssim10^{10}M_{\odot}$). However, the population of BHs in low-mass galaxies has the potential to place constraints on the mechanisms by which the seeds of present day BHs formed. The occupation fraction (i.e., the fraction of galaxies containing BHs) is expected to differ depending on the seed formation mechanisms at play (see reviews by \citealt{2012NatCo...3E1304G, 2014GReGr..46.1702N}). In particular, the occupation fraction is sensitive to the seed formation mechanism for galaxies with stellar masses $M_{\ast}<10^{10}~M_{\odot}$.

Detecting BHs in low-mass galaxies poses unique observational challenges. A BH with $M_{\rm BH}=10^{5}~M_{\odot}$ has a gravitational sphere of influence of just a few pc, i.e., largely unresolvable outside the Local Group even with the \textit{Hubble Space Telescope}. In recent years, an increasing number of actively accreting massive black holes have been discovered in low-mass galaxies, particularly using X-ray emission and optical spectroscopic signatures (\citealt{2004ApJ...610..722G, 2007ApJ...670...92G, 2008AJ....136.1179B,Reines:2011fr, Reines:2013fj, 2014ApJ...787L..30R, 2014AJ....148..136M, 2015ApJ...798...38S, 2015ApJ...805...12L, 2016ApJ...831..203P, 2016ApJ...817...20M, 2018MNRAS.478.2576M}). However, emission line ratio diagrams commonly used to identify AGN were developed using samples of massive galaxies and do not necessarily apply for lower-mass, lower-metallicity systems \citep{2006MNRAS.371.1559G, 2018arXiv180510874B}. Moreover, at low galaxy stellar masses, star formation can dilute the AGN emission-line signal, resulting in AGN potentially being missed \citep{2015ApJ...811...26T}. While a sufficiently bright, hard X-ray point source can be a relatively unambiguous signature of an AGN, X-ray imaging down to the relevant luminosities is observationally expensive for large samples. 

Motivated by the potential for identifying systems missed by other selection techniques, we take the approach of searching for AGN via low-level optical variability. AGN are known to vary at all wavelengths, and searching for optical variability has been a rather prolific tool for identifying quasars (e.g., \citealt{2003AJ....125....1G, 2007AJ....134.2236S, 2010ApJ...714.1194S, 2011ApJ...728...26M, 2011A&A...530A.122P, 2014ApJ...782...37C, 2014AJ....147...12B}). The origin of the variability remains uncertain, but is potentially related to thermal instabilities in the accretion disk (e.g., \citealt{1984ARA&A..22..471R, 2009ApJ...698..895K}). Since the advent of the Sloan Digital Sky Survey (SDSS), great advances have been made in understanding the characteristics of AGN variability, as well as in the identification of AGN through the presence of variability. Optical variations at the 0.03 magnitude level have been observed in at least $90\%$ of quasars in the SDSS Stripe 82 \citep{2007AJ....134.2236S}. Variable AGN can also be distinguished from other variable objects (such as variable stars) based on their variability properties \citep{2011AJ....141...93B}.

In the last several years, an increasing number of time domain surveys have come online; an incomplete list includes the Palomar Transient Factory \citep{2009PASP..121.1334R}, Zwicky Transient Facility \citep{2014htu..conf...27B}, Pan-STARRS \citep{2016arXiv161205560C, 2016arXiv161205243F}, La Silla-QUEST \citep{2015ApJ...810..164C} and GAIA \citep{2016A&A...595A...1G}. Additionally, the Large Synoptic Survey Telescope (LSST), which will image the entire visible sky every three nights, is scheduled to begin operations in 2023. Given the wealth of available and upcoming time domain data, it is important to assess the utility of variability for identifying AGN in low-mass galaxies. 

Using \textit{g-}band imaging data from the SDSS Stripe 82, we construct light curves for $\sim28,000$ galaxies in the NASA-Sloan Atlas, with stellar masses spanning from $10^{7}-10^{12}~M_{\odot}$. Section 2 describes the sample and data. Section 3 describes the difference imaging analysis and selection criteria for AGN candidates. In Section 4, we present the full sample of variability-selected AGN. In Section 5, we discuss the low-mass systems with AGN variability, and present an analysis of the detection limits and expected number of detections for the low-mass end. 

\section{Sample}

Our sample is comprised of galaxies in the NASA-Sloan Atlas (NSA) which fall within the area of Stripe 82. Stripe 82 is a $\sim300~\rm{deg}^{2}$ field near the celestial equator that has been imaged repeatedly by the 2.5m SDSS telescope. Observations between the years 2000 and 2004 were taken in optimal (photometric) conditions; observations from 2004-2008 were taken in a variety of conditions. 

The NSA is a reprocessing of the SDSS photometry using the SDSS five-band imaging combined with Galaxy Evolution Explorer (GALEX) imaging in the ultraviolet (see \citealt{2011ApJS..193...29A, 2011AJ....142..153Y, 2007AJ....133..734B, 2011AJ....142...31B} for details of the SDSS data re/processing techniques). Additionally, the NSA uses an improved background subtraction technique \citep{2011AJ....142...31B}, resulting in significantly improved photometry relative to the SDSS DR8 photometric catalog. We use the \textbf{nsa\_v1\_0\_1.fits} catalog \footnote{http://www.sdss.org/dr13/manga/manga-target-selection/nsa/ {}}, which was released with the SDSS DR13. The v1 catalog extends the NSA out to $z=0.15$, and adds elliptical Petrosian aperture photometry, using the Petrosian radius from the r band imaging. Aperture corrections have also been applied to the SDSS and GALEX photometry. Relevant quantities given in the NSA include spectroscopic redshifts, Petrosian radius, ellipticity, and position angle, and stellar mass. 

\begin{figure}
\centering
\includegraphics[width=0.45\textwidth]{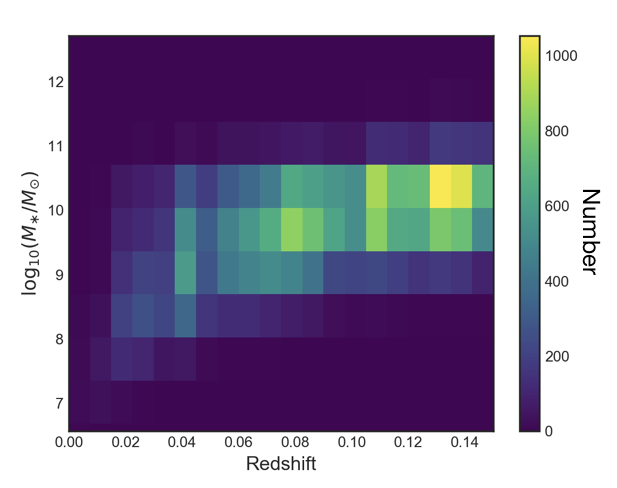}
\caption{Two-dimensional histogram showing the sample of 28062 galaxies in both Stripe 82 and the NSA for which we generate light curves, in bins of redshift and stellar mass. Stellar masses are taken from the second release of the NSA catalog.}
\label{mstar_dist}
\end{figure}

Using the DR13 context in CasJobs, we download all \textit{g}-band images of fields within Stripe 82 ($-1.28~{\rm deg} <{\rm Dec}<1.28~{\rm deg} $; $\rm{R.A.}< 60~{\rm deg} $ or $\rm{R.A.}>329~{\rm deg} $). There are 32206 galaxies in the NSA within the specified RA and Dec constraints. As we are trying to detect subtle variations in the nuclear light curves, we only use photometric observations (i.e., SDSS imaging score $>0.6$). For our analysis, we restrict our sample to galaxies with more than 10 data points in their light curves (see Section 3). We are unable to generate light curves for $4414/32206$ galaxies, due to either an insufficient number of observations, or poor image subtractions (due to e.g., an extremely bright foreground star). For the 28062 for which we construct light curves, the median sample stellar mass is $M_{\ast}\approx7\times10^{9}M_{\odot}$ (assuming $h=0.70$) and the median sample redshift is $z=0.09$. The stellar mass and redshift distribution is shown in Figure~\ref{mstar_dist}. 

\section{Analysis}

\subsection{Difference Imaging and Light Curve Construction}

Straightforward aperture photometry is insufficient for detecting low-level variations in the light output of a galaxy nucleus. Given that each night of observation has different seeing, simply measuring the flux within a constant aperture means that each observation contains a different fraction of the galaxy starlight in that aperture. The counts measured within a constant aperture depend strongly on the seeing. 

Thus, we use difference imaging to construct light curves for each galaxy. Difference imaging, or image subtraction, involves convolving a template image with a kernel to match the seeing and background of an individual observation, and then subtracting one from the other to create a diff.pdference image. Specifically, we use the difference imaging software \textit{Difference Imaging Analysis Pipeline 2} (DIAPL2)  \footnote{https://users.camk.edu.pl/pych/DIAPL/}, which is a modified version of the Difference Imaging Analysis software introduced by \cite{2000AcA....50..421W}. Both are implementations of the basic algorithm introduced by \cite{1998ApJ...503..325A} and \cite{2000A&AS..144..363A}. The basic steps of our difference imaging procedure are listed below; steps 2 through 4 are carried out within DIAPL2. 

\begin{enumerate}
\item Crop the SDSS field images to cutouts of 800x800 pixels (5.3$'$ x 5.3$'$) centered around the NSA galaxy. Since fields were in different epochs were centered differently, adjacent fields sometimes must be stitched together and then cropped.

\item Align the images, correcting for small pixel shifts between individual frames. 

\item Construct a template image consisting of the frames with the best seeing and lowest background. We chose 30\% of the frames to be used in the template. 

\item For each individual exposure, convolve the template image with a best-fit kernel to match the seeing of the exposure. The kernel is a sum of 2D Gaussians of different widths, and the best-fit parameters are found through chi-squared minimization. The convolved template is then subtracted from the exposure to create the difference image. 

\item Construct a light curve using forced photometry of template and difference images. Since we are searching for variability from a point source, photometry is done on a 2.5$''$ circle at the position of the galaxy nucleus as defined in the NSA. The flux value for each epoch is the value measured in the template plus the value measured in the difference image.

\end{enumerate} 

\subsection{Selection of AGN candidates}

AGN light curves are well-modeled by a damped random walk \citep{2009ApJ...698..895K, 2010ApJ...721.1014M, 2011ApJ...728...26M, 2011AJ....141...93B}. The damping functions to push deviations back towards a mean value, as opposed to the variance simply continuing to increase over time. 

We use the \cite{2011AJ....141...93B} QSO fitting software \textbf{qso\_fit} \footnote{http://butler.lab.asu.edu/qso\_selection/index.html} to determine whether the light curves are variable, and whether the observed variability is characteristic of an AGN. \cite{2011AJ....141...93B} define a covariance matrix $C$ which well-describes AGN variability as a function of time. It is then possible to write down a probability of observing some data $x$ given the AGN variability model $C$. Specifically, the \textbf{qso\_fit} code creates a model light curve by modeling each point given the other points and the covariance matrix describing the variability. It then assesses how well the best-fit damped random walk model describes the data. 

We compute several fit statistics, including $\chi^{2}/\nu$ (standard measure of variability), $[\chi^{2}/\nu]_{\rm QSO}$ (the fit statistic for the fit of the data to a damped random walk model), and $[\chi^{2}/\nu]_{\rm null}$ (the expected fit statistic for a non-AGN variable source). We also compute $\sigma_{\rm var}$ (the significance that the source is variable), $\sigma_{\rm QSO}$ (the significance that $[\chi^{2}/\nu]_{\rm QSO} < [\chi^{2}/\nu]_{\rm null}$), and $\sigma_{\rm notQSO}$ (the significance the source variability is not-AGN like). 

\begin{figure*}
\centering
\includegraphics[width=0.2\textwidth]{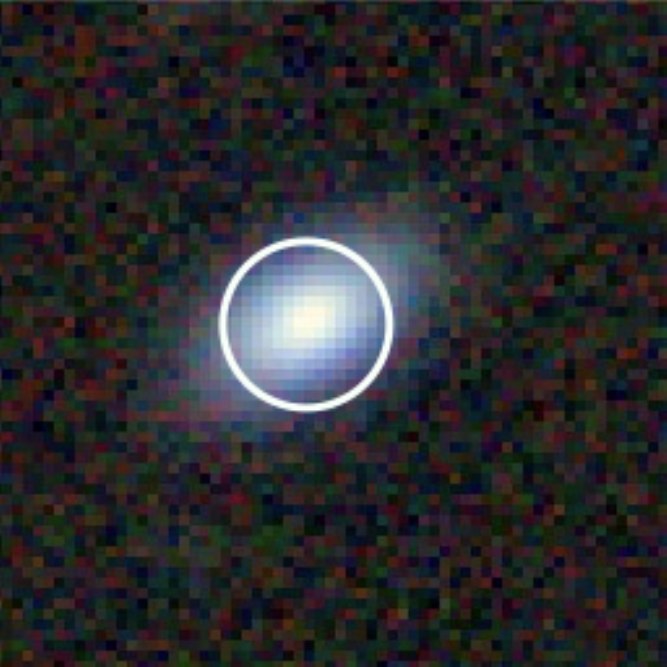}\hspace{0.8in}
\includegraphics[width=0.2\textwidth]{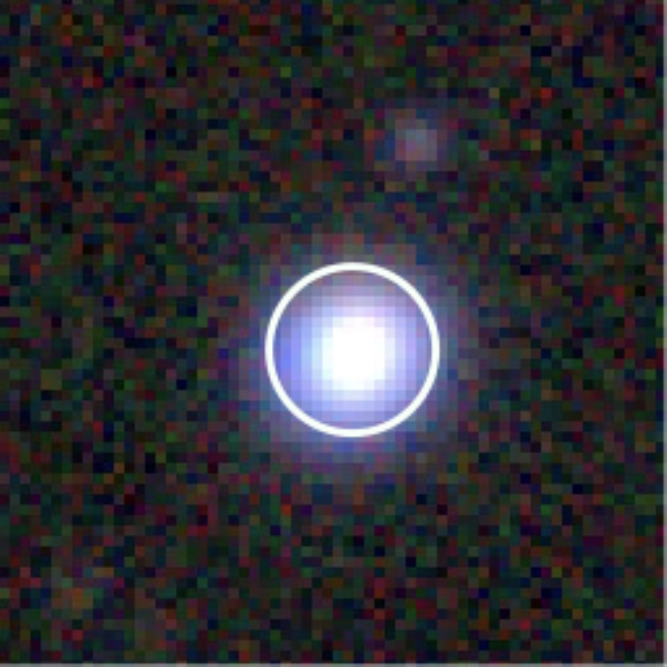}\hspace{0.8in}
\includegraphics[width=0.2\textwidth]{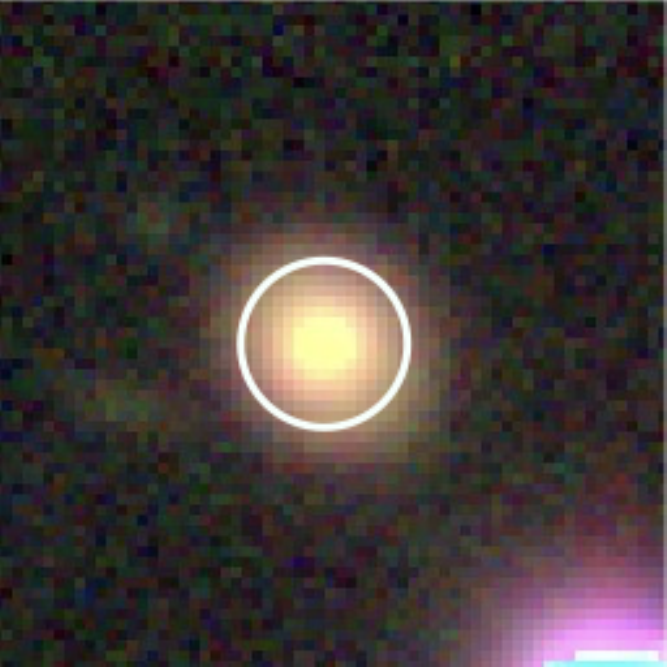}\\
\includegraphics[width=0.31\textwidth]{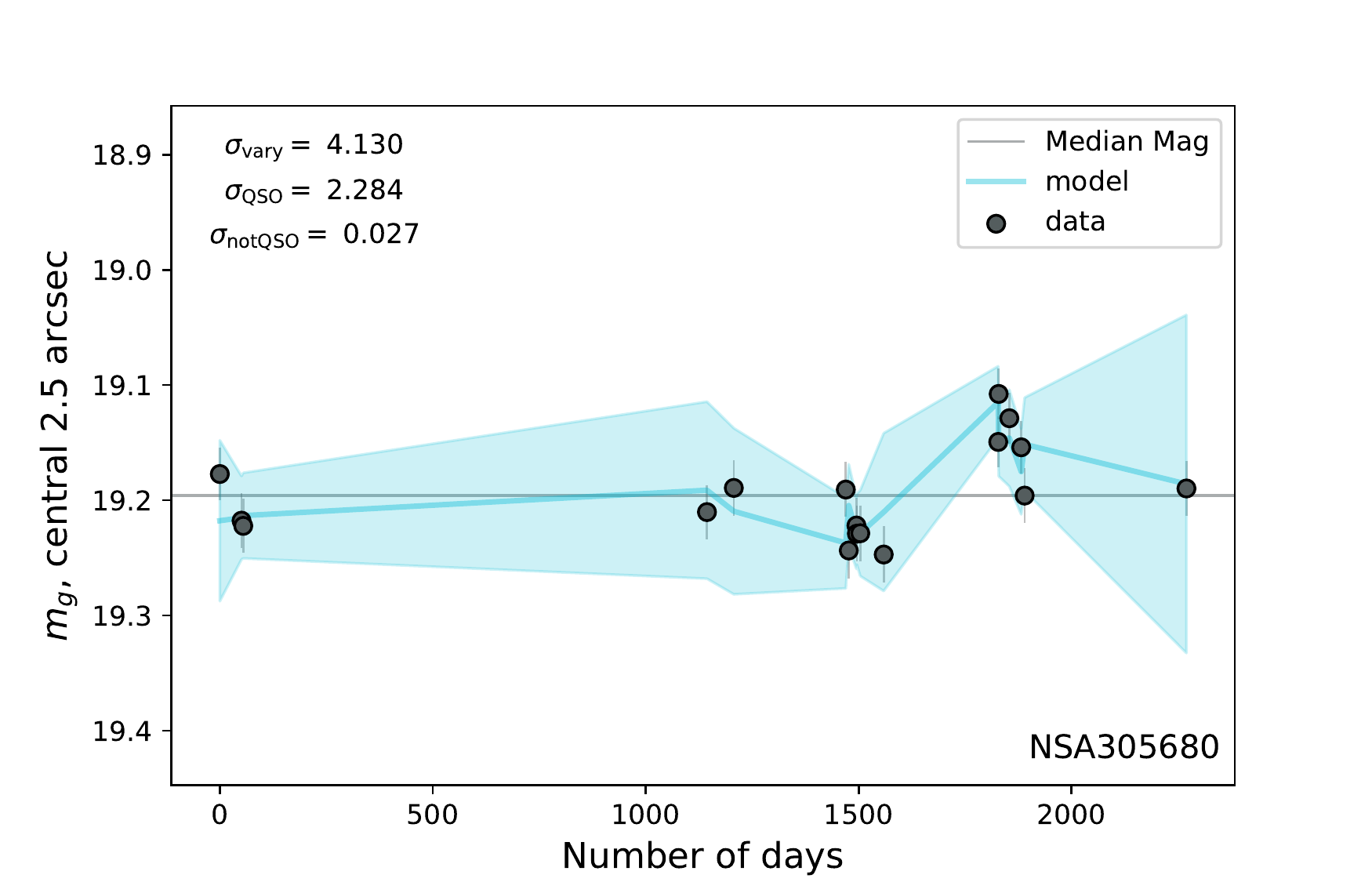}
\includegraphics[width=0.31\textwidth]{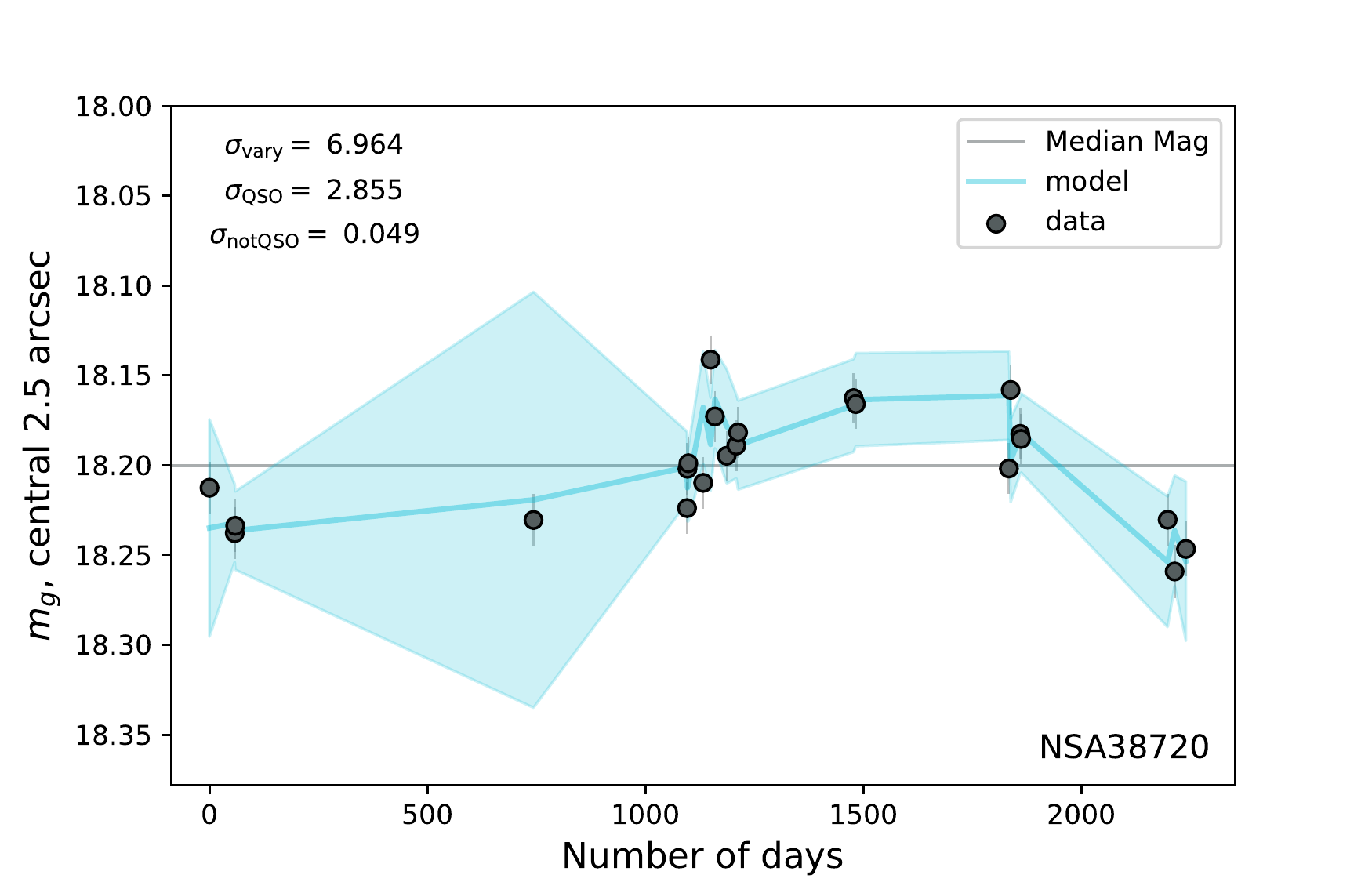}
\includegraphics[width=0.31\textwidth]{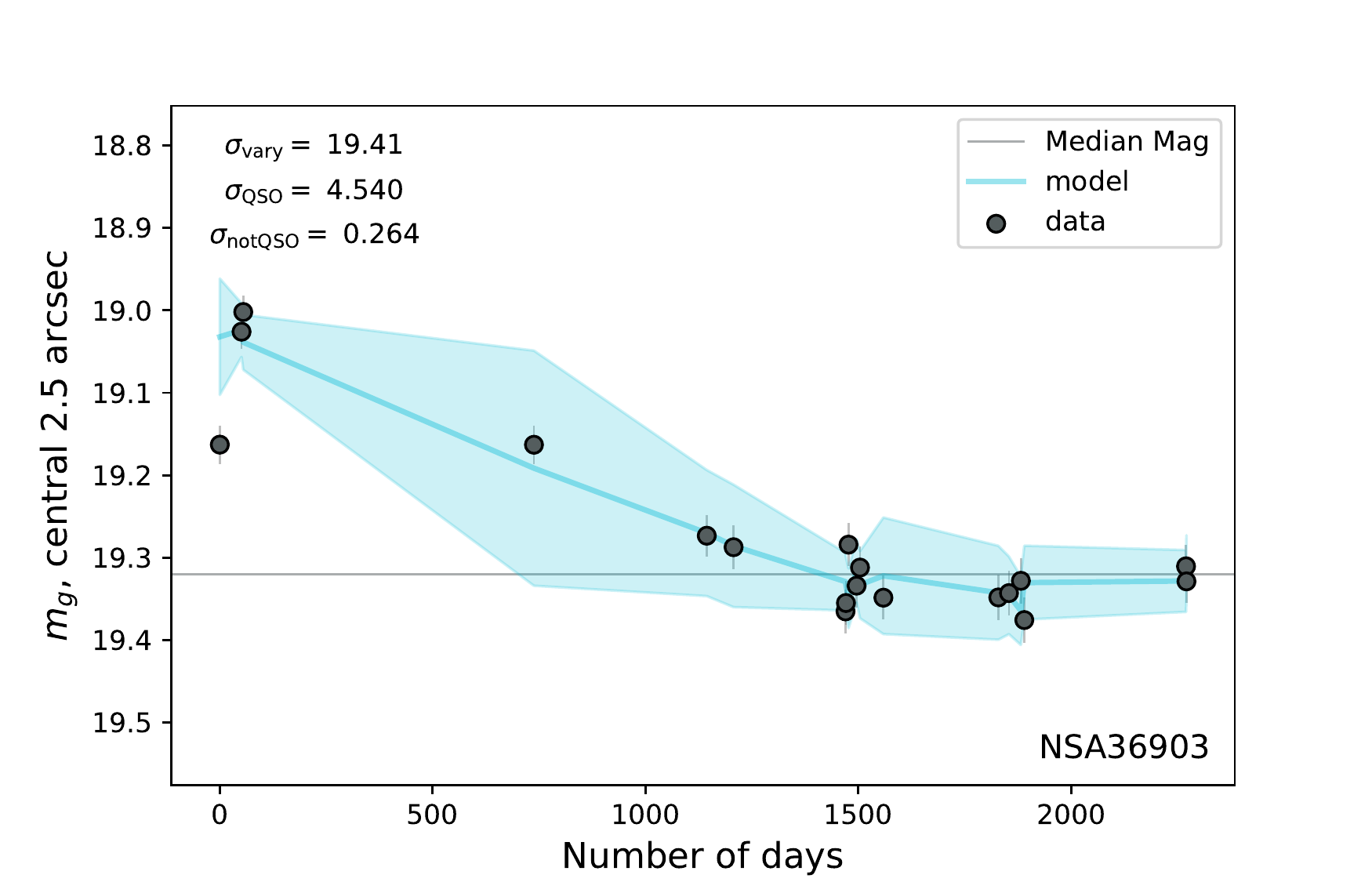}
\caption{Examples of galaxies which meet our AGN selection criteria ($\sigma_{\rm var}>2$ and $\sigma_{\rm QSO}>2$). The top panel shows DECaLs images (http://legacysurvey.org/viewer) with a circle of radius 2.5$''$ over the nucleus as given in the NSA. The images are 20$''$ across. The lower panel shows the nuclear \textit{g}-band light curve. The grey points are the observed nuclear \textit{g}-band magnitudes with corresponding errors. The blue solid line shows the best fit damped random walk model from \textbf{qso\_fit}, and the light blue shaded region shows the model uncertainties.}
\label{agn_lcs}
\end{figure*}

\begin{figure*}
\centering
\includegraphics[width=0.2\textwidth]{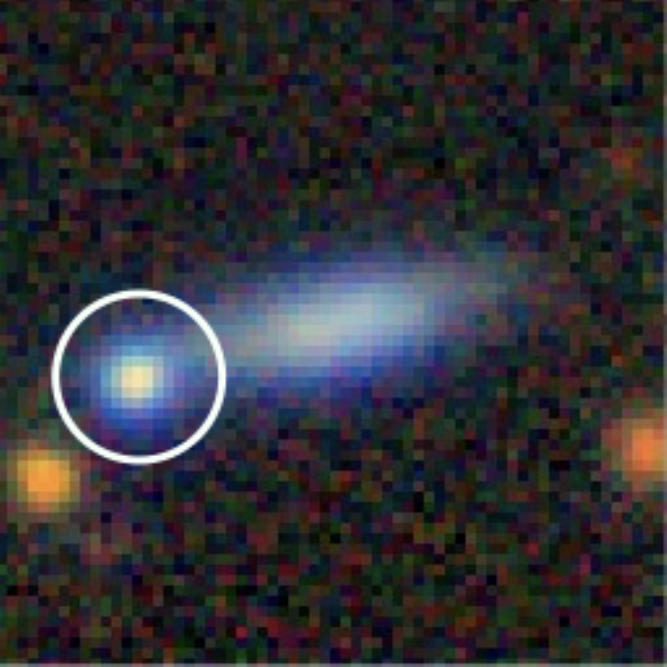} \hspace{0.7in}
\includegraphics[width=0.2\textwidth]{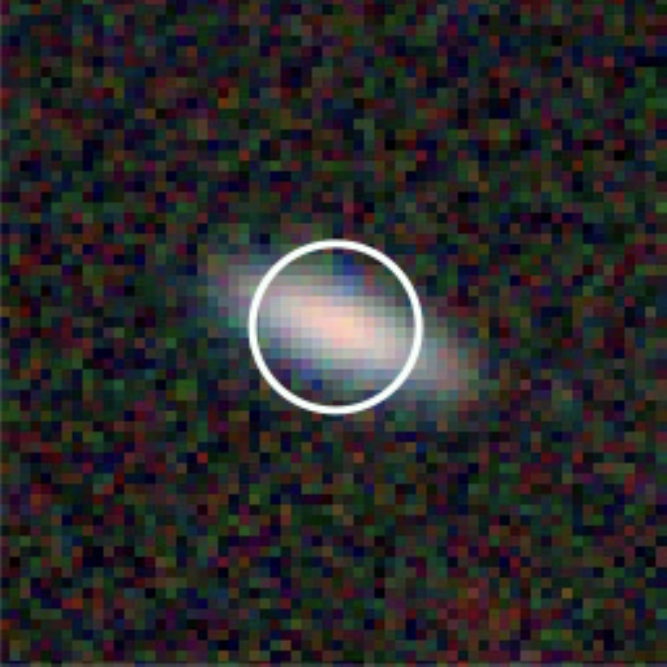}\hspace{0.9in}
\includegraphics[width=0.2\textwidth]{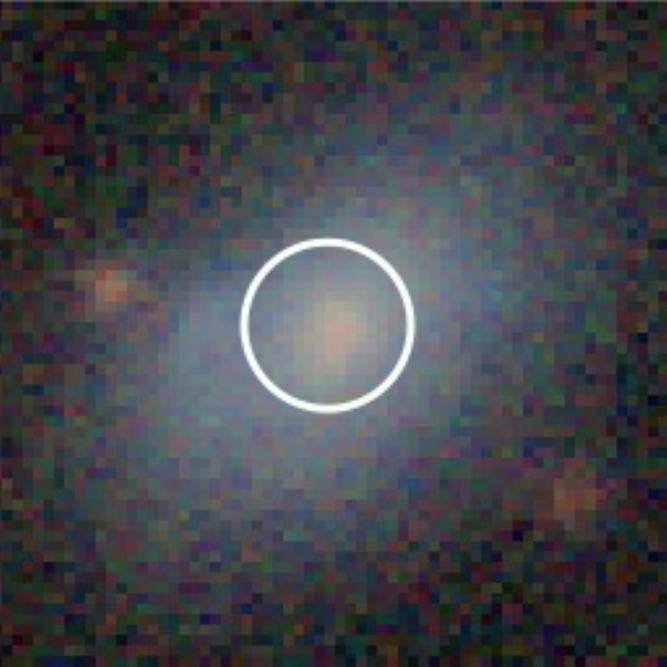}\\
\includegraphics[width=0.31\textwidth]{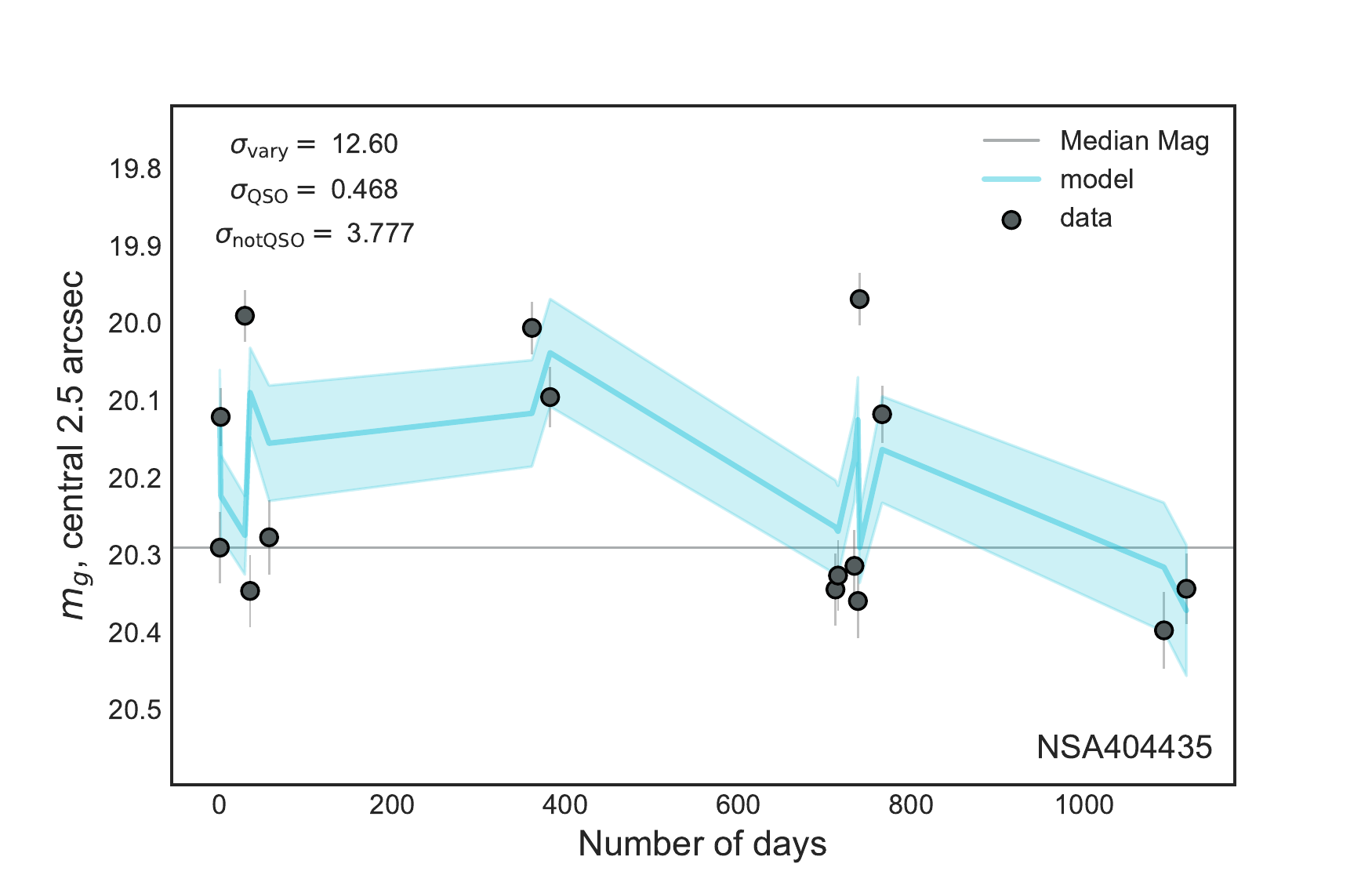}
\includegraphics[width=0.31\textwidth]{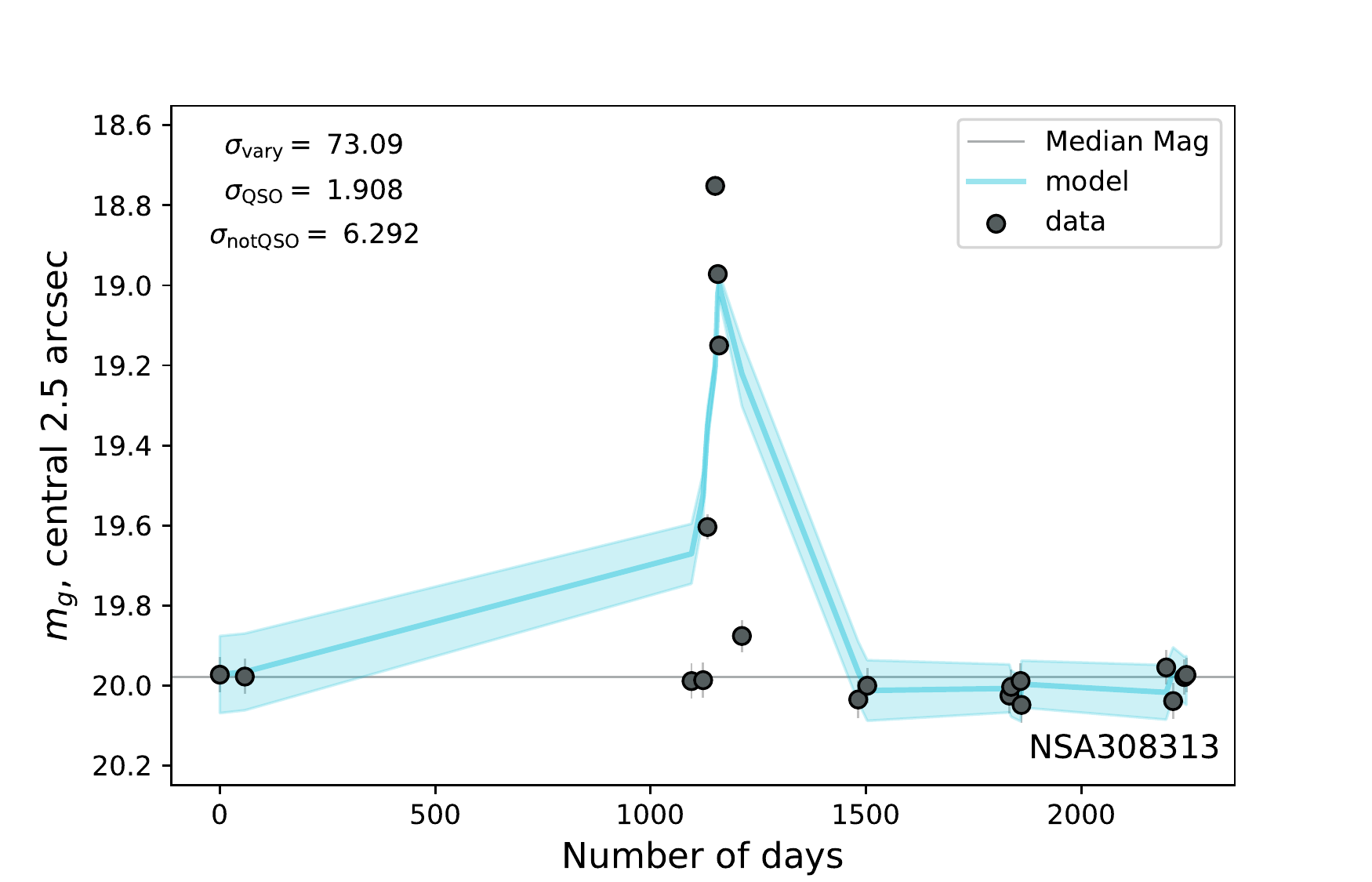}
\includegraphics[width=0.31\textwidth]{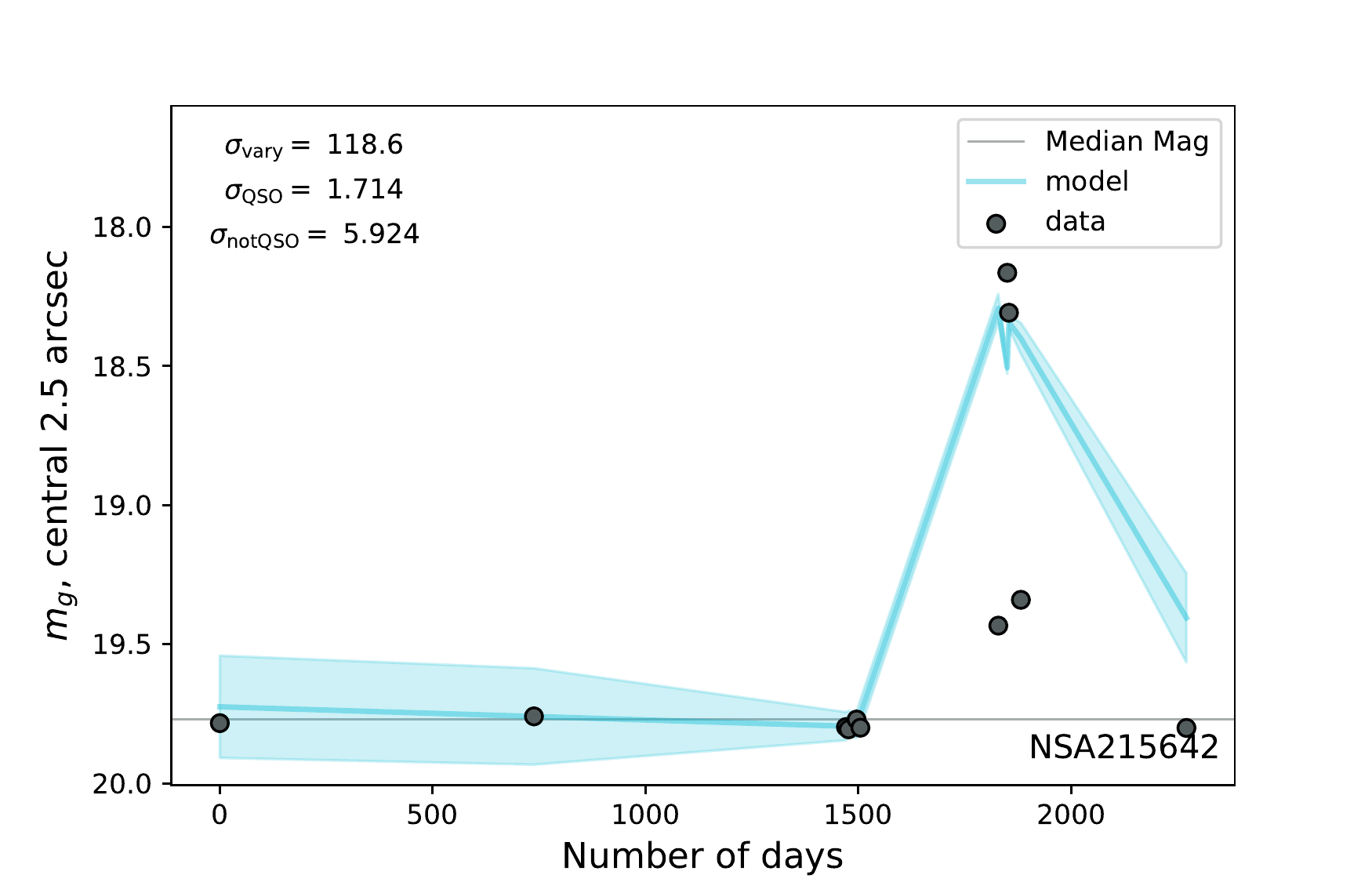}
\caption{Same as Figure~\ref{agn_lcs}, but with examples of galaxies which have variable light curves which are classified as uncharacteristic of an AGN ($\sigma_{\rm var}>2$ and $\sigma_{\rm notQSO}>2$). The top panel shows DECaLs images with a 2.5$''$ circle over the nucleus. The lower panel shows the nuclear \textit{g}-band light curve. }
\label{nonagn_lcs}
\end{figure*}

First, we select objects with a variability significance $\sigma_{\rm var}>2$ and a QSO significance $\sigma_{\rm qso}>2$ (i.e., the significance that the source is variable and that the fit to the damped random walk model is better than the fit to a randomly variable model are $>2\sigma$). We then inspect light curves by eye to remove objects that are identified as variable due to poor image subtractions or likely supernova missed by the variance and QSO light curve cuts. See Figure~\ref{agn_lcs} for examples of light curves with AGN-like variability, and Figure~\ref{nonagn_lcs} for examples of objects with variable light curves uncharacteristic of an AGN.

\begin{figure}
\includegraphics[width=0.5\textwidth]{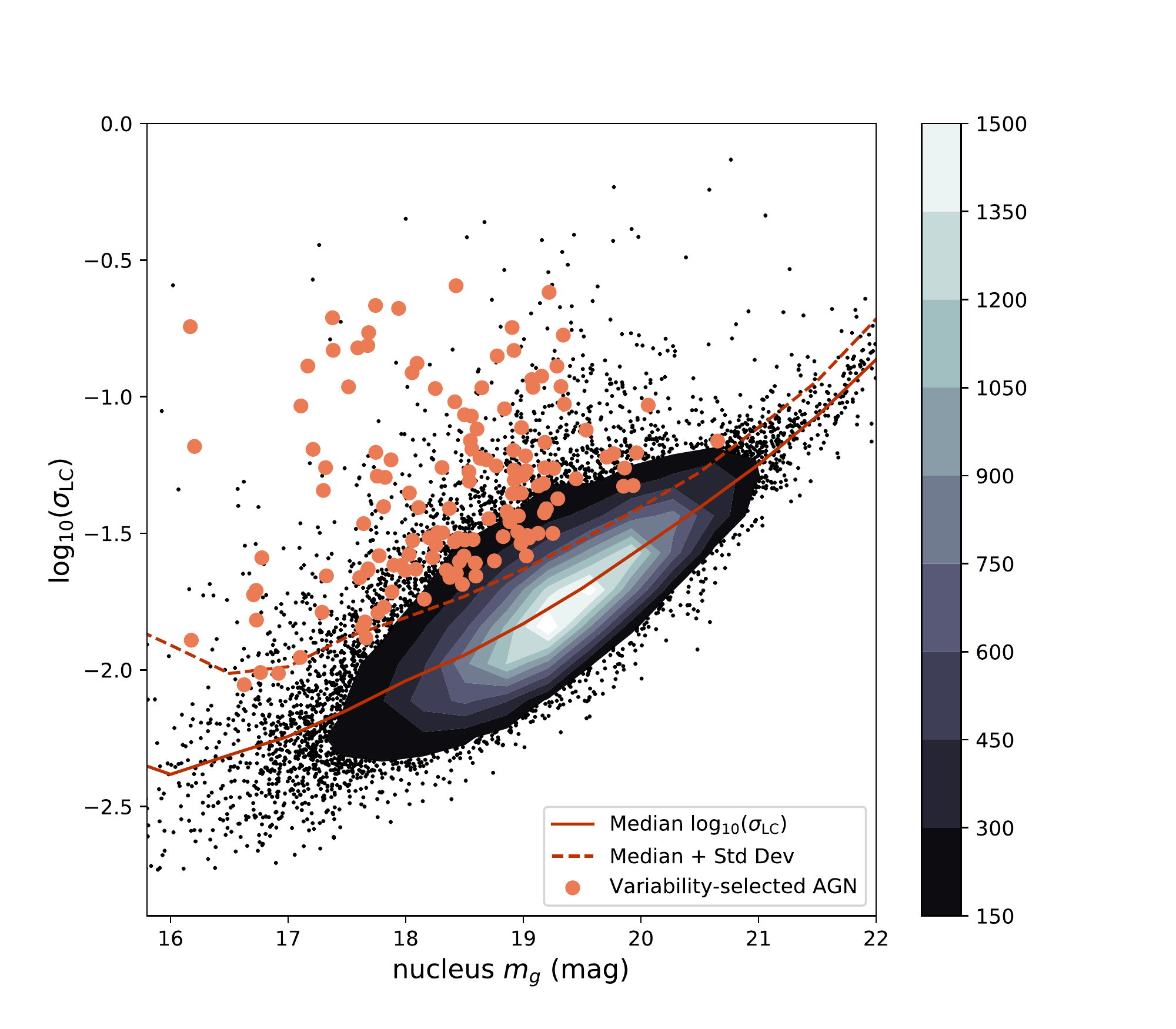}
\caption{RMS scatter about the light curve median value ($\sigma_{\rm LC}$) versus the median nucleus \textit{g}-band magnitude. The contour shading corresponds to the number of objects. The scatter for a given light curve increases for fainter galaxies due to larger flux uncertainties. Galaxies selected as having AGN-like variability are shown as orange circles. The median $\sigma_{\rm LC}$ value for a given magnitude is shown by the solid red line, and the values of $\sigma_{\rm LC}$ which are one standard deviation from the median are shown by the dashed red line. Note that the variability-selected AGN have $\sigma_{\rm LC}$ values at least $\sim1\sigma$ above the typical value for their magnitude bin. }
\label{rmsvmag}
\end{figure}

Figure~\ref{rmsvmag} shows a two-dimensional histogram of light curve standard deviation ($\sigma_{\rm LC}$) versus median nucleus magnitude (with the nucleus defined as the central 2.5$''$). The light curve standard deviation corresponds to the 1-$\sigma$ scatter about a light curve median value. All of the objects identified as variable AGNs by our criteria as variable AGN also happen to have $\sigma_{\rm LC}$ more than one standard deviation above the typical $\sigma_{\rm LC}$ for nuclei of the same magnitude.  

\begin{figure*}
\centering 
\includegraphics[width=0.32\textwidth]{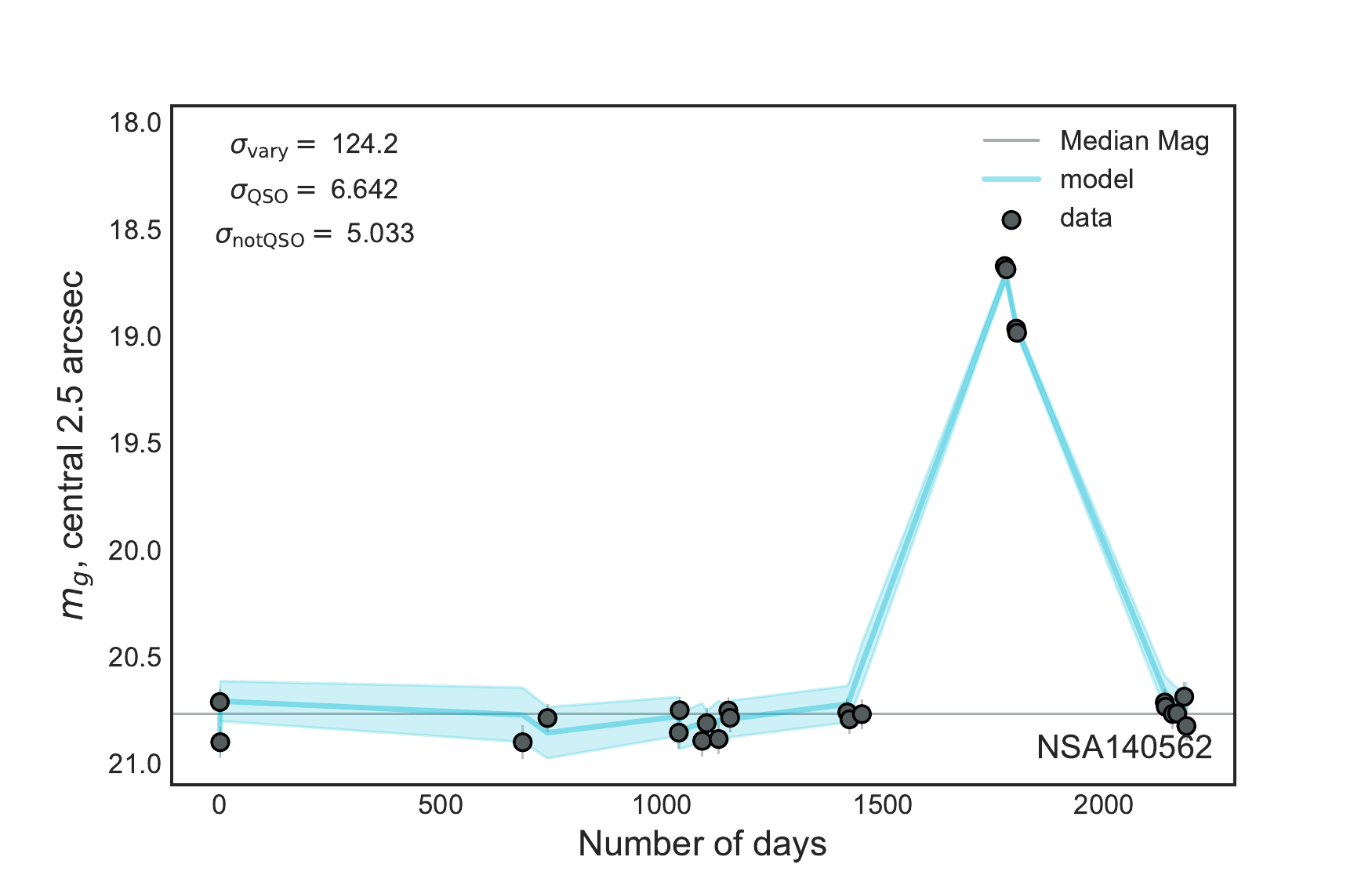}
\includegraphics[width=0.32\textwidth]{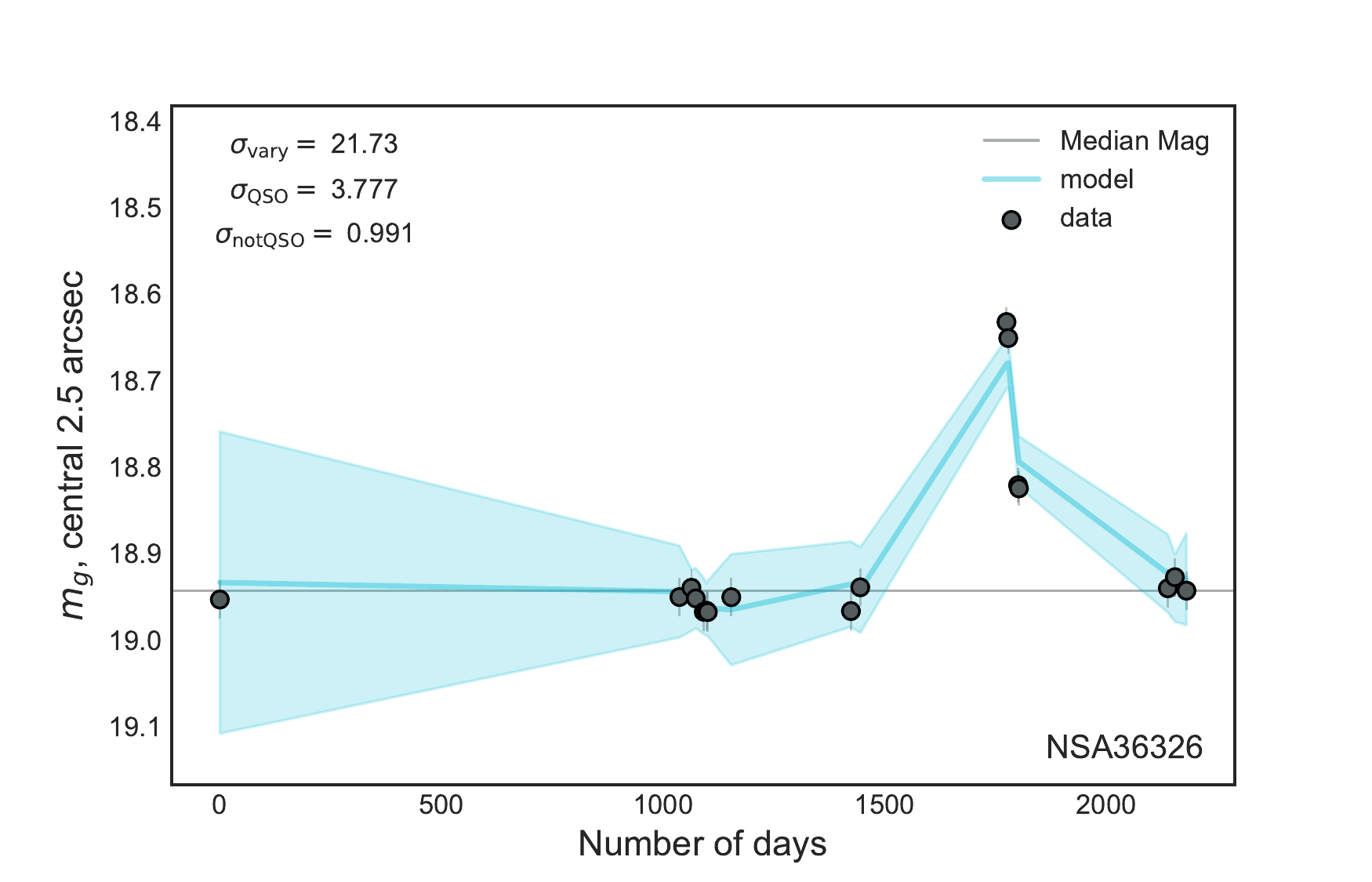}
\includegraphics[width=0.32\textwidth]{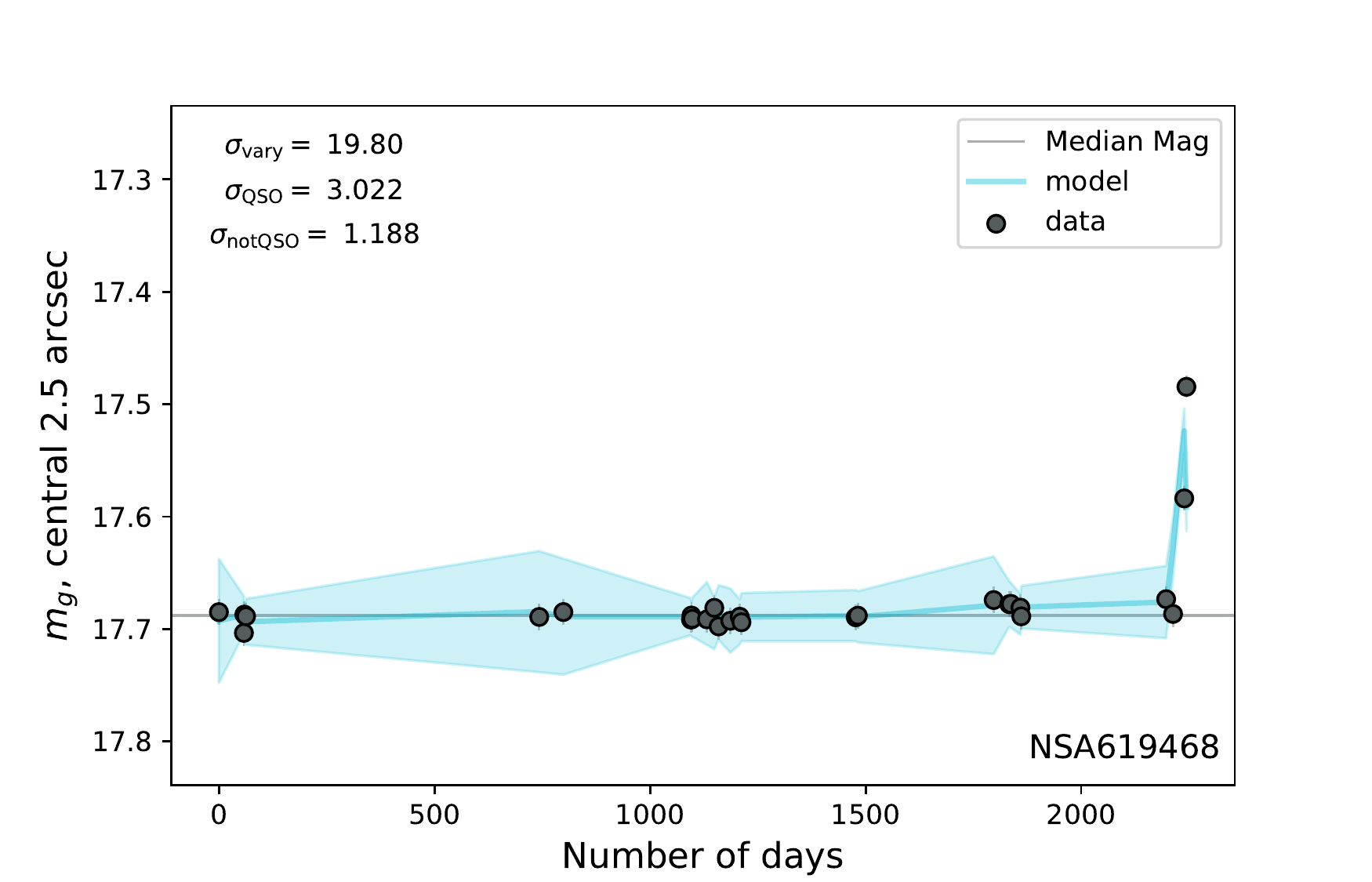}
\caption{Examples of galaxies with SN-like light curves which were classified as having AGN-like variability by our selection criteria. These light curves are relatively constant but for a few elevated points which rise and/or fall on SN-like timescales. }
\label{sn_cand}
\end{figure*}

We inspect each of the light curves selected as having AGN-like variability. Of the 28062 total galaxies with light curves, 201 meet our AGN variability selection criteria. After inspection of individual light curves, we find that 29 of them are SN-like (i.e., show one frame with a bright nucleus, or an obvious rise and fall on a SN-like timescale); 15 are mischaracterized in some way in the NSA (i.e., are either stars or higher redshift AGN); and 21 have problematic image subtractions, such as interference from a nearby saturated star. One system - NSA 29189 - has AGN-like variability but is classified in the literature to contain a luminous blue variable star \citep{2009ApJ...690.1797I}.
This leaves 135 variability selected AGN. Figure~\ref{sn_cand} shows objects with by-eye selected SN-like light curves which were initially categorized as variable AGN.

\subsubsection{Other variable objects}
In additional to the variable AGN, there are 2650 galaxies that have $\sigma_{\rm var}>2$ and $\sigma_{\rm QSO}<2$. These objects tend to fall in one of three categories: stars misclassified in the NSA; genuinely variable objects that are not AGN; and objects with poor image subtractions. When comparing the distribution of the number of data points for the full sample to that for objects with $\sigma_{\rm var}>2$ and $\sigma_{\rm QSO}<2$, the latter skews towards a lower number of data points. This is likely due to the fact that fewer frames generally results in a worse image subtraction. However, it is possible that some are variable AGN that are missed due to poor light curve sampling.

We explore the effect of reducing the number of light curve points on our ability to detect variability by taking variable light curves with more than 20 data points, removing a specified number of data points, and then re-running the light curve analysis. We randomly remove data points to create light curves with 10 and 15 data points, and re-run the light curve analysis. This is done 2000 times each for light curves containing 10 and 15 points. The results vary widely depending on the AGN light curve. For example, for NSA 37000 (23 data points; $\sigma_{\rm var}$=25.4; $\sigma_{\rm QSO}$=5.6), it is classified as a variable AGN 98.5\% of the time with 15 data points, and 87\% of the time with 10 data points. For NSA 40018 (24 data points; $\sigma_{\rm var}$=31.5; $\sigma_{\rm QSO}$=4.2), it is classified as a variable AGN 81\% of the time with 15 data points, and 65\% of the time with 10 data points. However, they are virtually always found to have $\sigma_{\rm var}$>2; the AGN significance tends to fall with decreasing numbers of data points. We require a minimum of 10 data points for our light curve analysis; the median number of data points in a given light curve in our sample is 17. None of our conclusions are affected by the loss of variable AGN due to sampling, though our sample of variability-selected AGN is certainly not complete. 

\subsection{Spectroscopic analysis}

We analyze SDSS spectroscopy for our variability-selected AGN. If a galaxy has more than one spectrum, we use the primary spectrum designated by SDSS. In particular we measure emission line fluxes and search for broad H$\alpha$ emission potentially indicative of dense gas orbiting around a central black hole. Here, we describe the emission line modeling procedure (see also similar fitting procedures described in \citealt{1997ApJS..112..391H, 2004ApJ...610..722G, Reines:2013fj, 2015ApJ...809L..14B, 2016ApJ...829...57}). 

We first create a model for the narrow-line emission using the intrinsically narrow [S II] $\lambda\lambda$ 6713 and 6731 lines. As these are forbidden transitions, they are guaranteed not to be produced in the denser broad line region gas. This narrow-line model is then used to fit the narrow H$\alpha$ emission and the [NII] $\lambda\lambda$6548,6684 lines simultaneously. We allow the width of narrow H$\alpha$ to increase by up to $25\%$ and fix the relative amplitudes of the [NII] lines to laboratory values. Next, an additional Gaussian component representing the broad H$\alpha$ emission is added to the model. If the $\chi^{2}$ value of the fit improves by 20\%, the broad component is kept in the model. We allow up to two Gaussian components to be used to model the broad emission. 

We also fit H$\beta$, [OIII] $\lambda$5007, [OIII]$\lambda$4959, and the [OII]$\lambda\lambda$3726,3729 doublet. We also allow H$\beta$ to be fit with an additional (broad) Gaussian component. In all of our emission line fitting, the continuum is modeled simply as a line across the relevant spectral region.

\section{Variability selected AGNs}

We find 135 galaxies with AGN-like optical photometric variability (see Section 3.2 for selection criteria). The variability amplitudes range from $\sim0.01$ to $0.25$ magnitudes, with a median $1-\sigma$ variability of 0.05 mag. Apparent \textit{g}-band magnitudes of the nuclei range from 15 to 20.5 mag. The detectable level of variability depends on the apparent magnitude of the nucleus, as shown in Figure~\ref{rmsvmag}. In the following, we discuss the host galaxy properties and spectroscopic properties. 

\subsection{Host galaxy properties}

\begin{figure}
\centering
\includegraphics[width=0.45\textwidth]{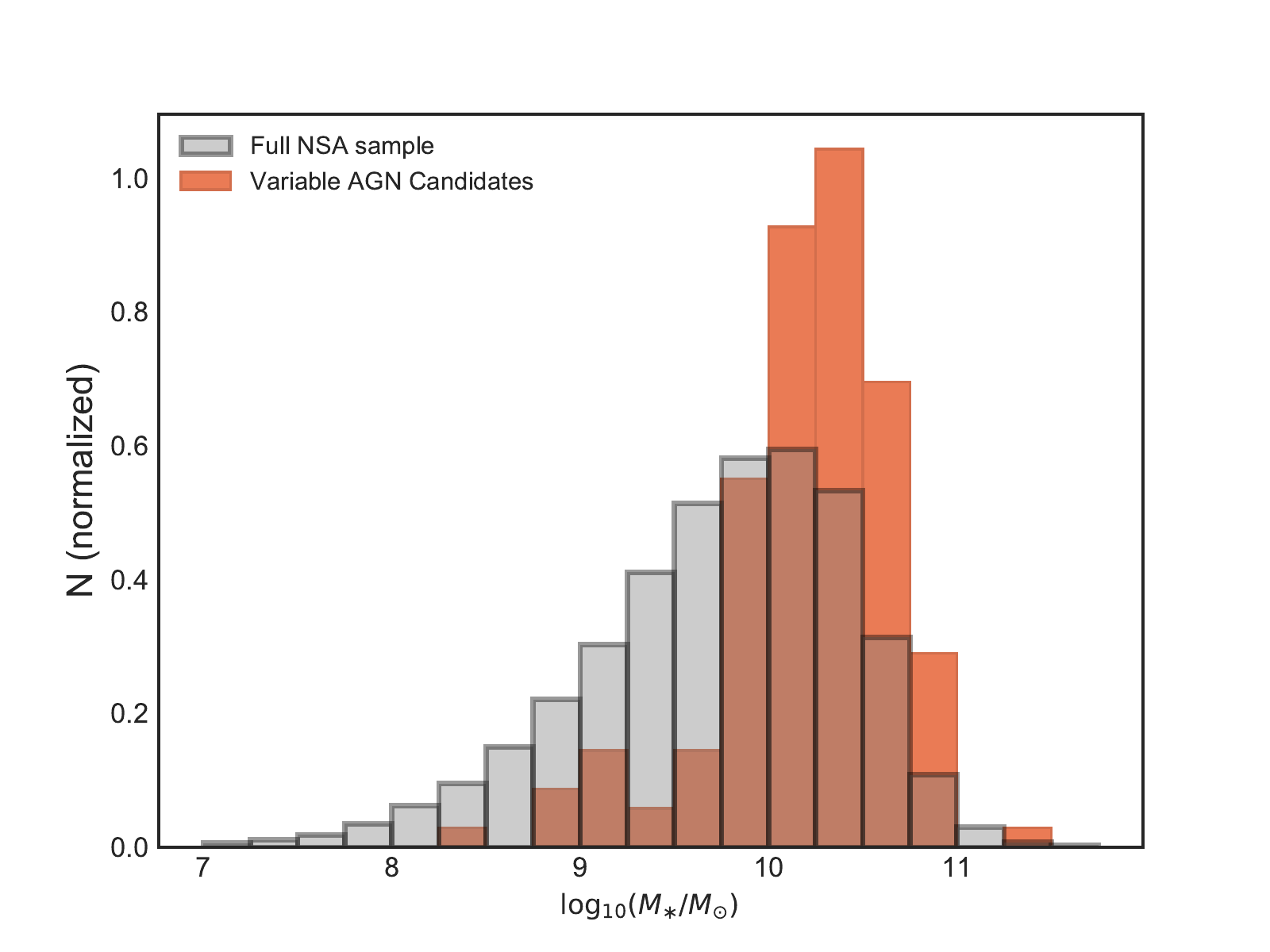}
\caption{Normalized histogram showing the stellar mass distribution of the full NSA parent sample (grey), and of the variability-selected AGNs (orange). The sample of variability-selected AGNs skews towards a higher stellar mass than the underlying parent sample. }
\label{mstarhist}
\end{figure}

\begin{figure}
\centering
\includegraphics[width=0.5\textwidth]{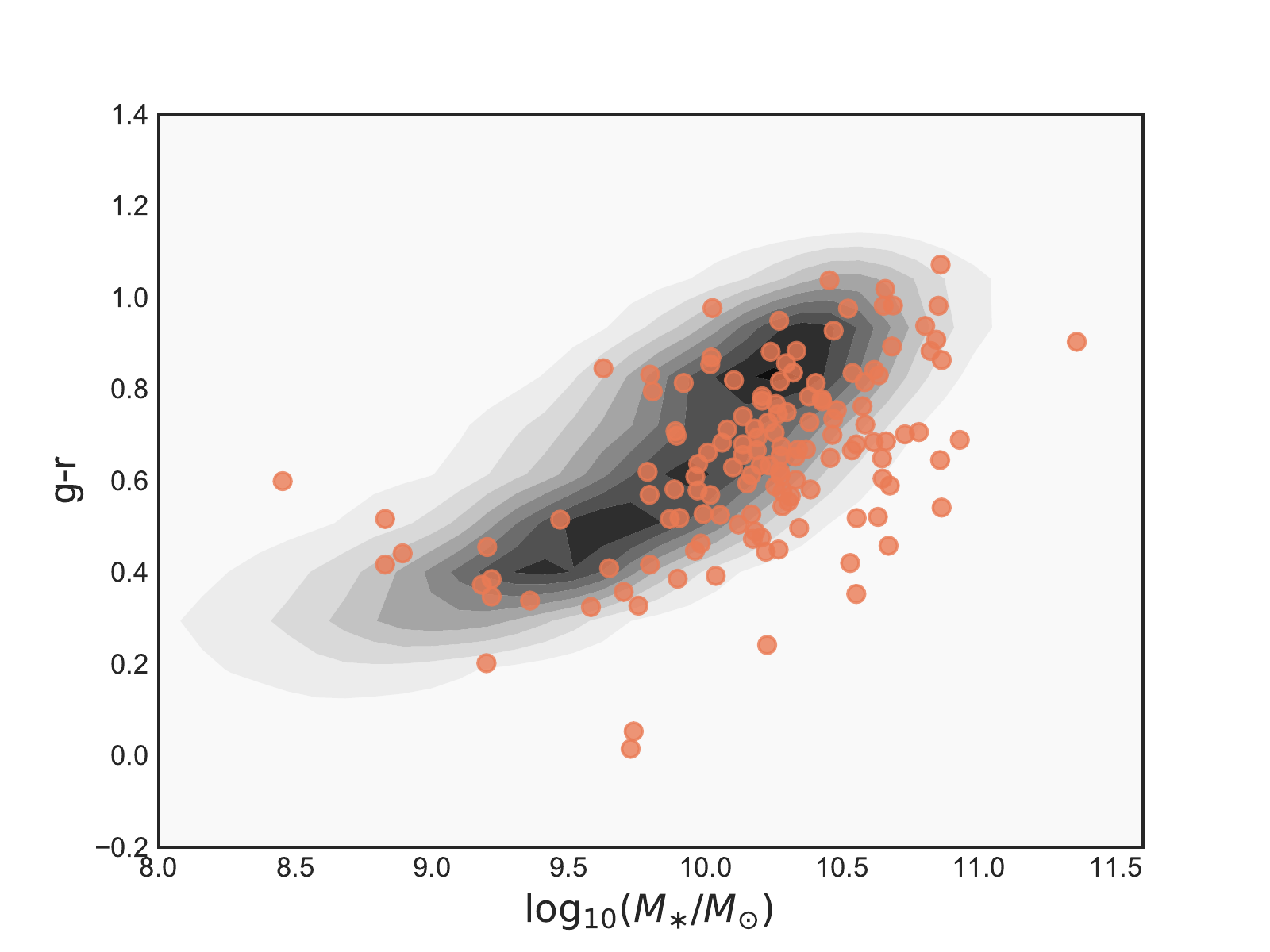}
\caption{Host galaxy $g-r$ color versus stellar mass. Colors were computed using the elliptical Petrosian fluxes given in the NSA, and have been corrected for extinction. The grey contours show the entirety of the NSA sample for which we were able to construct light curves. The variability-selected AGNs are shown as orange data points. Overall, the variability-selected AGNs tend to be bluer than non-variability selected galaxies at a given stellar mass.   }
\label{grfull}
\end{figure}

The host galaxies of the variability-selected AGNs range from $\sim2\times10^{8} - 5\times10^{11}~M_{\odot}$ in stellar mass. Figure~\ref{mstarhist} shows the stellar mass distribution of the 135 variability selected AGN hosts as compared to the parent sample of $\sim28000$ galaxies in the NSA for which we construct light curves. The variability selected AGNs tend towards higher stellar mass; the median stellar mass of the variability selected AGNs is $1.8\times10^{10}~M_{\odot}$, while the parent sample has a median stellar mass of $6.8\times10^{9}~M_{\odot}$. We show the host galaxy $g-r$ colors in Figure~\ref{grfull}. The $g-r$ colors of the variability-selected AGNs range from 0.01 to 1.1, and as a sample tend to be bluer than the parent population (see also \citealt{2014ApJ...780..162M}).

\subsection{Spectroscopic properties}

\begin{figure*}
\centering
\includegraphics[width=0.8\textwidth]{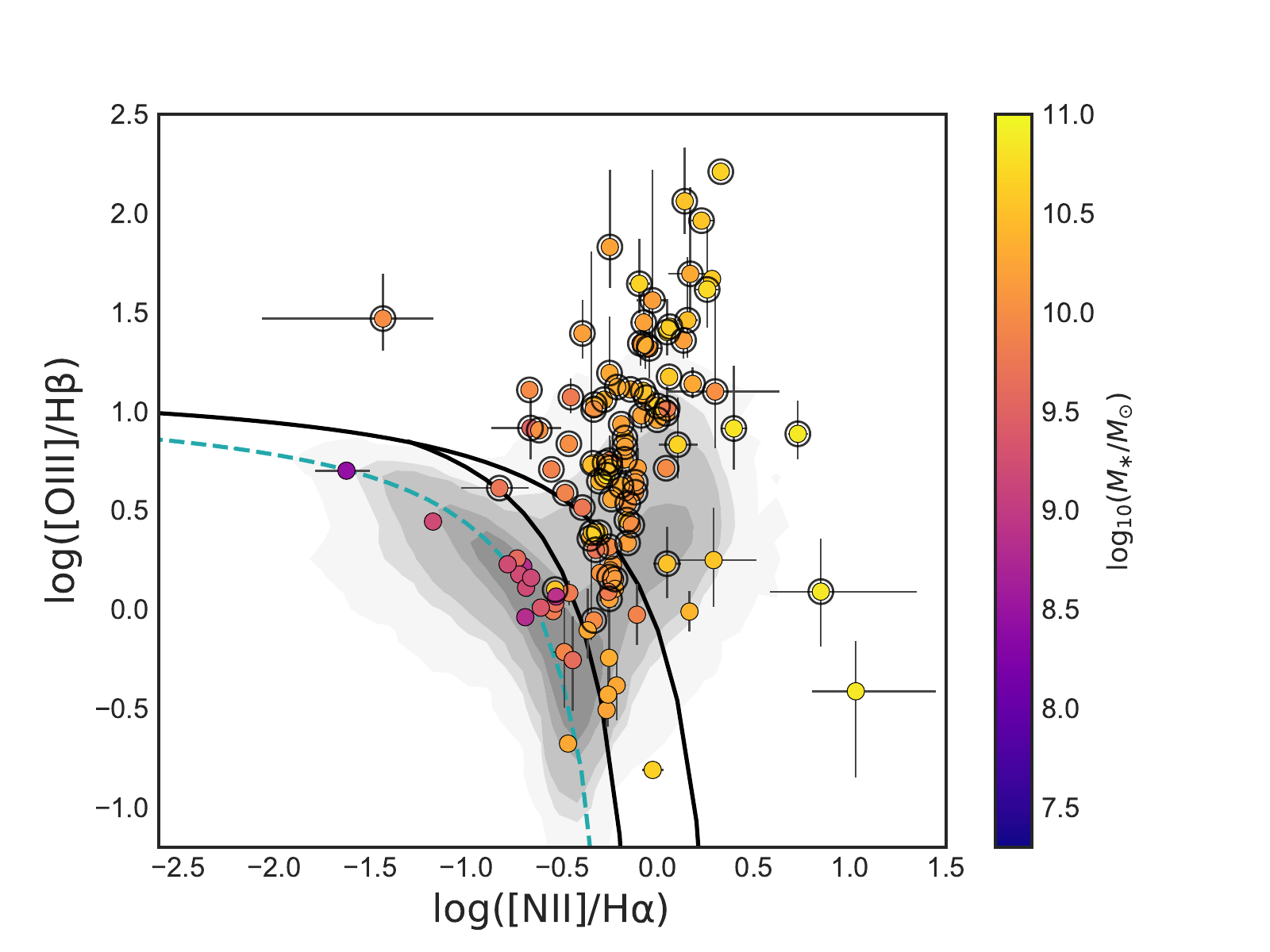}
\caption{BPT diagram showing the positions of the variability-selected AGN. The black lines show the traditional BPT diagram separation lines \citep{1981PASP...93....5B, 2001ApJ...556..121K, 2006MNRAS.372..961K}, and the blue dashed line shows the star forming main sequence \citep{2013ApJ...774..100K}. Data points are color-coded by galaxy stellar mass, obtained from the NSA. 100 galaxies reside in the AGN or composite regions of the BPT diagram, while 20 are in the star-forming region.  }
\label{bpt_full}
\end{figure*}

We show where the variability-selected AGNs sit on the BPT diagram \citep{1981PASP...93....5B} in Figure~\ref{bpt_full}. Of the full sample, 14 could not be plotted due to either a poor quality SDSS spectrum or not having one or more of the relevant emission lines. Of the 121 galaxies for which we could measure H$\beta$, [OIII]$\lambda 5007$, H$\alpha$, and [NII]$\lambda 6584$, most ($85\%$) have narrow emission line ratios placing them in the AGN or composite regions of the BPT diagram (22 BPT composites and 78 BPT AGN). Interestingly, 18 galaxies (or $\sim15\%$ of the sample) have narrow emission line ratios dominated by star formation. These galaxies tend to be the lowest-mass systems in our sample. Only 2/18 in the star forming region have stellar masses greater than $10^{10}~M_{\odot}$; the median stellar mass for these 18 objects is $1.7\times10^{9}~M_{\odot}$. 

\begin{figure}
\centering
\includegraphics[width=0.5\textwidth]{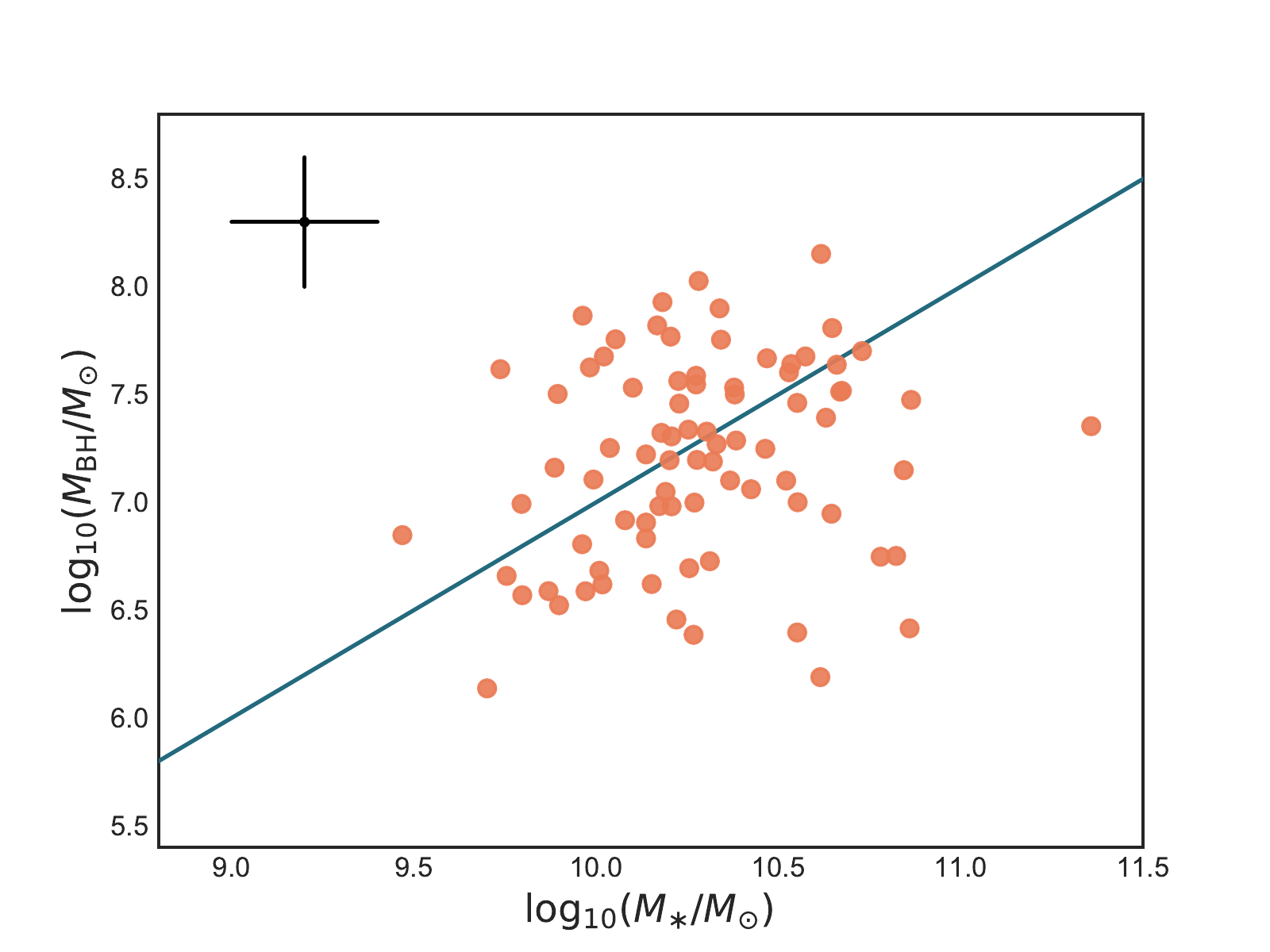}
\caption{Black hole mass versus stellar mass for objects with broad H$\alpha$ emission. The black point in the upper left hand corner shows the typical BH mass uncertainty ($\sim0.3$ dex) and stellar mass uncertainty ($\sim0.2$ dex). The solid blue line corresponds to the canonical BH-to-stellar mass ratio of 0.001 \citep{Kormendy:2013ve}. }
\label{bhmass_mstar}
\end{figure}

We also search for broad H$\alpha$ emission indicative of dense gas orbiting around the central BH. We find that 82/135 of the galaxies with AGN-like variability ($61\%$) have broad H$\alpha$ emission lines. Of the objects with narrow lines in the composite/AGN regions, $\sim80\%$ also show broad H$\alpha$. We estimate single-epoch virial BH masses using the broad H$\alpha$ FWHM and luminosity \citep{2005ApJ...630..122G}. The FWHM of the broad component gives a characteristic velocity of gas in the broad line region, and the distance to the broad line region is correlated with the continuum luminosity at 5100~\AA~ \citep{2009ApJ...705..199B, 2010ApJ...716..993B, 2013ApJ...767..149B}, which in turn is correlated with the luminosity of broad H$\alpha$. BH masses estimated in this way have systematic uncertainties of roughly $0.3$ dex (or a factor of $\sim2$). Specifically, we use the BH mass formula given in \cite{Reines:2013fj}. The broad H$\alpha$ luminosities range from $\sim1\times10^{39} - 3\times10^{42}~\rm{erg~s^{-1}}$ and the FWHM range from $\sim900 - 6500~\rm{km~s^{1}}$. We calculate BH masses ranging from from $1\times10^{6} - 1\times10^{8}~M_{\odot}$, and the median $M_{\rm BH}$ - to - $M_{\ast}$ ratio is $8\times10^{-4}$, close to the canonical value of 0.001 \citep{Kormendy:2013ve}. In Figure~\ref{bhmass_mstar} we show the BH mass versus host stellar mass for the objects with broad H$\alpha$ emission.

\subsection{Variable AGN fraction versus stellar mass}

\begin{figure*}
\centering
\includegraphics[width=\textwidth]{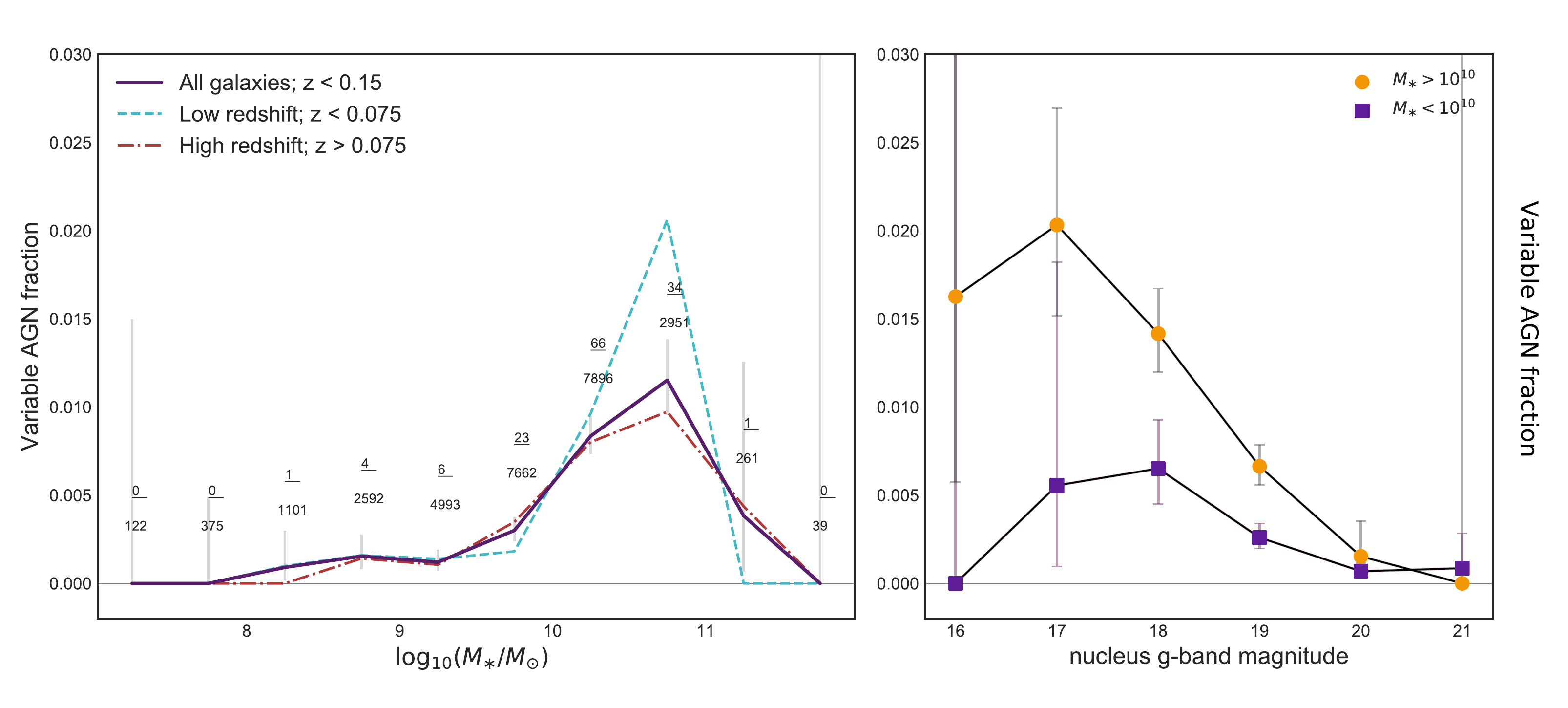}
\caption{\textit{Left:} Variable AGN fraction versus stellar mass. The active fraction corresponds to the fraction of galaxies with nuclear variability consistent with the presence of an AGN (see text for details). The solid purple line shows the full sample with $z>0.15$, while the dashed blue and dashed-dot red lines show the sub-samples with $z<0.075$, and $z>0.075$, respectively. Also shown are the number of detections over the total number of galaxies in each bin, for the full sample out to z=0.15. \textit{Right:} Active fraction versus median light curve \textit{g}-band magnitude for galaxies with $M_{\ast}>10^{10}~M_{\odot}$ (blue circles) and $M_{\ast}<10^{10}~M_{\odot}$ (orange squares). The active fraction is computed in bins one magnitude wide; the plotted magnitude value is the center of each bin. Uncertainties are computed assuming a binomial distribution and correspond to a $1-\sigma$ confidence limit. }
\label{agnvmstar}
\end{figure*}

We present the measured active fraction (i.e., the fraction of galaxies with AGN-like variability) as a function of stellar mass in the left panel of Figure~\ref{agnvmstar}. At face value, Figure~\ref{agnvmstar} suggests that there is a steep increase in the fraction of variable AGN as a function of stellar mass. However, there are a number of observational biases which must be taken into account. As shown in Figure~\ref{rmsvmag}, lower levels of variability are detectable in brighter nuclei. The median nucleus \textit{g}-band magnitude for galaxies $M_{\ast}>10^{10}~M_{\odot}$ sample is $\sim18.5$, while for galaxies with $M_{\ast}<10^{10}~M_{\odot}$ the median magnitude is $\sim19.5$.
The right side of Figure~\ref{agnvmstar} shows the active fraction as a function of nucleus \textit{g}-band magnitude for two different mass bins. At a given nucleus magnitude, the active fraction tends to be lower for low-mass galaxies than for higher-mass galaxies (see Section 6 for a discussion of potential explanations). At $m_{g}\approx18$, the variable AGN fraction is roughly twice as high in the sample with $M_{\ast}>10^{10}~M_{\odot}$ than for the low-mass sample. 

\begin{deluxetable}{ccc}   
\tablecaption{Active fraction across apparent magnitude bins\label{tab:activefrac_mag_split}}
\tablecolumns{3}
\tablenum{3}
\tablewidth{0pt}
\tablehead{
\colhead{Mag. bin } &
\colhead{$\rm{F_{AGN}}$ ($M_{\ast}<10^{10}M_{\odot}$) } &
\colhead{$\rm{F_{AGN}}$ ($M_{\ast}>10^{10}M_{\odot}$) } 
}
\startdata
16 & 0/22 & 2/123 \\
17 & 1/180 & 15/738 \\
18 & 10/1535 & 41/2895  \\
19 & 17/6535 & 39/5878 \\
20 & 6/7292 & 2/1305 \\
21 & 1/1163 & 0/55 \\
\enddata
\tablecomments{This table gives the number of high and low mass galaxies in a given magnitude bin (denominator) and the number of objects with AGN-like variability in that bin. }
\end{deluxetable}

\section{Variability as a tool for identifying AGNs in low-mass galaxies} 

Above stellar masses of $~10^{10}~M_{\odot}$, virtually all of the variability-selected AGNs are also narrow-line AGN. However, at $M_{\ast}<10^{10}~M_{\odot}$, roughly half of the sample has narrow emission line ratios dominated by star formation (see Figure~\ref{bpt_full}). Motivated by the potential for using low-level optical variability to identify low-mass AGNs, in this section we examine further the 35 ``low-mass" ($M_{\ast}<10^{10}~M_{\odot}$) galaxies with AGN-like variability and explore how they compare to low-mass AGNs identified via other techniques. We caution that these objects should generally be regarded as candidates; deeper, higher-spatial resolution spectroscopic follow-up is underway for this sample to better isolate emission from the nucleus. All variability-selected galaxies with $M_{\ast}<10^{10}~M_{\odot}$ are listed in Table~\ref{tab:active_samp}. The galaxies range in $g-r$ color from $\sim0-1$, with a median $g-r$ = 0.46. See the Appendix for images and light curves for all objects with $M_{\ast}<10^{10}~M_{\odot}$.

\begin{deluxetable*}{cccccccccccc} 
\tablecaption{Candidate variability-selected AGN in low-mass galaxies \label{tab:active_samp}}
\tablecolumns{11}
\tablenum{1}
\tablewidth{0pt}
\tabletypesize{\footnotesize}
\tablehead{
\colhead{NSAID \tablenotemark{a} } &
\colhead{RA} &
\colhead{Dec } &
\colhead{Redshift} &
 \colhead{$\log_{10}(\frac{M_{\ast}}{M_{\odot}})$} &
 \colhead{$m_{g}$} & 
\colhead{RMS} & 
\colhead{$\sigma_{var}$}  &
\colhead{$\sigma_{QSO}$}  & 
 \colhead{$\sigma_{noQSO}$} &
 \colhead{BPT Class}  & 
 \colhead{$\log_{10}(\frac{M_{\rm{BH}}}{M_{\odot}})$ } }
\startdata
31861 & 358.35750 & -0.506680 & 0.0234 & 6.73E+08 & 18.574 & 0.0300 & 3.753 & 2.297 & 0.025 & SF & -- \\
32083 & 0.47615 & 0.125693 & 0.1395 & 7.69E+09 & 19.072 & 0.1155 & 17.503 & 2.951 & 1.447 & AGN & 7.2   \\
32653  & 2.36003 & 0.325606 & 0.1138 & 2.94E+09 & 19.963 &  0.0623 & 3.823 & 2.705 & 0.216 & AGN & 6.8  \\
33205 & 4.27033 & -0.493348 & 0.0672 & 9.35E+09 & 17.653 & 0.0150 & 2.246 & 2.346 & 0.001 & Comp & 6.6 \\
33611 & 6.96605 & 0.794382 & 0.0821 & 6.26E+09 & 17.994 & 0.0231 & 4.25 & 2.674 & 0.004 & SF& --  \\
33835 & 8.39416 & 0.048925 & 0.1067 & 8.03E+09 & 19.179 & 0.0376 & 3.87 & 2.753 & 0.004 & SF & -- \\
33851 & 8.15918 & -1.009817 & 0.0917 & 9.15E+09 & 17.826 & 0.0507 & 14.746 & 3.762 & 0.196 & Comp & 6.8  \\
34071 & 11.38946 & -0.969121 & 0.1376 & 7.40E+09 & 18.920 & 0.1477 & 28.678 & 4.81 & 0.928 & AGN &  6.6 \\
34281 & 11.51401 & 0.636591 & 0.1246 & 3.81E+09 & 18.346 & 0.0232 & 2.853 & 2.055 & 0.012 & SF & -- \\
34728 & 14.01926 & 0.592550 & 0.0560 & 7.76E+09 & 17.884 & 0.0193 & 2.721 & 2.122 & 0.004 & Comp& -- \\
35245 & 16.14294 & -1.042719 & 0.0746 & 1.65E+09 & 18.923 & 0.0495 & 8.225 & 3.044 & 0.118 & SF & -- \\
35813 & 18.74797 & 0.245735 & 0.0426 & 6.73E+08 & 18.953 & 0.0320 & 3.457 & 3.167 & 0.0 & SF & -- \\
35920 & 19.87115 & -0.144376 & 0.0901 & 7.91E+09 & 17.875 & 0.0587 & 18.336 & 2.54 & 1.491 & Comp & 6.5 \\
36541 & 21.64659 & 0.144874 & 0.1194 & 6.14E+09 & 19.349 & 0.0940 & 15.997 & 4.408 & 0.093 & SF & -- \\
38331 & 33.44361 & -0.831365 & 0.0933 & 4.44E+09 & 18.315 & 0.0318 & 6.151 & 4.267 & 0.0 & SF & -- \\
38720 & 36.53416 & -0.888560 & 0.1065 & 5.68E+09 & 18.200 & 0.0304 & 6.965 & 2.856 & 0.05 & AGN & 6.7 \\
39920 & 44.96551 & -0.256340 & 0.1014 & 9.40E+09 & 19.219 & 0.2410 & 46.447 & 5.929 & 1.741 & -- & 7.8  \\
40018 & 45.70326 & 0.358798 & 0.1070 & 7.84E+09 & 19.340 & 0.1680 & 31.455 & 4.169 & 2.938 & AGN & 7.5  \\
40170 & 48.25895 & -0.686379 & 0.1312 & 6.24E+09 & 19.535 & 0.0755 & 9.191 & 3.884 & 0.079  & AGN & 7.0 \\
40182 & 47.61596 & -0.830791 & 0.0802 & 5.46E+09 & 16.168 & 0.180 & 109.36 & 4.362 & 4.66 &  Comp & 7.6  \\
40677 & 49.42965 & 0.326904 & 0.0687 & 5.02E+09 & 17.961 &  0.024 & 6.432 & 2.851 & 0.0133 & Comp & 6.1   \\
112799 & 346.84687 & 0.285572 & 0.1128 & 9.83E+09 & 18.686 & 0.0588 & 12.278 & 3.206 & 0.587 & AGN & 7.1  \\
113167 & 354.59937 & 1.226183 & 0.0637 & 2.28E+09 & 18.375 & 0.0218 & 2.897 & 3.604 & 0.0 &  SF& -- \\
114482 & 17.68218 & 0.423656 & 0.0823 & 8.32E+09 & 18.958 & 0.0366 & 3.109 & 2.402 & 0.004 & -- & -- \\
114617 & 20.15146 & 0.601762 & 0.1249 & 7.80E+08 & 19.861 & 0.0548 & 2.911 & 2.615 & 0.005 & SF & -- \\
115553 & 43.73940 & 0.432075 & 0.1077 & 1.58E+09 & 19.251 & 0.0316 & 2.314 & 3.278 & 0.0  & SF & -- \\
141699 & 47.72545 & 1.030173 & 0.1233 & 6.41E+09 & 19.934 & 0.0473 & 2.032 & 2.102 & 0.008 & Comp &  \\
204319 & 330.36643 & -0.800648 & 0.1102 & 9.18E+09 & 19.083 & 0.1083 & 15.312 & 2.924 & 0.8 & AGN & 7.9 \\
214788 & 27.36857 & 1.028639 & 0.0581 & 1.59E+09 & 19.182 & 0.0549 &  7.657 & 2.835 & 0.088 & SF & -- \\
215903 & 8.240055 & -0.454011 & 0.1070 & 1.65E+09 & 19.852 & 0.0470 & 2.204 & 2.021 & 0.02  & SF & -- \\
217124 & 331.50298 & 0.913803 & 0.0411 & 2.84E+08 & 20.062 & 0.0931 & 5.603 & 3.06 & 0.028 & SF & -- \\
305046 & 19.50434 & -0.496008 & 0.0911 & 6.27E+09 & 19.293 & 0.0423 & 3.738 & 2.765 & 0.01 & AGN  & 6.6 \\
305680 & 29.26929 & -1.179740 & 0.0823 & 1.52E+09 & 19.196 & 0.0389 &  4.13 & 2.284 & 0.028 & SF & -- \\
310445 & 46.12571 & 0.344850 & 0.1260 & 4.23E+09 & 20.652 & 0.0688 & 2.502 & 3.712 & 0.0 & SF & -- \\
583482 & 336.14706 & -0.184434 & 0.0580 & 9.61E+09 & 17.168 & 0.1295 & 61.768 & 3.413 & 3.495 & AGN & 7.6  \\
\enddata
\tablenotetext{a}{NSA ID from \textbf{nsa\_v1\_0\_1.fits} catalog. }
\tablecomments{Galaxies with light curves that meet our selection criteria for having an AGN, i.e., $\sigma_{\rm var}>2$ and a QSO significance $\sigma_{\rm qso}>2$. Galaxies are ordered from lowest to highest stellar mass. }
\end{deluxetable*}

\subsection{SDSS detection limit}

In order to determine what population of BHs we are sensitive to, in this section we explore the AGN detection limit for the low-mass sample, given the depth of the SDSS observations and the difference imaging technique. For each light curve, we compute the median nuclear \textit{g}-band magnitude and the standard deviation of the individual measurements about the median magnitude ($\sigma_{\rm LC}$). Thus for a given magnitude bin, we can compute the median light curve standard deviation of objects in that bin. For example, the median $\sigma_{\rm LC}$ for nuclei with an apparent \textit{g}-band magnitude of 18 is $0.01$ mag. For nuclei with apparent \textit{g}-band magnitude of 20, the median $\sigma_{\rm LC}$ is $0.035$ mag. The nuclei selected as having AGN-like variability all have light curve standard deviations $\gtrsim1\sigma$ above the median $\sigma_{\rm LC}$ \textit{for their magnitude bin} (see Figure~\ref{rmsvmag}). 

We next compute the typical galaxy magnitude within the $2.5''$ nuclear aperture as a function of redshift. For a given redshift bin, we use the median S{\'e}rsic profile index (as given by the NASA-Sloan Atlas) to determine the typical fraction of galaxy light contained within 2.5$''$. We then use this value to compute the typical \textit{g}-band magnitude within 2.5$''$. As mentioned above, each magnitude has a light curve standard deviation value above which the objects classified as AGN are found.  For each redshift, we convert the variations in the apparent magnitude to variations in the luminosity. Then, assuming that the observed variations are on order 10\% or 50\% of the total AGN luminosity, we compute the total \textit{g}-band luminosity an AGN would need to have to be detectable in a typical galaxy of a given stellar mass as a function of redshift (top panels of Figure~\ref{det_limits}). 

We stress that this analysis comprises a rough estimate of our detection limits -- at any stellar mass, there exist galaxies with a wide range of S{\'e}rsic indices and luminosities and thus a range of AGN luminosities/BH masses that could be detected. Nevertheless, based on this calculation, we can roughly expect to detect a $\sim10^{5}~M_{\odot}$ BH in a low-mass galaxy accreting at its Eddington luminosity and varying at the $10\%$ level out to $z\approx0.05$. For luminosity variations on the 50\% level, the same should be detectable to  $z\approx0.1$.

\subsection{Expected number of detections}

We next explore how many variable AGN we could expect to detect in our $M_{\ast}<10^{10}~M_{\odot}$ sample, given local scaling relations between BH mass and the host galaxy. Using the relations between BH mass and stellar mass from \cite{2015ApJ...813...82R}, we assign each galaxy a BH mass given its stellar mass in the NSA (and given the intrinsic scatter in the $M_{\rm BH}-M_{\ast}$ relation). \cite{2015ApJ...813...82R} find different $M_{\rm BH}-M_{\ast}$ relations for inactive elliptical/S0 galaxies and local active broad line AGN; we test both relations. For each of the $M_{\rm BH}-M_{\ast}$ relations, we then compute bolometric luminosities by applying the Eddington ratio distribution given by \cite{2018MNRAS.476..436B} for low-mass galaxies ($10^{8}-10^{10}~M_{\odot}$). We can then determine how many galaxies would be detected, assuming the \textit{g}-band luminosity is 5\% of the total bolometric luminosity, and luminosity variations at the 10\% or 50\% level. 

\begin{figure*}
\centering
\includegraphics[width = 0.45\textwidth]{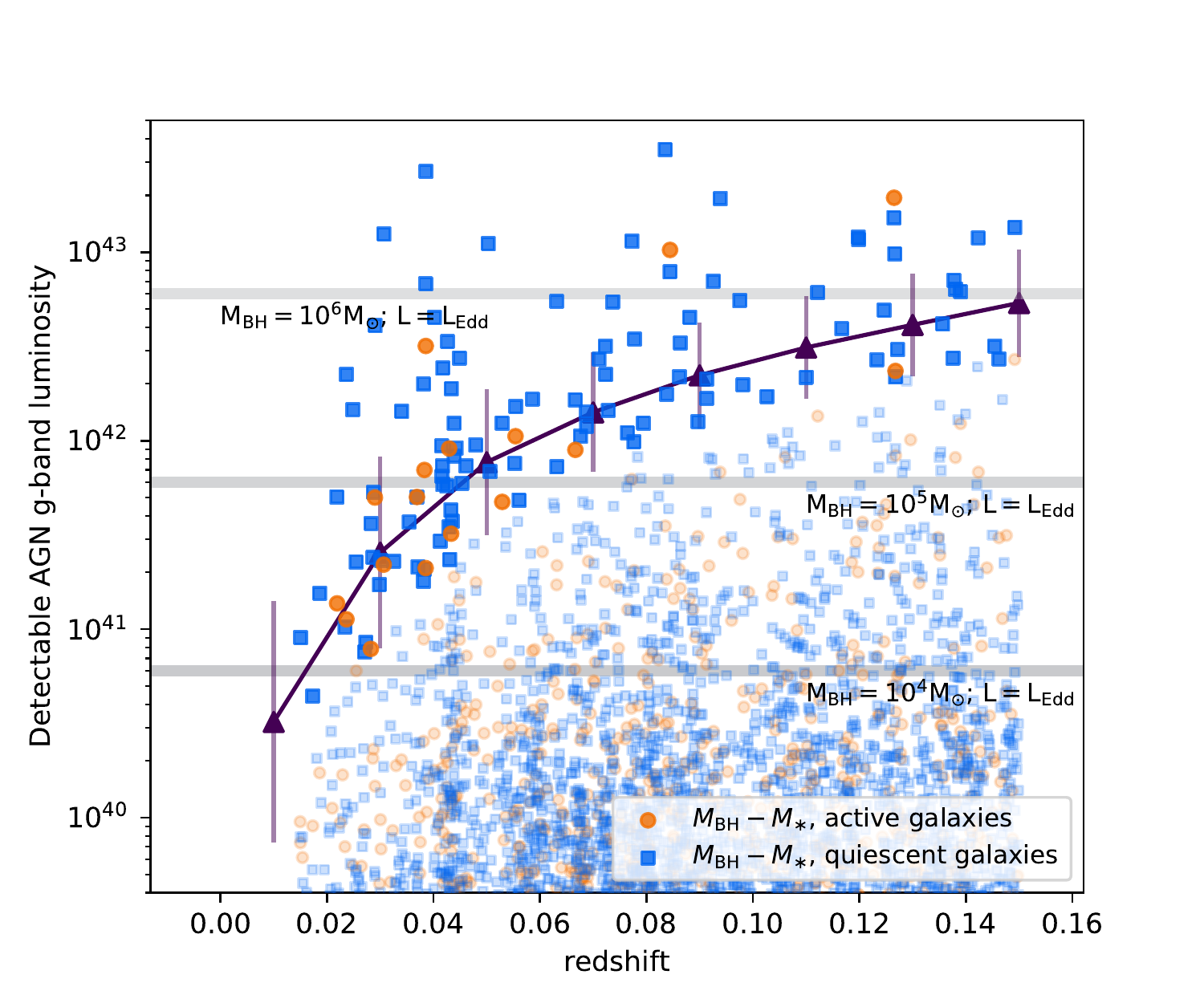}
\includegraphics[width = 0.45\textwidth]{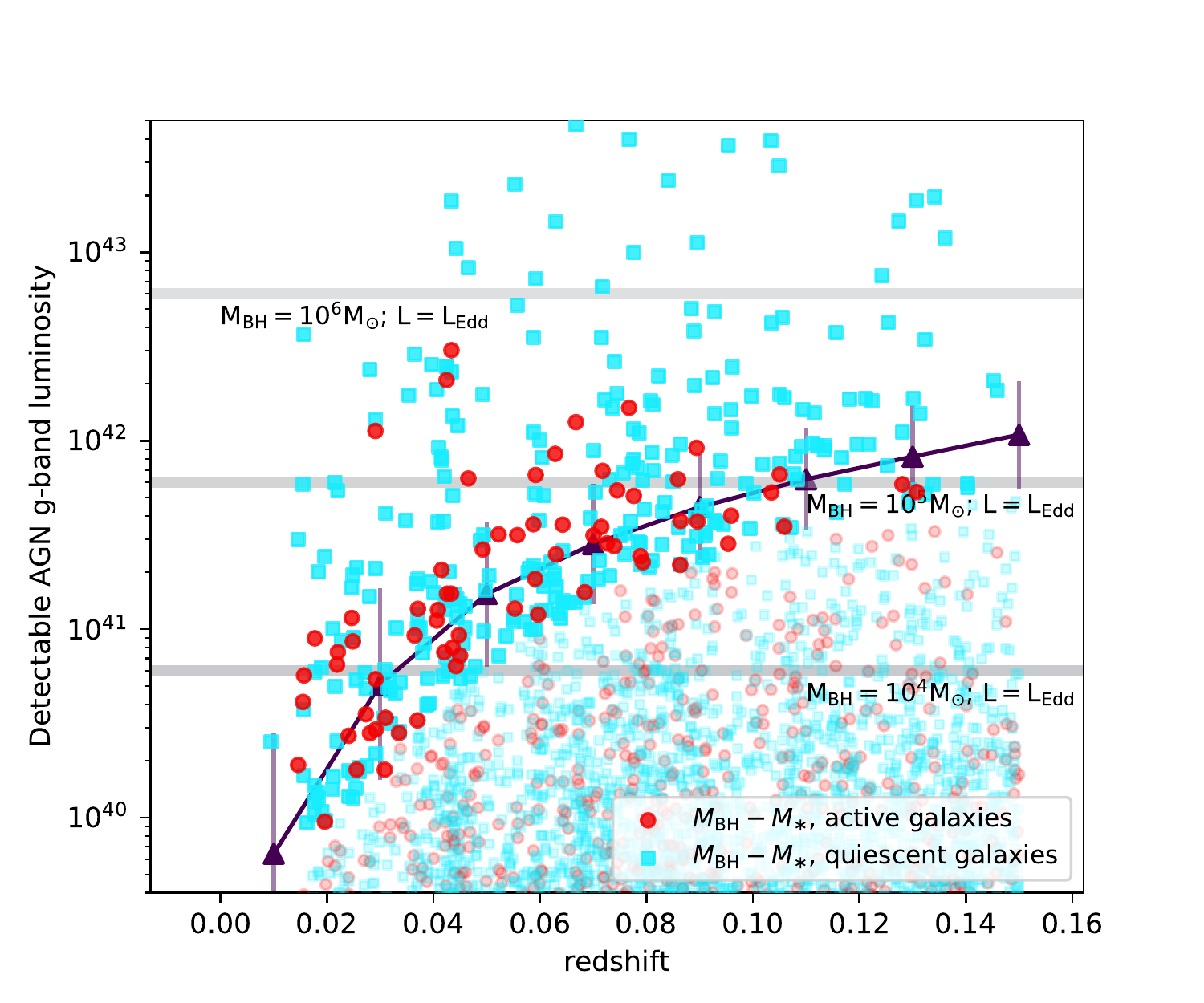}\\
\includegraphics[width = 0.4\textwidth]{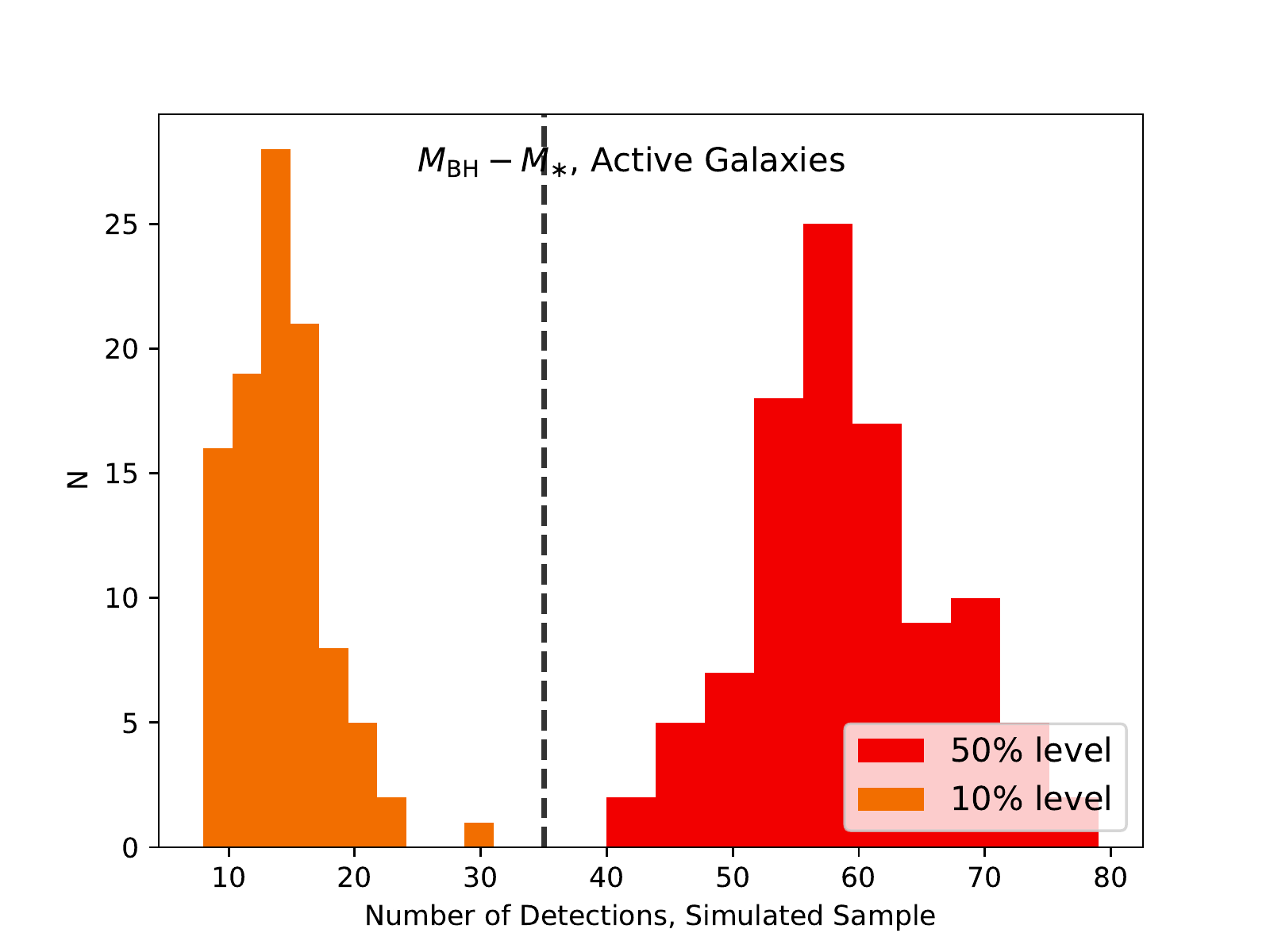}
\includegraphics[width = 0.4\textwidth]{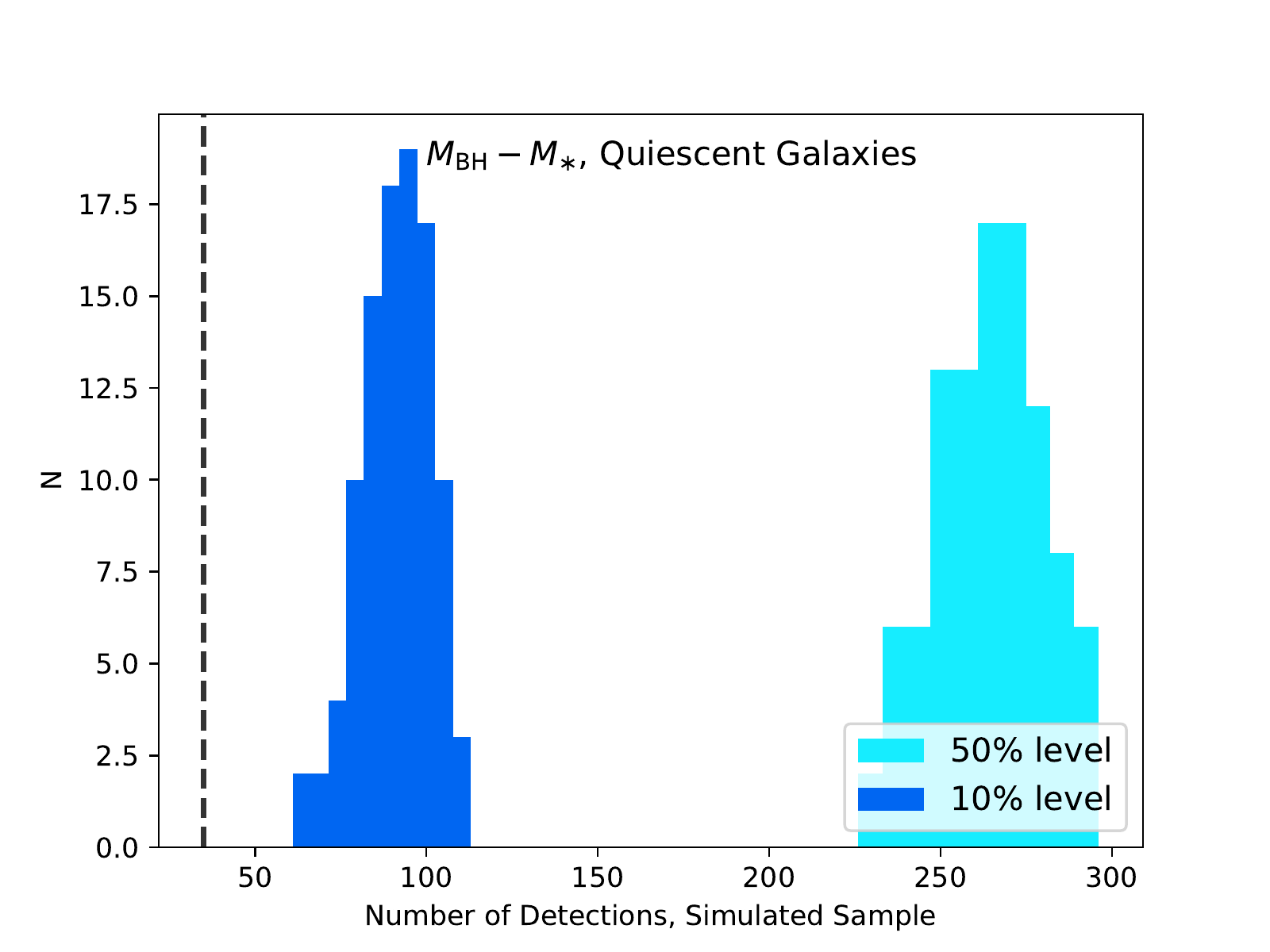}
\caption{\textit{Top row:} Detectable AGN \textit{g}-band luminosity versus redshift. The purple points show the minimum \textit{g}-band luminosity an AGN would need to have to be detectable as a function of redshift, if the luminosity is varying at the 10\% level (left) and 50\% level (right). Gray horizontal lines mark the \textit{g}-band luminosities of BHs ranging from $10^{4-6}~M_{\odot}$, if they are accreting at their Eddington luminosity (and given that $\sim5\%$ of the luminosity is emitted in the \textit{g}-band, based on AGN SEDs from \citealt{2006ApJS..166..470R}). In each plot we also show simulated samples of BHs, given the the $M_{\rm BH}-M_{\ast}$ relations in \cite{2015ApJ...813...82R} and an Eddington ratio distribution \citep{2018MNRAS.476..436B}. The faded points would not be detected based on the minimum detection luminosity, while the darker points would theoretically be detectable. See text for more details on the computation of the minimum detectable luminosity. 
\textit{Bottom row:} We create 100 simulated BH samples for each combination of $M_{\rm BH}-M_{\ast}$ and luminosity variation level. These histograms show the number of detectable objects per simulated sample. On the left, we show the number of detections in simulated BH samples based on the $M_{\rm BH}-M_{\ast}$ for local active galaxies, for luminosity variations at the 10\% (orange) or 50\% (red) level. On the right, we show the number of detections in simulated BH samples based on the $M_{\rm BH}-M_{\ast}$ for quiescent elliptical/S0 galaxies, for luminosity variations at the 10\% (dark blue) or 50\% (turquoise) level. The dashed black line marks our actual number of detected variable AGNs for the low-mass sample.  }
\label{det_limits}
\end{figure*}

In Figure~\ref{det_limits}, we show the minimum detectable AGN luminosity (as discussed in the previous section), along with examples of simulated BH samples given both $M_{\rm BH}-M_{\ast}$ relations and luminosity variation level. We also show the histograms of the number of detected AGN for 100 simulated samples using each combination of $M_{\rm BH}-M_{\ast}$ and luminosity variation level. 
For BH masses populated according to the $M_{\rm BH}-M_{\ast}$ relation for elliptical/S0 galaxies, we expect to detect $92\pm10.4$ variable AGN for variations at the 10\% level, and $264\pm15.6$ for 50\% luminosity variations. For BH masses populated according to the $M_{\rm BH}-M_{\ast}$ relation for local active galaxies, we expect to detect $13.5\pm3.6$ for 10\% variations, and $58\pm7.6$ for 50\% variations. Our detected number of low-mass galaxies with AGN-like variability -- 35 -- is more closely in line with the number expected based on the $M_{\rm BH}-M_{\ast}$ relation for active galaxies.

\subsection{Comparison to other low-mass AGN selection techniques} 

One of the most fruitful techniques for identifying AGNs in low-mass galaxies is optical spectroscopy. \cite{Reines:2013fj} searched for AGNs in dwarf galaxies (defined as galaxies with $M_{\ast}<3\times10^{9}~M_{\odot}$) in the NASA-Sloan Atlas using optical narrow emission line ratio selection criteria. Of $\sim25,000$ dwarf galaxies, 136 had narrow emission line signatures indicative of AGN activity, for a detection fraction of 0.5\%. Ten of them also have broad H$\alpha$ emission, with BH masses ranging from $8\times10^{4}-1\times10^{6}~M_{\odot}$. Our sample contains 8941 galaxies with $M_{\ast}<3\times10^{9}~M_{\odot}$, 12 of which have AGN-like variability. This gives a variability detection fraction of 0.1\%. Of the 12 dwarf galaxies with AGN-like variability, one (NSA 32653) has narrow-emission line ratios in the AGN region of the BPT diagram and shows broad H$\alpha$ emission ($M_{\rm BH} = 6\times10^{6}~M_{\odot}$). It was not selected in \cite{Reines:2013fj}, as that work used the NSA v0, which only extends out to z=0.055. 

\cite{2014AJ....148..136M} also used optical spectroscopy to search for AGNs in low-mass galaxies. Specifically, they searched for AGN signatures in galaxies with $M_{\ast}<10^{10}~M_{\odot}$ within 80 Mpc. Overall, they detected AGNs in 28/9526 galaxies, for a detection fraction of 0.29\% (similar to our detection fraction of 0.22\%). Of their 28 galaxies, seven have stellar masses in the regime considered by \cite{Reines:2013fj}, five of which are also classified as AGN in that work. When considering just galaxies with stellar masses between $4\times10^{9}$ and $10^{10}~M_{\odot}$, their detection fraction is 2.7\%. The \cite{2014AJ....148..136M} AGN host galaxies have a median $g-r$ color of 0.62, slightly redder than our sample. 

There are six dwarf galaxies with narrow-emission line evidence for AGN activity from \cite{Reines:2013fj} which fall inside the area covered by the Stripe 82 survey (none of the \cite{2014AJ....148..136M} objects fall within Stripe 82). None of the six were classified as having nuclear variability according to our selection criteria. One of the six galaxies (RGG 1 in \citealt{Reines:2013fj}) has broad H$\alpha$ emission, from which we can obtain a BH mass estimate. RGG 1 is at a redshift of $0.046$ and has a stellar mass of $2.6\times10^{9}~M_{\odot}$. The BH mass reported in \cite{Reines:2013fj} is $5.0\times10^{5}~M_{\odot}$ (note that systematic uncertainties on BH masses derived from broad H$\alpha$ are on the order of $\sim0.3$ dex). Using \textit{Chandra} X-ray observations, \cite{2017ApJ...836...20B} find that this BH is accreting at $0.001~L_{\rm Edd}$. Using the same assumptions as the previous section, this BH would have a \textit{g}-band luminosity of $3.0\times10^{39}~{\rm erg~s^{-1}}$. As shown in Figure~\ref{det_limits}, at a redshift of 0.05, we are sensitive to AGNs with \textit{g}-band luminosities greater than $\sim10^{41}~{\rm erg~s^{-1}}$. Therefore, we would not expect to detect any variability from this AGN (see Figure~\ref{det_limits}).

\section{Discussion \& Conclusions}

We analyze the light curves of $\sim28,000$ nearby (z<0.15) galaxies in the SDSS Stripe 82 field for AGN-like nuclear variability. We concentrate our analysis on galaxies in NASA-Sloan Atlas, which have spectroscopic redshifts and stellar mass estimates. We construct light curves using difference imaging, which allows us to detect small variations superposed on top of the stellar light from the galaxy. We then determine whether variability is AGN-like by assessing how well individual light curves are described by a damped random walk.

We find 135 galaxies with AGN-like variability. The variability-selected sample spans roughly four orders of magnitude in stellar mass. Above $M_{\ast}=10^{10}~M_{\odot}$ (100 galaxies), almost all of the variability-selected AGNs also have narrow emission line ratios in the AGN or composite regions of the BPT diagram. Below $M_{\ast}=10^{10}~M_{\odot}$, half of the galaxies have narrow emission-line ratios dominated by star formation, indicating they may be AGNs missed by other selection techniques due to star formation dilution or metallicity effects, which are expected to be more important at lower stellar masses. We stress that the low-mass galaxies with AGN-like variability falling in the star forming region of the BPT diagram should be regarded as candidates; we are in the process of obtaining higher spatial resolution follow-up spectroscopy which better isolates emission from the nucleus. High spatial resolution X-ray and/or radio observations will also be valuable in the effort to confirm the presence of AGNs in these systems. 

Using this sample, we study the fraction of variable AGN as a function of stellar mass, and find that even when accounting for magnitude bias, there is a decline in the fraction of variable AGN with decreasing stellar mass. This could be attributed to a lower BH occupation fraction for galaxies with $M_{\ast}<10^{10}~M_{\odot}$, a change in the $M_{BH}-M_{\ast}$ relation for low mass galaxies, or a change in the variability properties of BHs in low-mass galaxies.

Figure~\ref{agnvmstar} shows that the active fraction is lower for galaxies with $M_{\ast}<10^{10}~M_{\odot}$ than for galaxies with $M_{\ast}>10^{10}~M_{\odot}$, even after taking into account the bias introduced by different apparent magnitude distributions for the high and low-mass subsamples. There are several possible explanations for a decline in the active fraction with stellar mass, which we discuss below. 

A relatively straightforward interpretation of our results is that there is a drop in the \textit{occupation fraction} (i.e., the fraction of galaxies with BHs) which leads naturally to a drop in the fraction which are active. While it is well established that the occupation fraction for massive galaxies is 100\% (or extremely close; \citealt{1998AJ....115.2285M}), it is unclear whether that holds for low-mass galaxies. There are several nearby galaxies for which there are stringent limits on the mass of a central BH based on dynamical modeling; for example, \cite{2001AJ....122.2469G} find that the upper limit on a central BH in M33 is just $1500~M_{\odot}$. There are ongoing efforts to constrain the BH occupation fraction at low masses using X-ray observations; current estimates place a firm lower limit of 20\% on the low-mass occupation fraction \citep{2015ApJ...799...98M}. 

Another possibility is that $M_{\rm BH}/M_{\ast}$ is lower for low-mass galaxies than for more massive galaxies. Massive galaxies show a fairly constant (with some scatter) ratio between the mass of the central BH and the mass of the galaxy ($\sim1/1000$ or 0.001). Recent works have shown that BHs in low-mass galaxies may be under-massive with respect to scaling relations between BH mass and bulge/galaxy stellar mass defined for more massive galaxies \citep{Greene:2008qy, 2015ApJ...813...82R, 2017ApJ...850..196B}. As mentioned above, we estimate BH masses for the 16 low-mass galaxies with broad H$\alpha$ emission. The BH masses range from $\log(M_{\rm BH}/M_{\odot})$ = 6.0 to 7.9 (with uncertainties of $\sim0.3$ dex). The median BH mass is $\log(M_{\rm BH}/M_{\odot})$ = 6.8. These BH masses correspond to $M_{\rm BH}/M_{\ast}$ ratios of 0.0002 to 0.007 (median $M_{\rm BH}/M_{\ast}$=0.0013). Thus, we may only be able to detect variable AGN (with this data set) in the low-mass galaxies that have unusually massive BHs (as compared to other galaxies of similar stellar mass). 

Finally, it is possible that AGNs in low-mass galaxies are intrinsically less variable, or are less variable in optical wavelengths. Monitoring campaigns of known low-mass galaxies with AGNs will be important for studying whether low-mass galaxies vary on similar timescales and/or with similar amplitudes as more massive AGNs.

The objects with AGN-like variability falling in the star forming region of the BPT diagram will require multi-wavelength follow-up to search for additional evidence for AGNs in these systems. Our analysis will also be extended to additional existing repeat imaging surveys to search for more low-mass systems with AGN-like variability. The Large Synoptic Survey Telescope will be ideal for continuing searches for low-level variability on days-to-months timescales and should be sensitive to less massive BHs and/or BHs accreting at lower Eddington fractions.

\acknowledgements

Support for VFB was provided by the National Aeronautics and Space Administration through Einstein Postdoctoral Fellowship Award Number PF7-180161 issued by the Chandra X-ray Observatory Center, which is operated by the Smithsonian Astrophysical Observatory for and on behalf of the National Aeronautics Space Administration under contract NAS8-03060. The authors thank Michael Tremmel and Michael Warrener for useful comments which have improved this paper.

\software{Astropy \citep{2018AJ....156..123T}, DIAPL2 \citep{2000AcA....50..421W}, QSO\_fit \citep{2011AJ....141...93B} }

\bibliographystyle{apj}

\clearpage
\begin{appendix}
\section{Light curves of low-mass galaxies with AGN-like variability}
In this section, we show light curves and DECaLs imaging for all objects with NSA stellar masses $M_{\ast}<10^{10}M_{\odot}$. The objects are in order of NASA-Sloan Atlas ID. \\

\begin{figure*}[b]
\centering
%
\raisebox{0.5cm}{\includegraphics[width=0.23\textwidth]{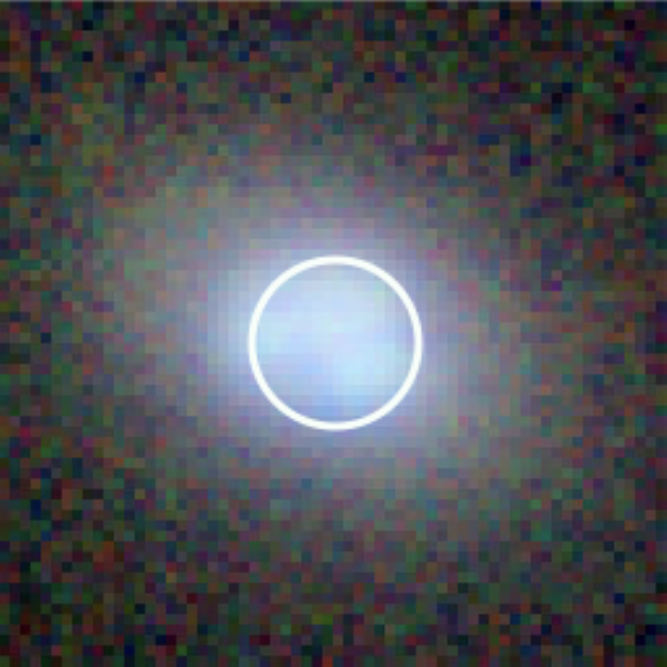}}
\includegraphics[width=0.45\textwidth]{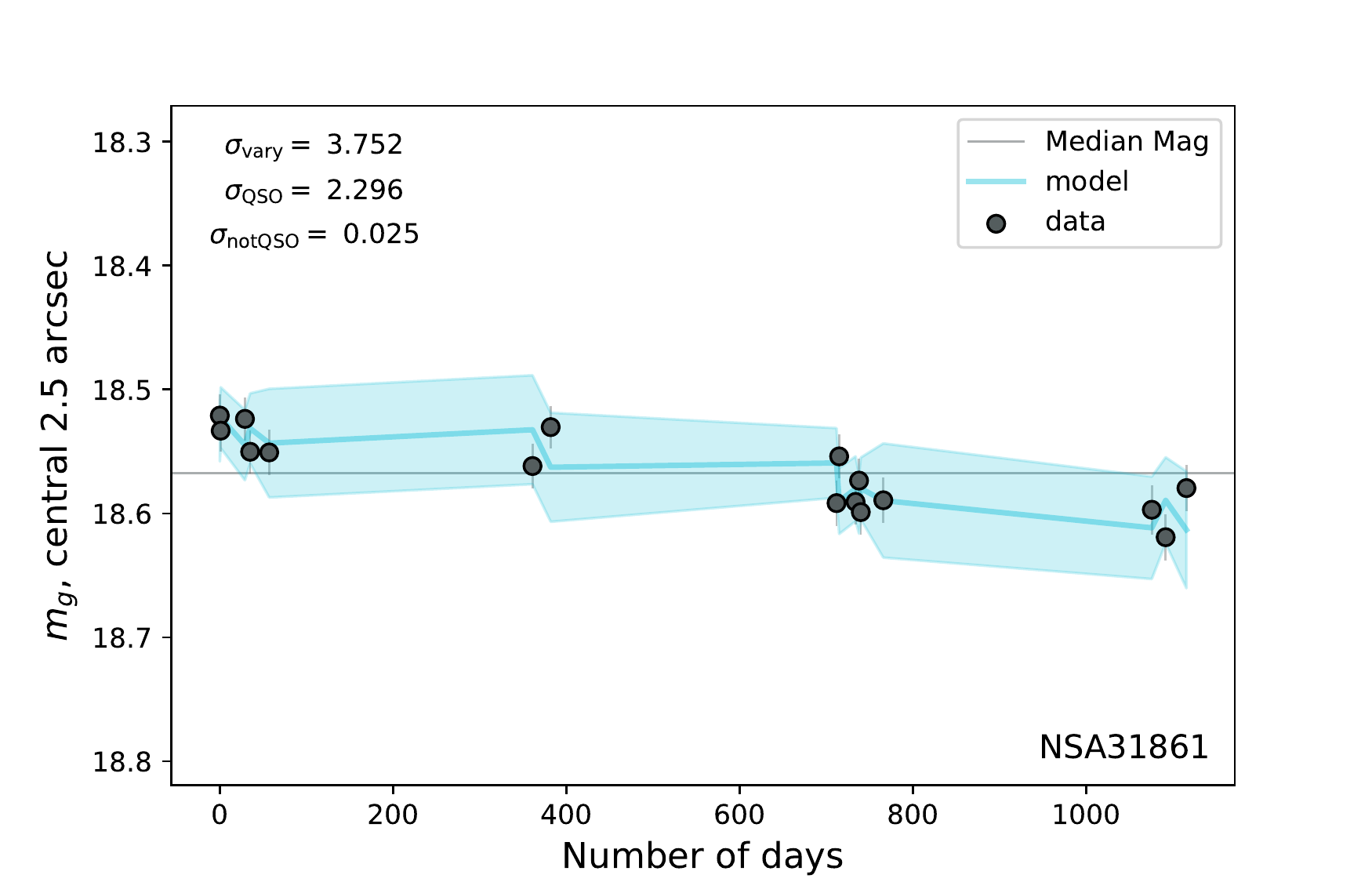}\\
\raisebox{0.5cm}{\includegraphics[width=0.23\textwidth]{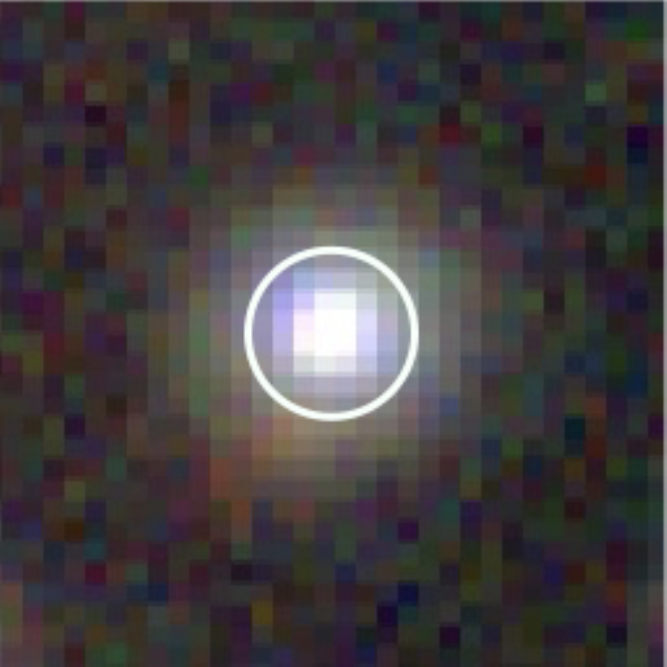}}
\includegraphics[width=0.45\textwidth]{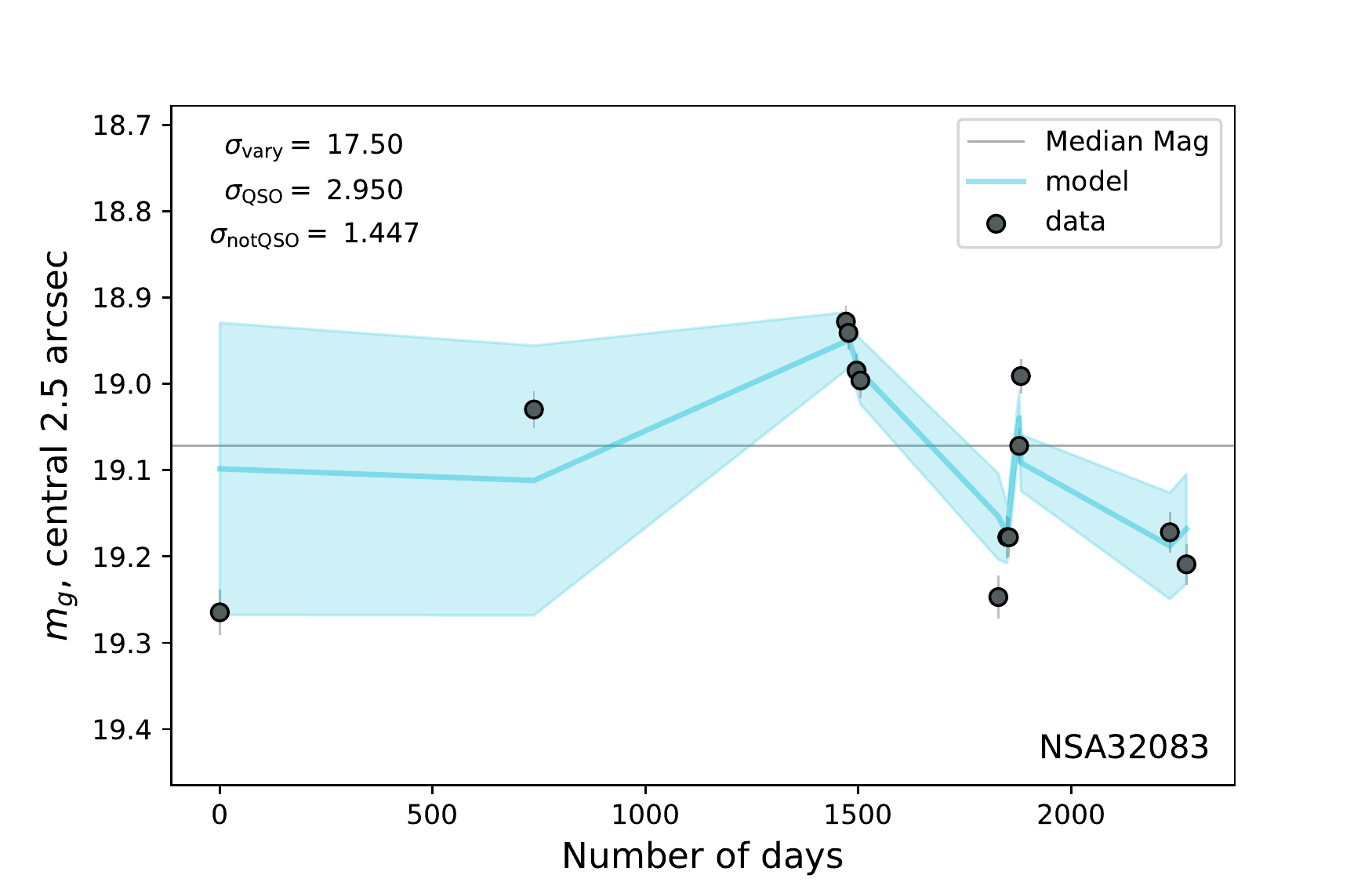}\\
\raisebox{0.5cm}{\includegraphics[width=0.23\textwidth]{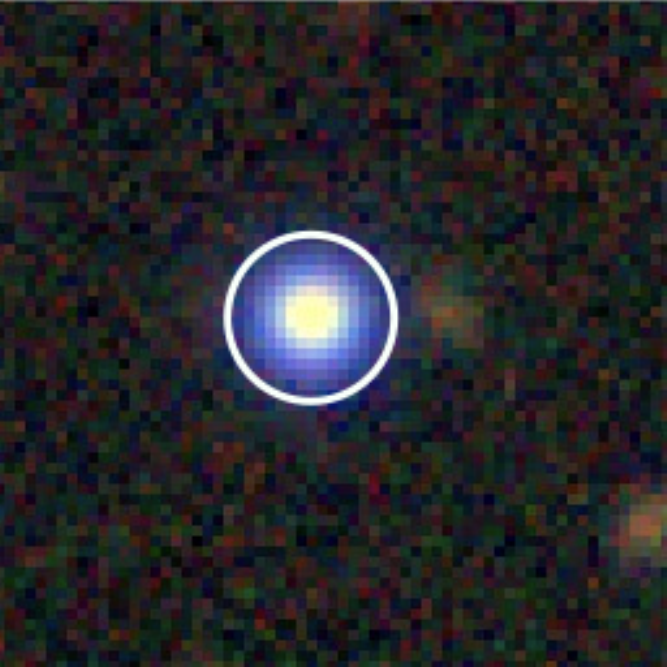}}
\includegraphics[width=0.45\textwidth]{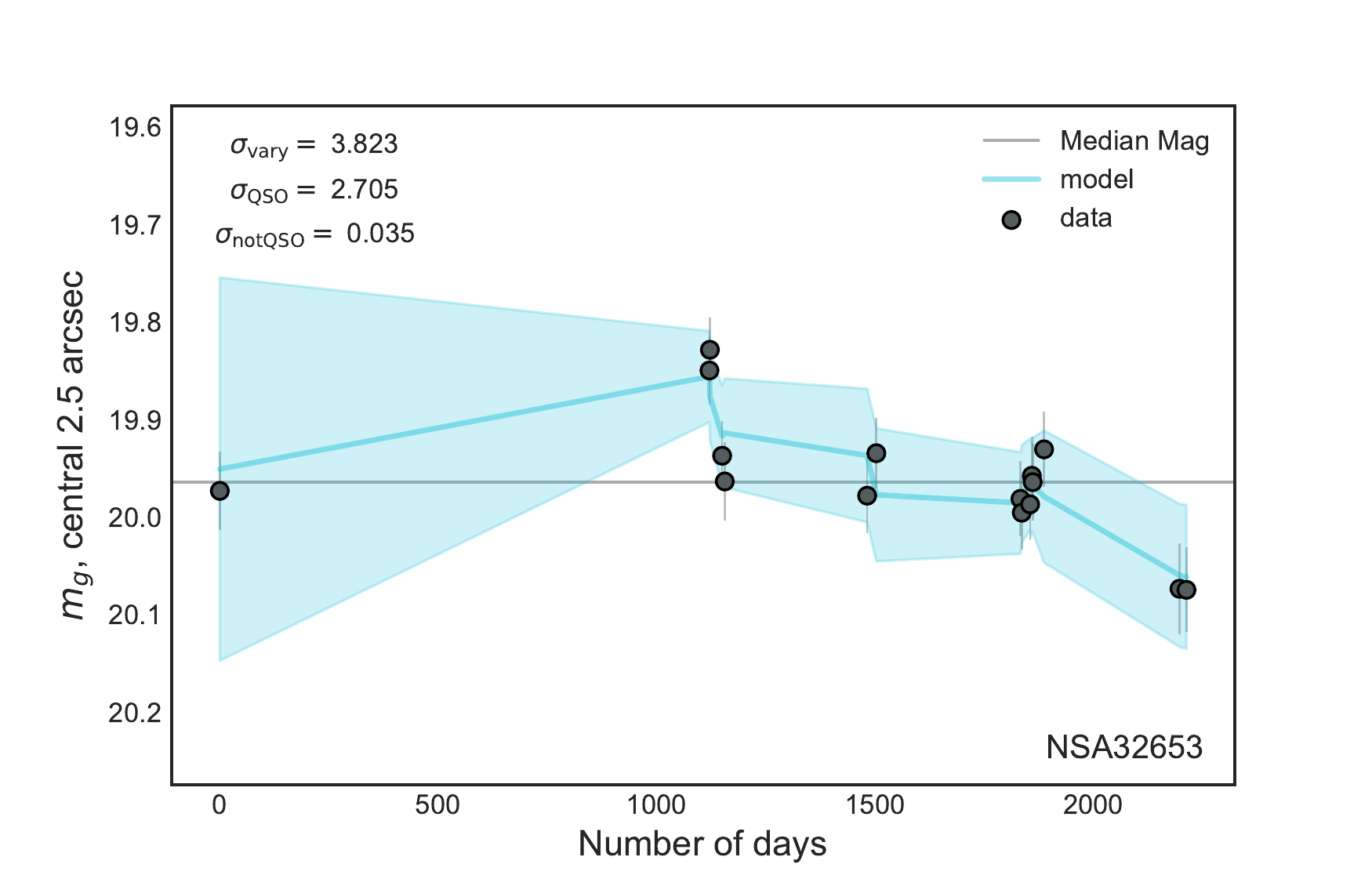}\\
\caption{DECaLs imaging and SDSS \textit{g}-band light curves of low-mass galaxies ($M_{\ast}<10^{10}M_{\odot}$) which meet our AGN variability selection criteria (continued in Figure~\ref{lc_pt2}). The grey points are the observed nuclear \textit{g}-band magnitudes with corresponding errors. The blue solid line shows the best fit damped random walk model from \textbf{qso\_fit}, and the light blue shaded region shows the model uncertainties.   }
\label{lc_pt1}
\end{figure*}

\begin{figure*}
\centering
\raisebox{0.5cm}{\includegraphics[width=0.23\textwidth]{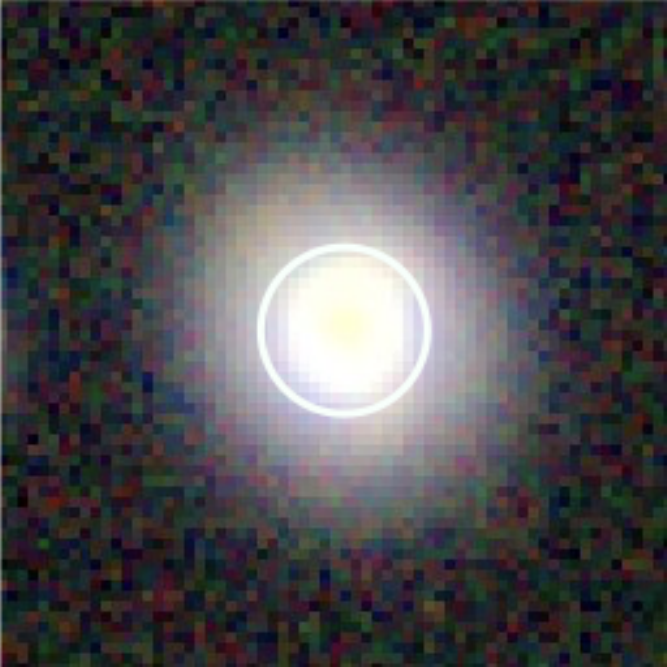}}
\includegraphics[width=0.45\textwidth]{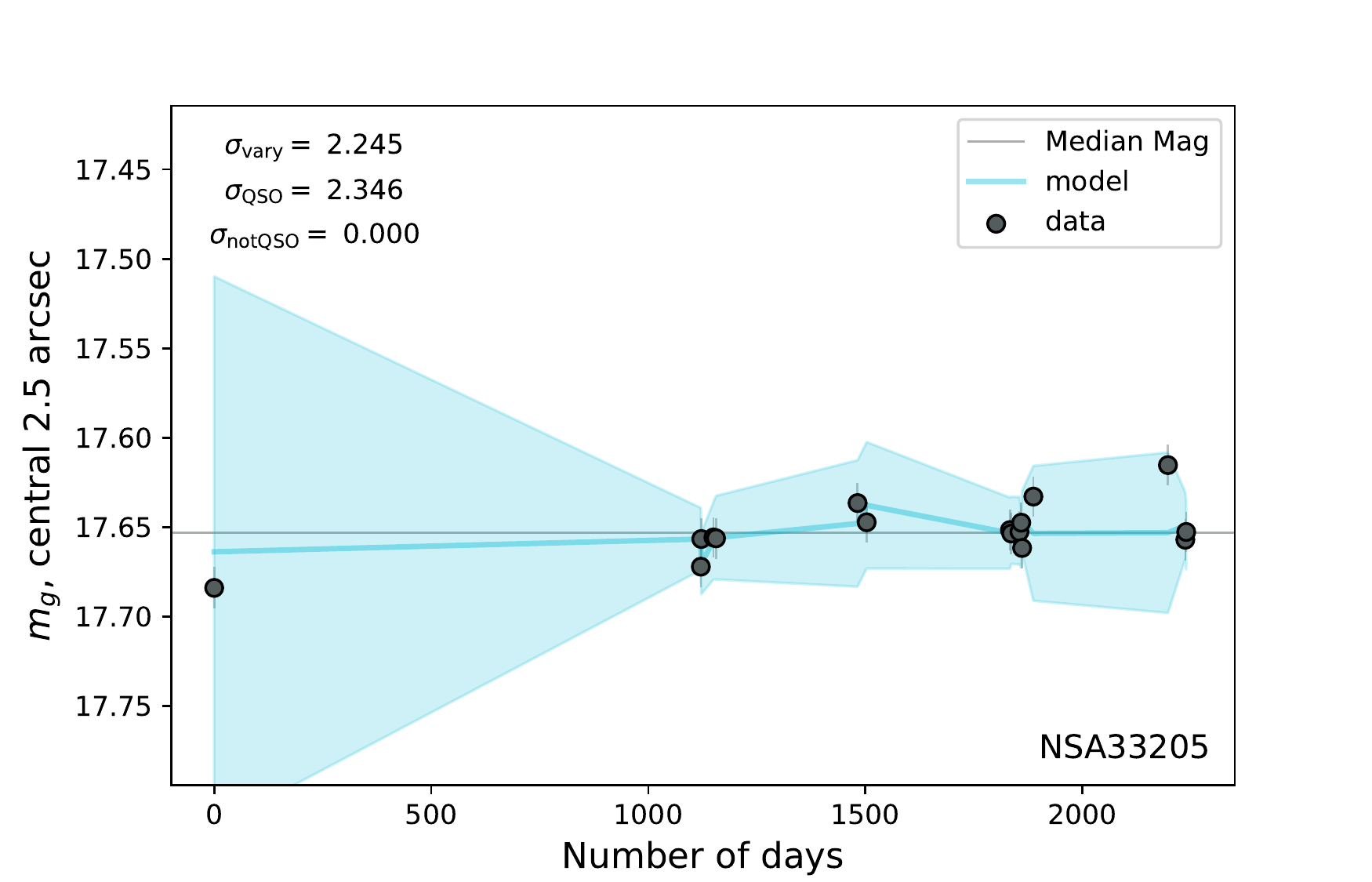}\\
\raisebox{0.5cm}{\includegraphics[width=0.23\textwidth]{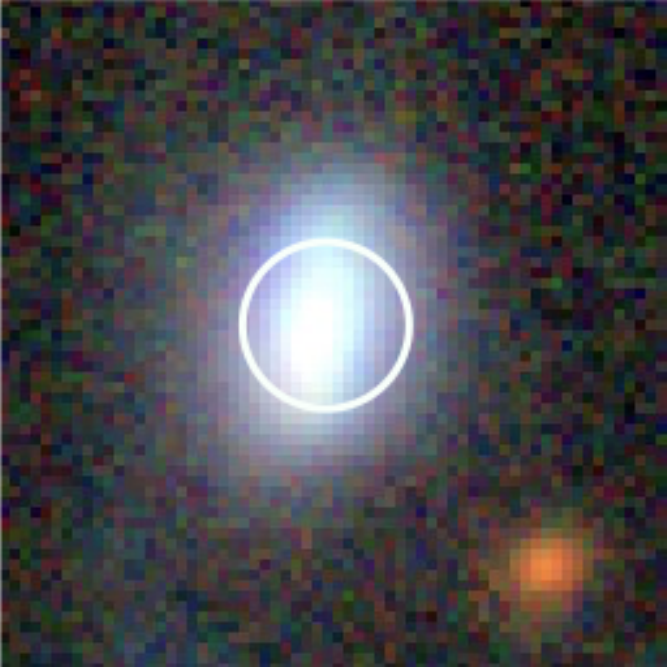}}
\includegraphics[width=0.45\textwidth]{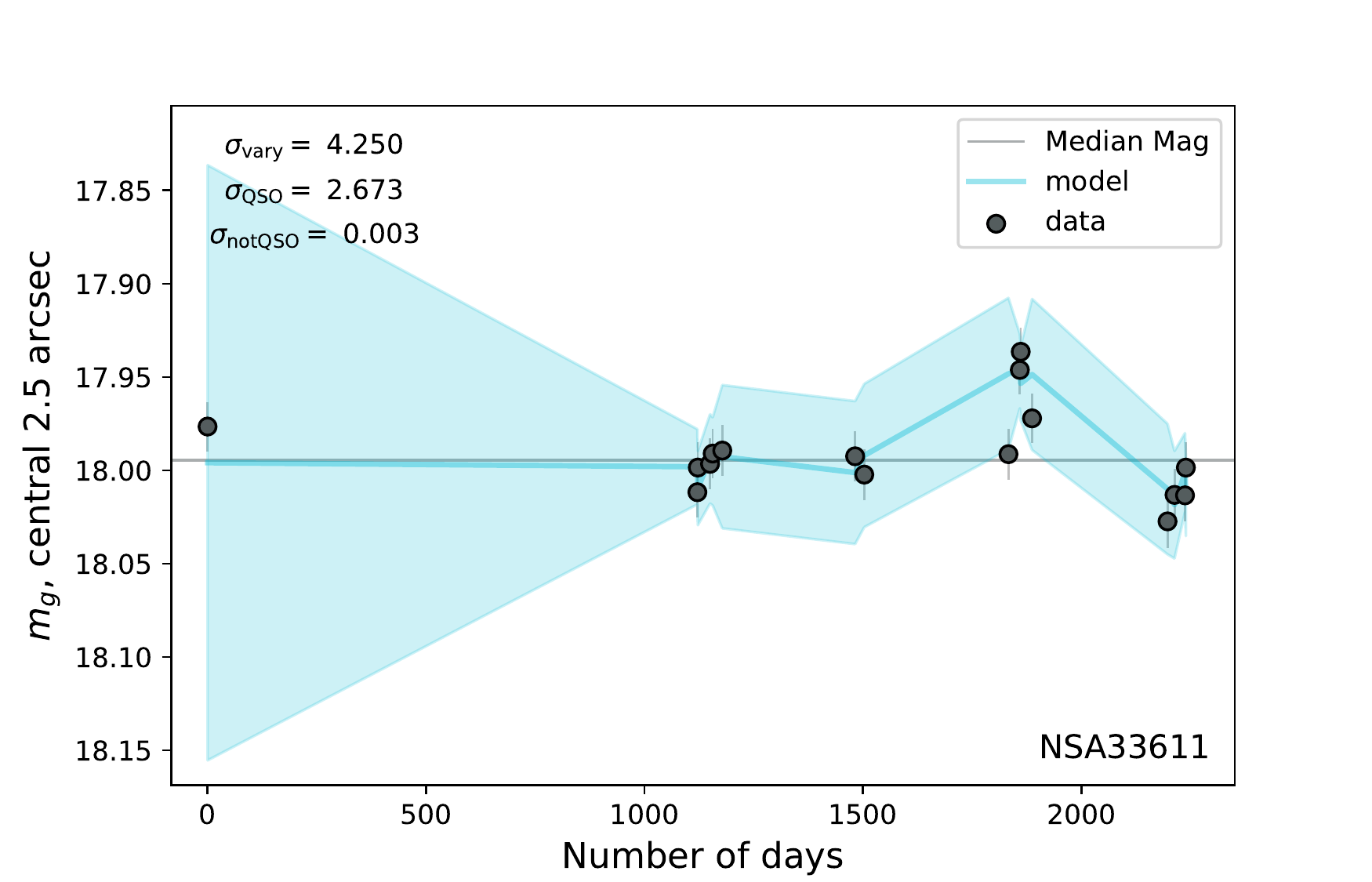}\\
\raisebox{0.5cm}{\includegraphics[width=0.23\textwidth]{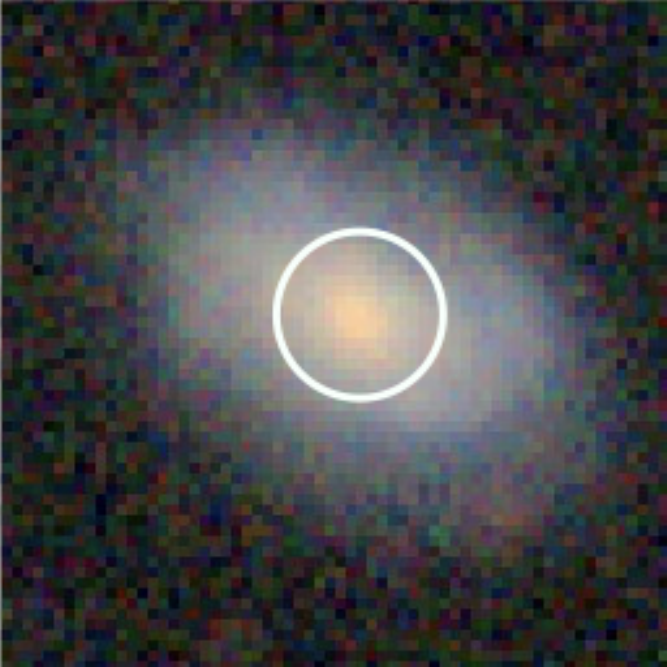}}
\includegraphics[width=0.45\textwidth]{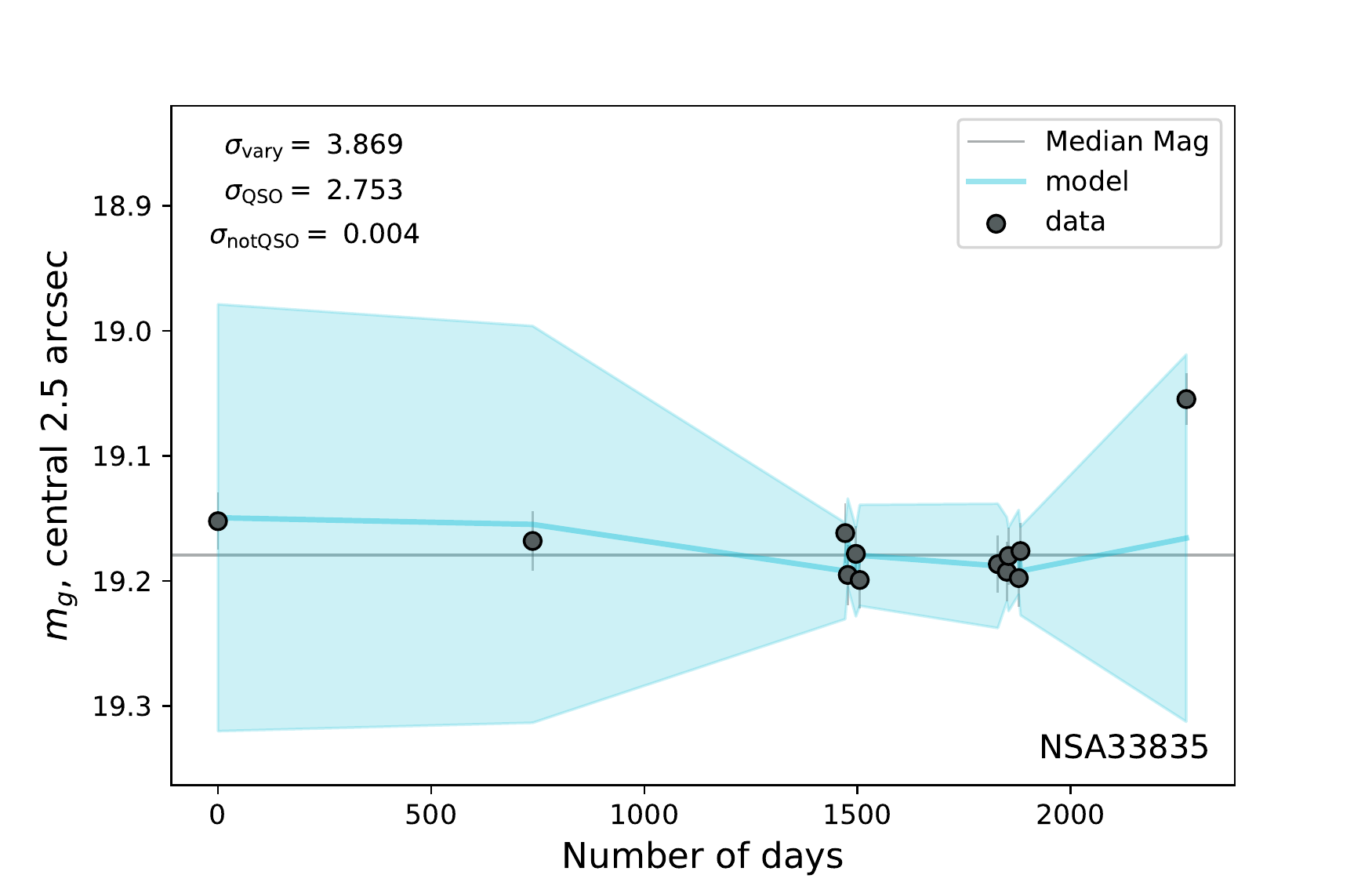}\\
\raisebox{0.5cm}{\includegraphics[width=0.23\textwidth]{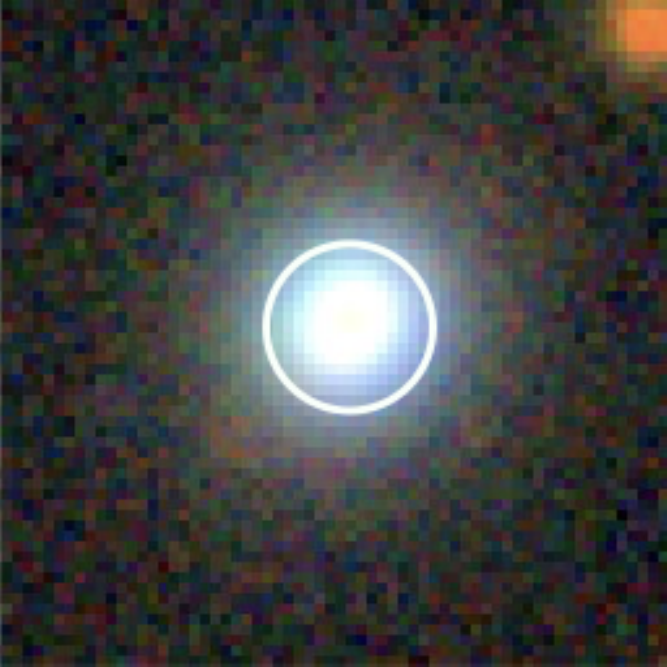}}
\includegraphics[width=0.45\textwidth]{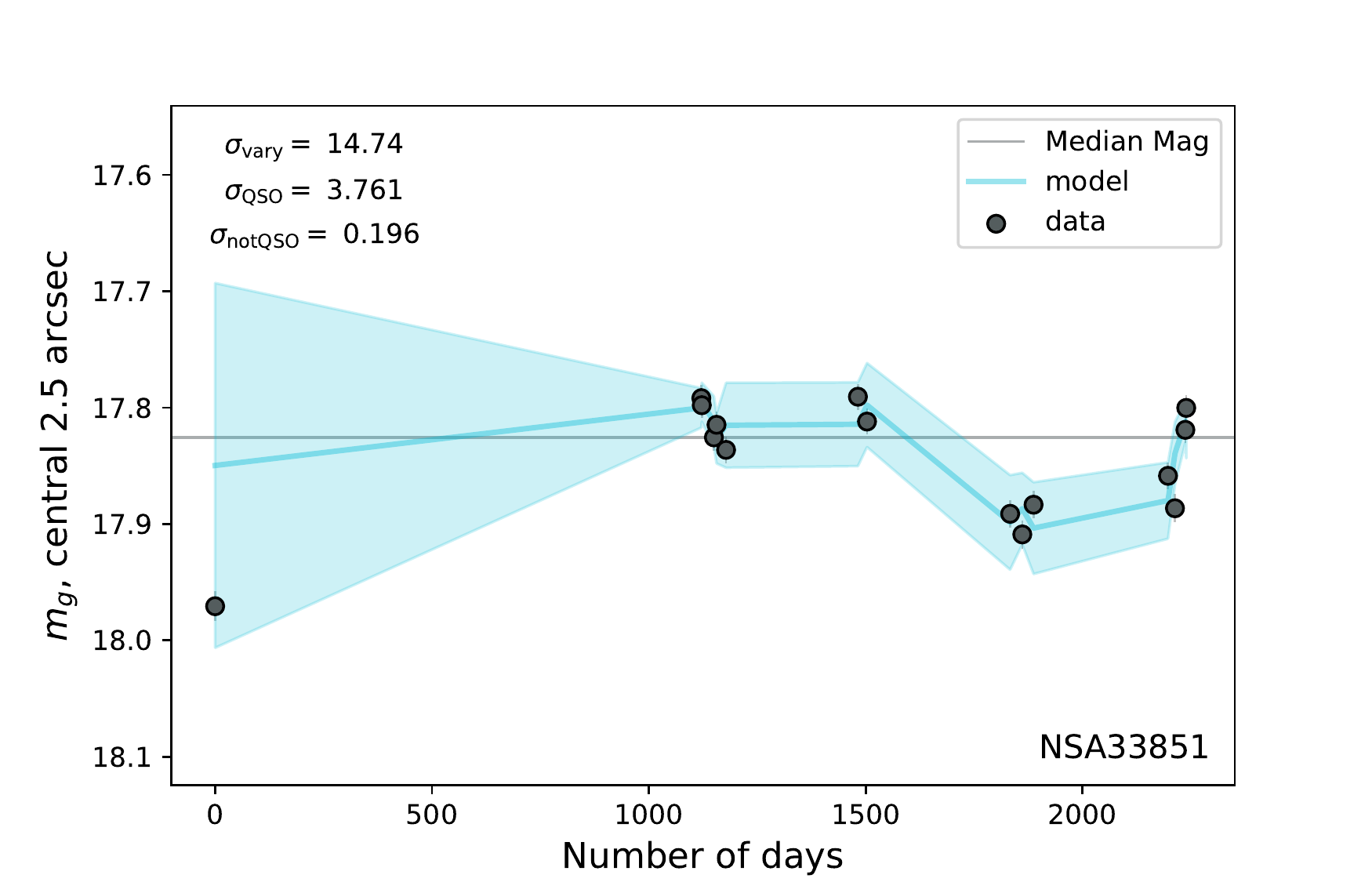}\\
\caption{SDSS \textit{g}-band light curves of low-mass galaxies ($M_{\ast}<10^{10}M_{\odot}$) which meet our AGN variability selection criteria (continued in Figure~\ref{lc_pt3}).  }
\label{lc_pt2}
\end{figure*}

\begin{figure*}
\centering
\raisebox{0.5cm}{\includegraphics[width=0.23\textwidth]{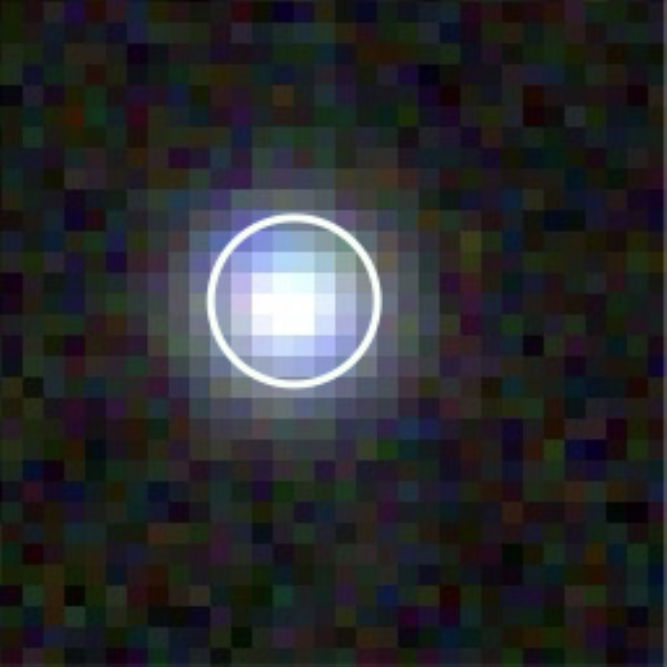}}
\includegraphics[width=0.45\textwidth]{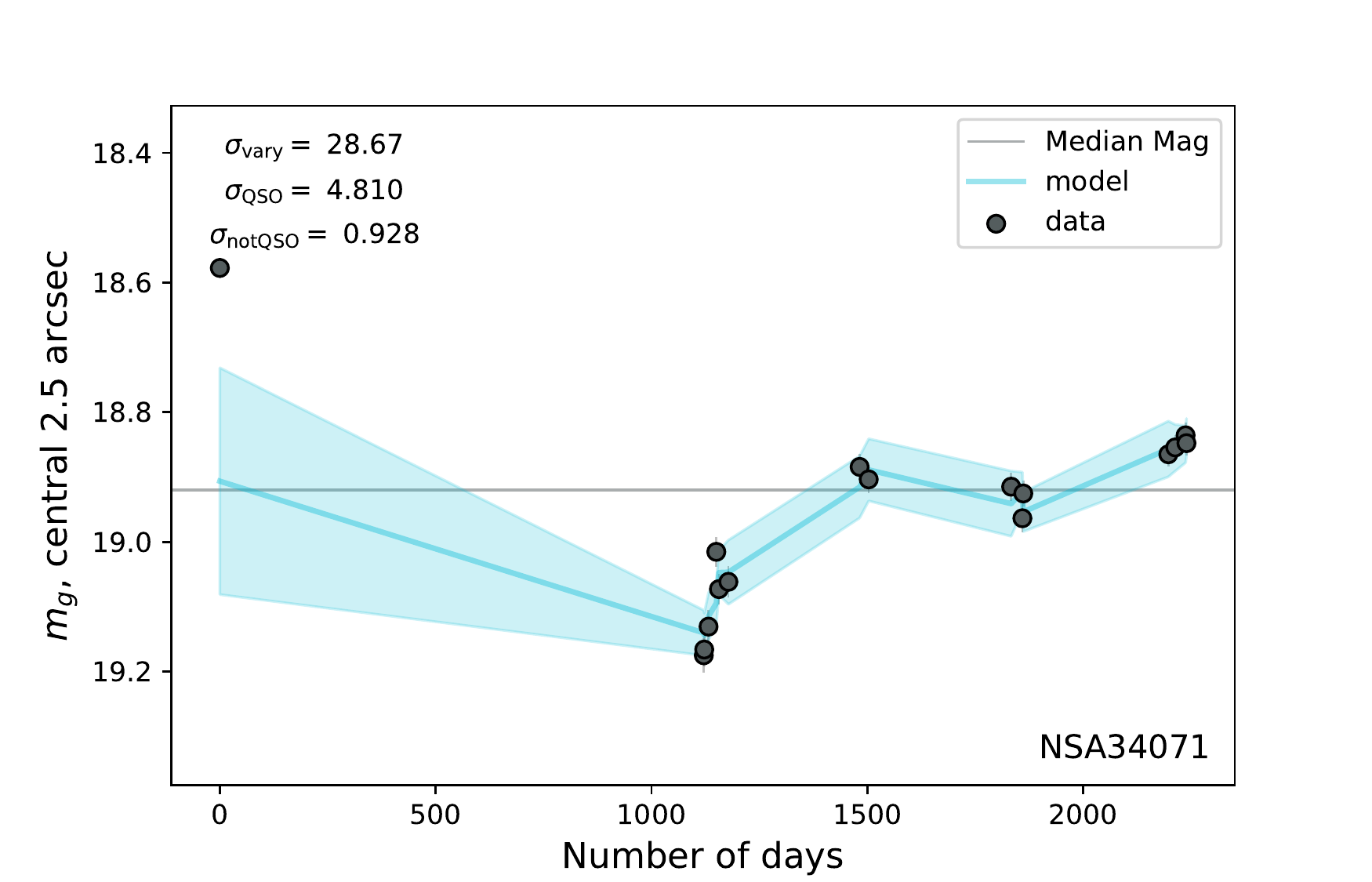}\\
\raisebox{0.5cm}{\includegraphics[width=0.23\textwidth]{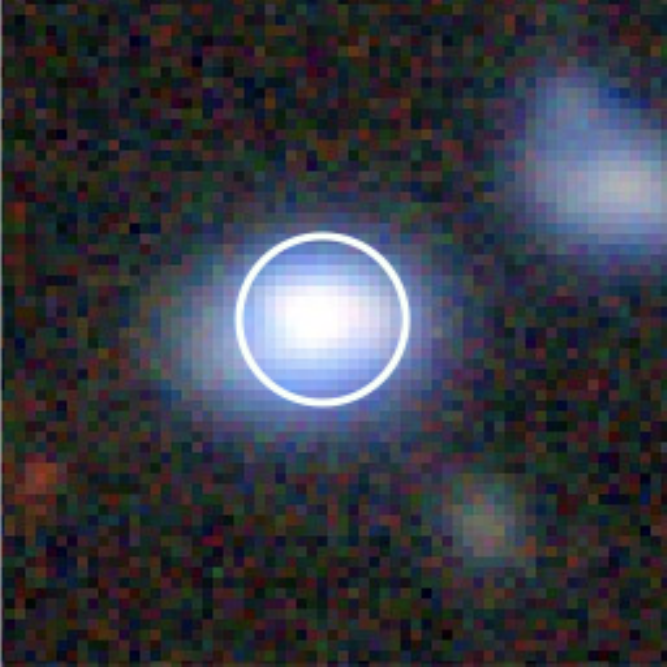}}
\includegraphics[width=0.45\textwidth]{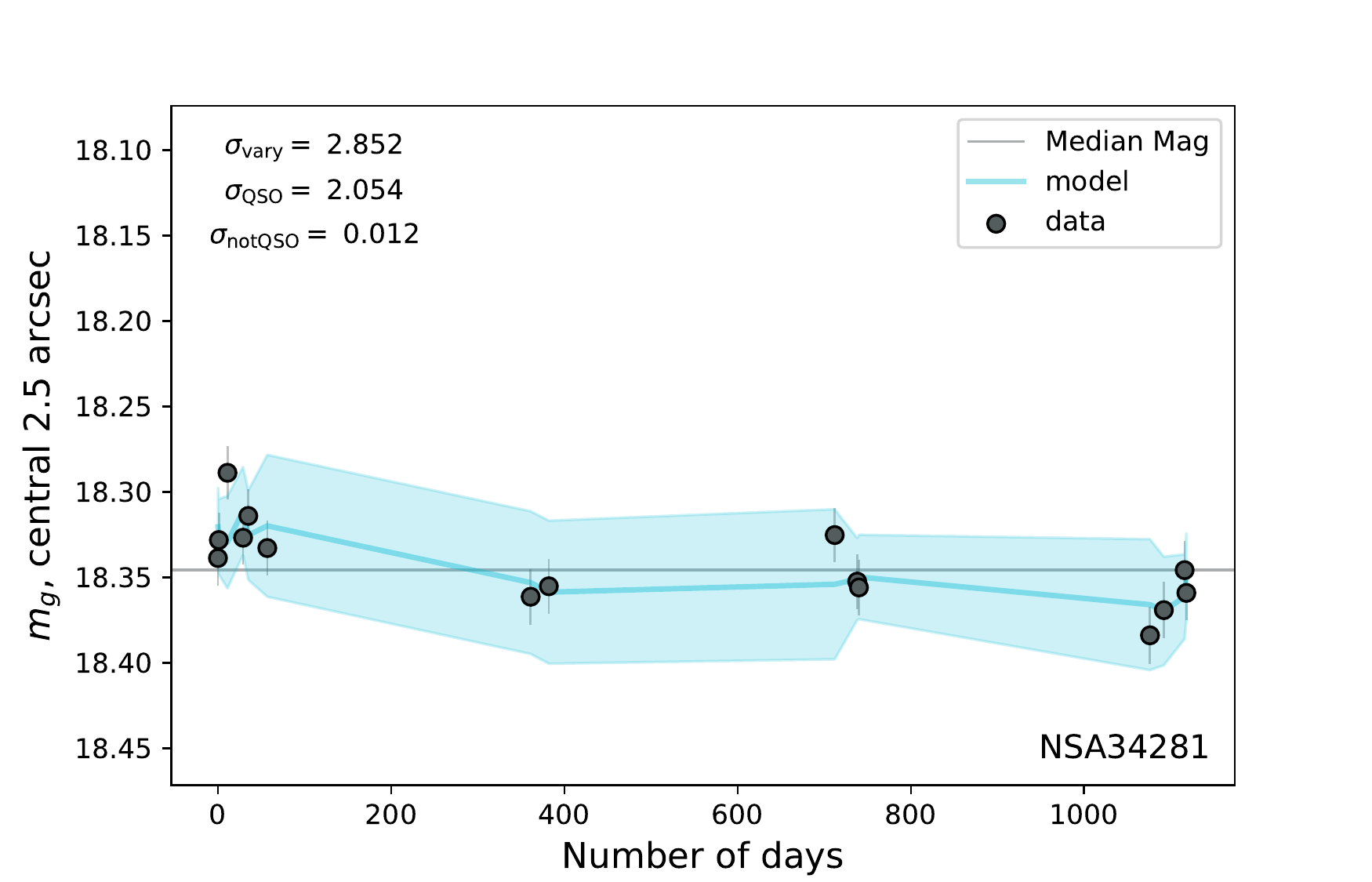}\\
\raisebox{0.5cm}{\includegraphics[width=0.23\textwidth]{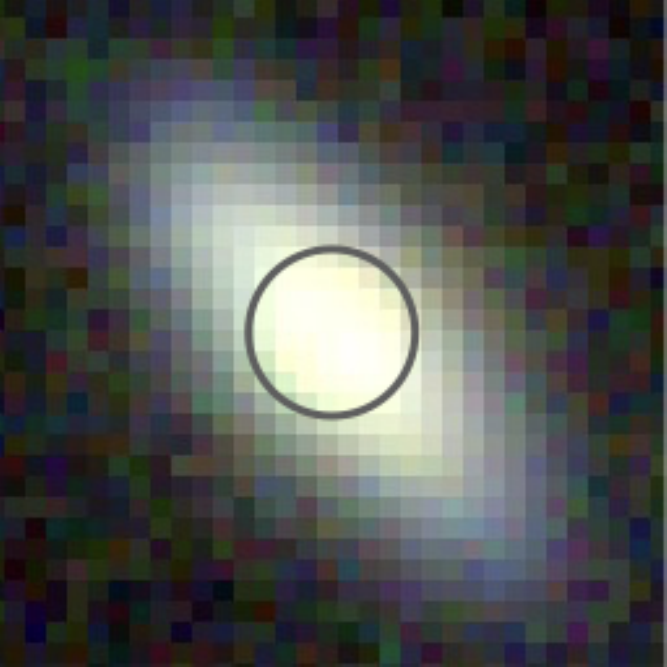}}
\includegraphics[width=0.45\textwidth]{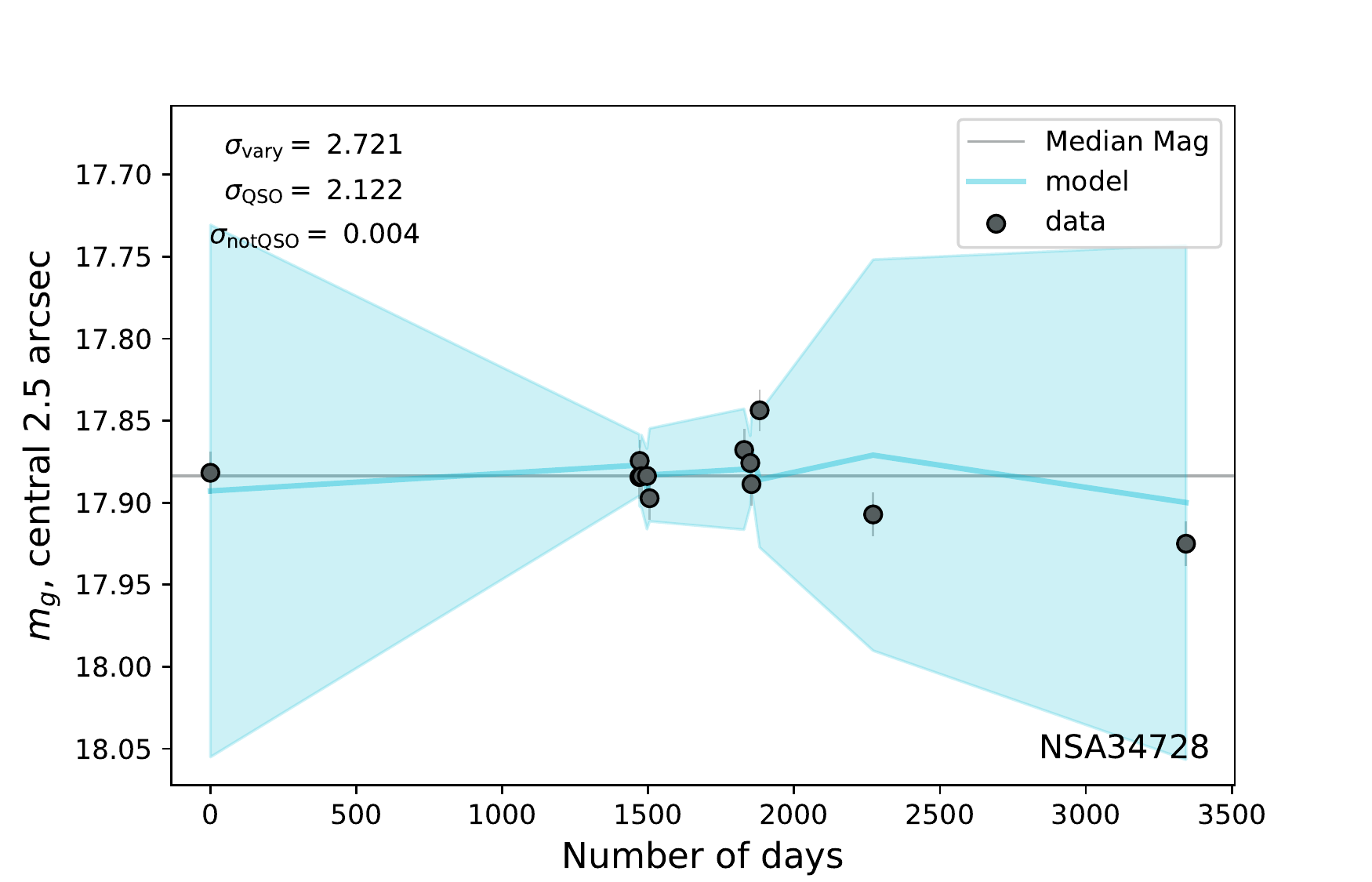}\\
\raisebox{0.5cm}{\includegraphics[width=0.23\textwidth]{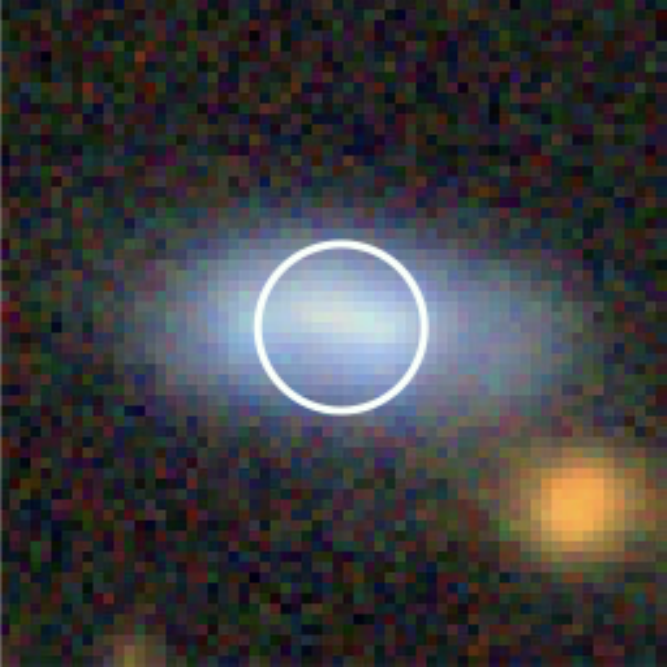}}
\includegraphics[width=0.45\textwidth]{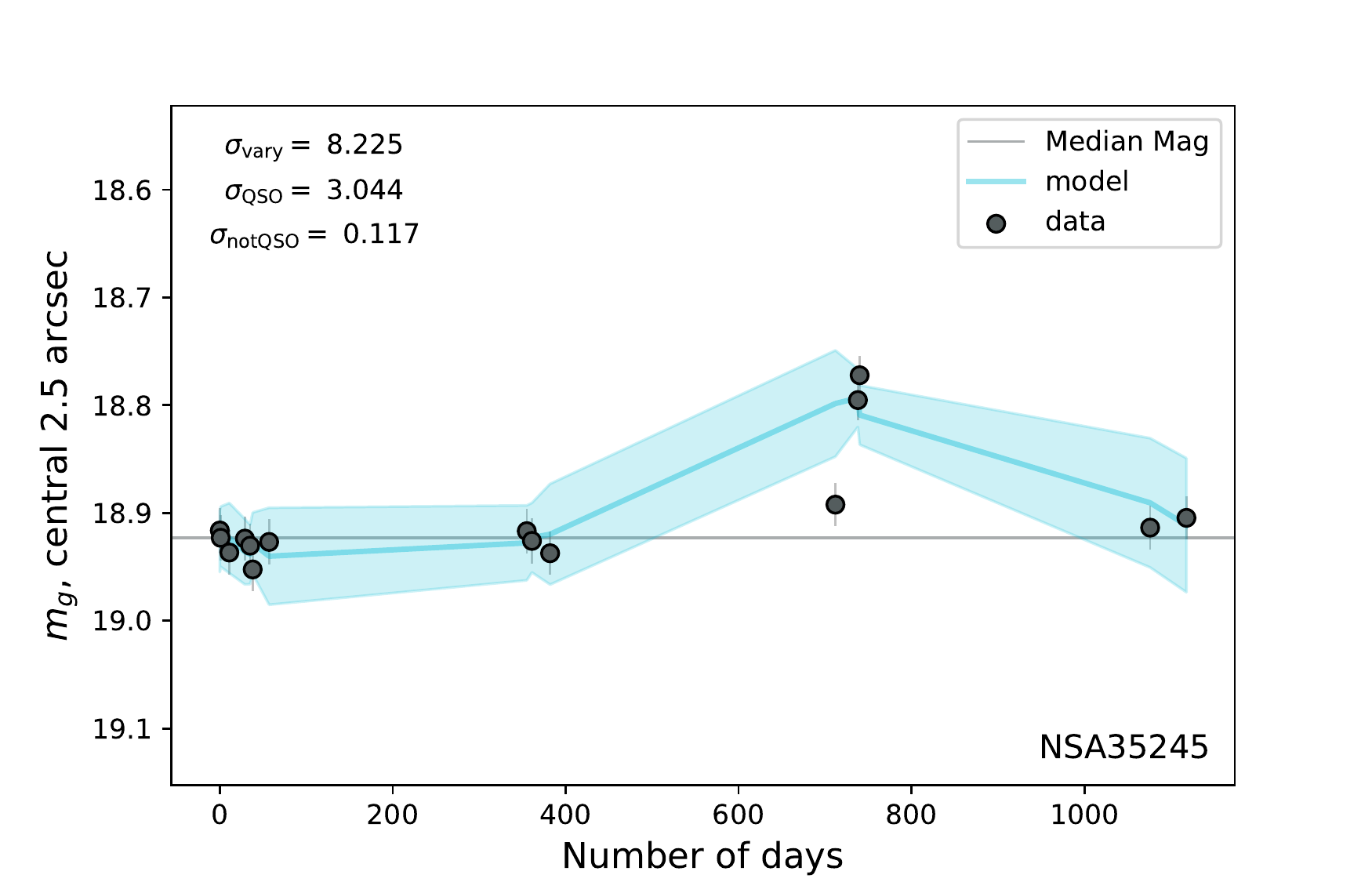}\\
\caption{SDSS \textit{g}-band light curves of low-mass galaxies ($M_{\ast}<10^{10}M_{\odot}$) which meet our AGN variability selection criteria (continued in Figure~\ref{lc_pt4}).  }
\label{lc_pt3}
\end{figure*}

\begin{figure*}
\centering
\raisebox{0.5cm}{\includegraphics[width=0.23\textwidth]{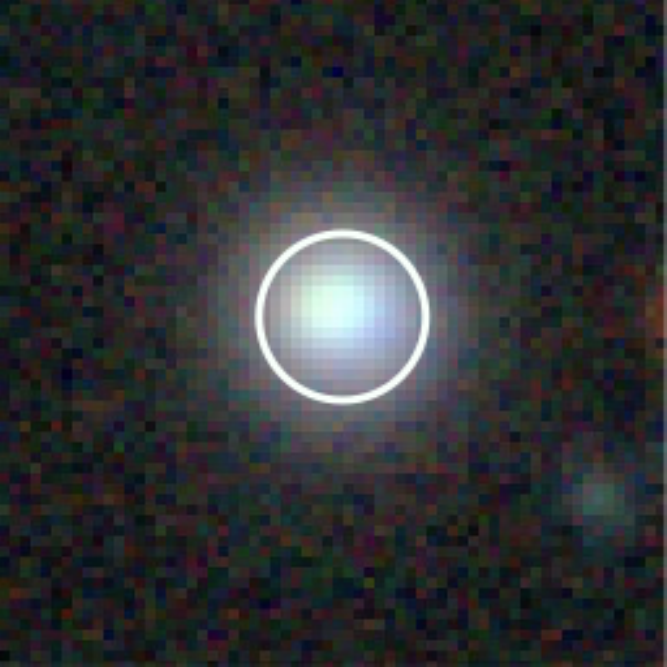}}
\includegraphics[width=0.45\textwidth]{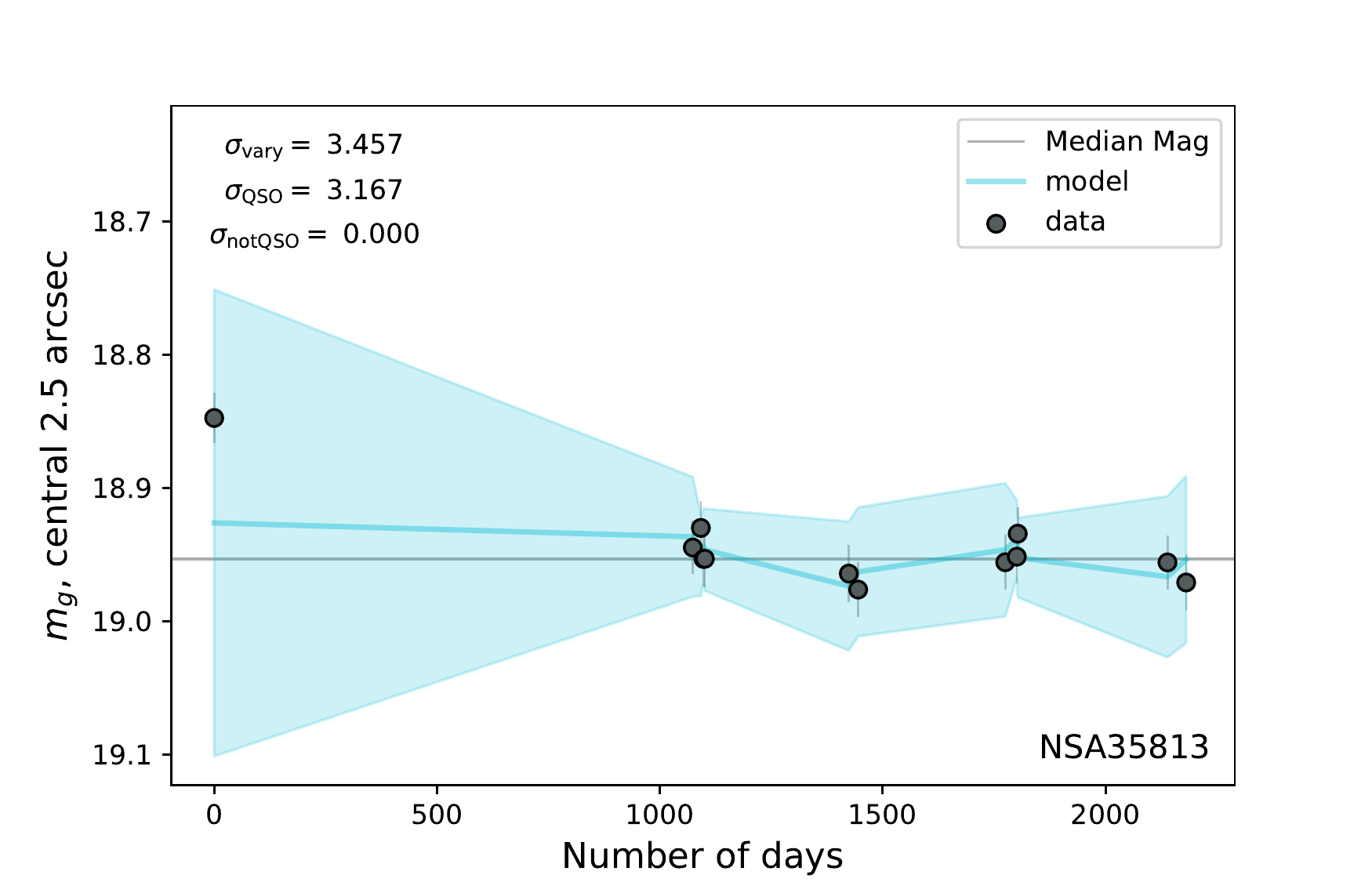}\\
\raisebox{0.5cm}{\includegraphics[width=0.23\textwidth]{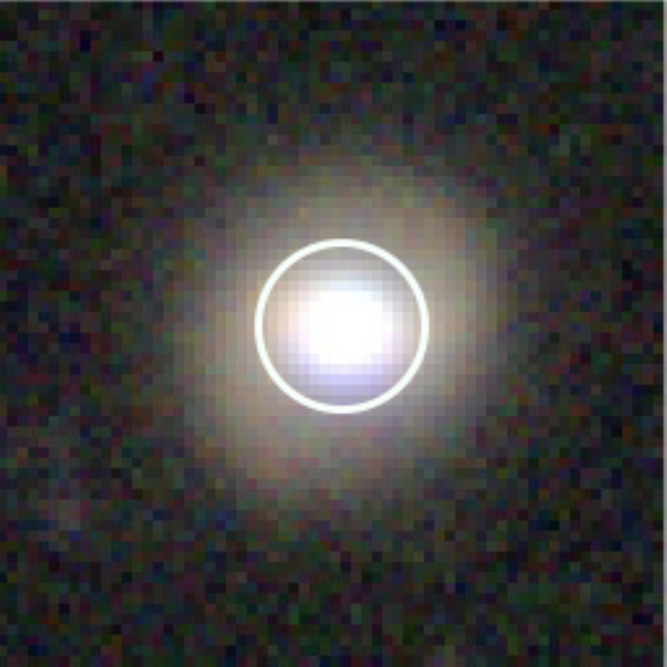}}
\includegraphics[width=0.45\textwidth]{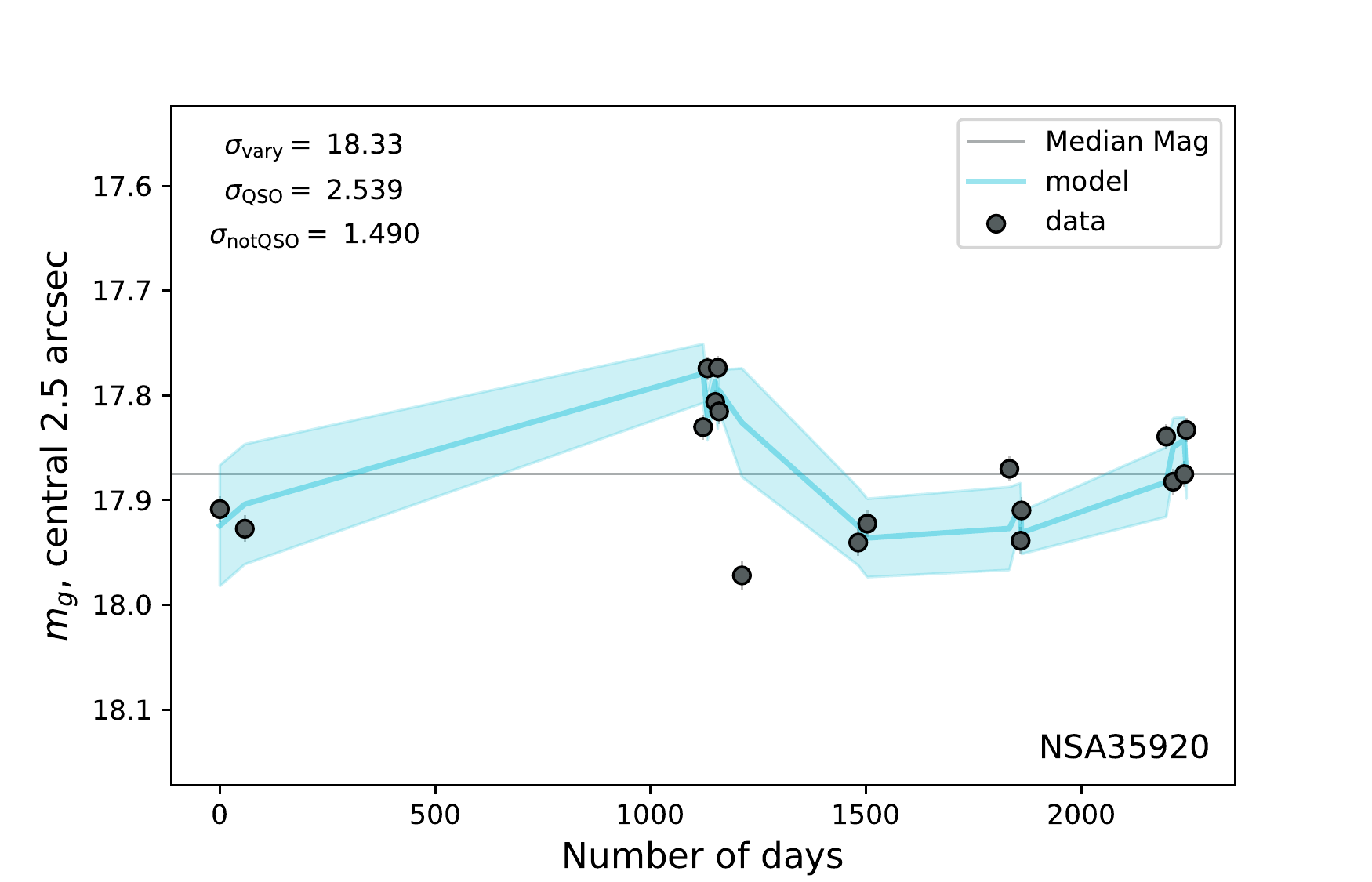}\\
\raisebox{0.5cm}{\includegraphics[width=0.23\textwidth]{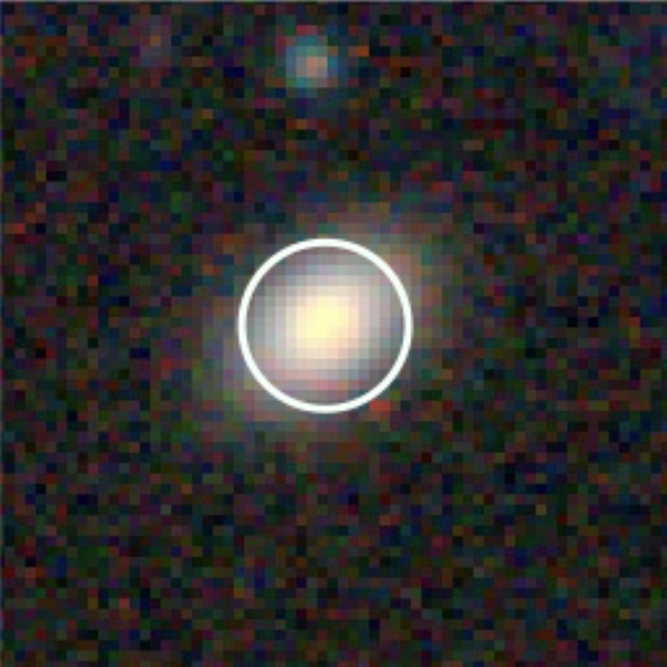}}
\includegraphics[width=0.45\textwidth]{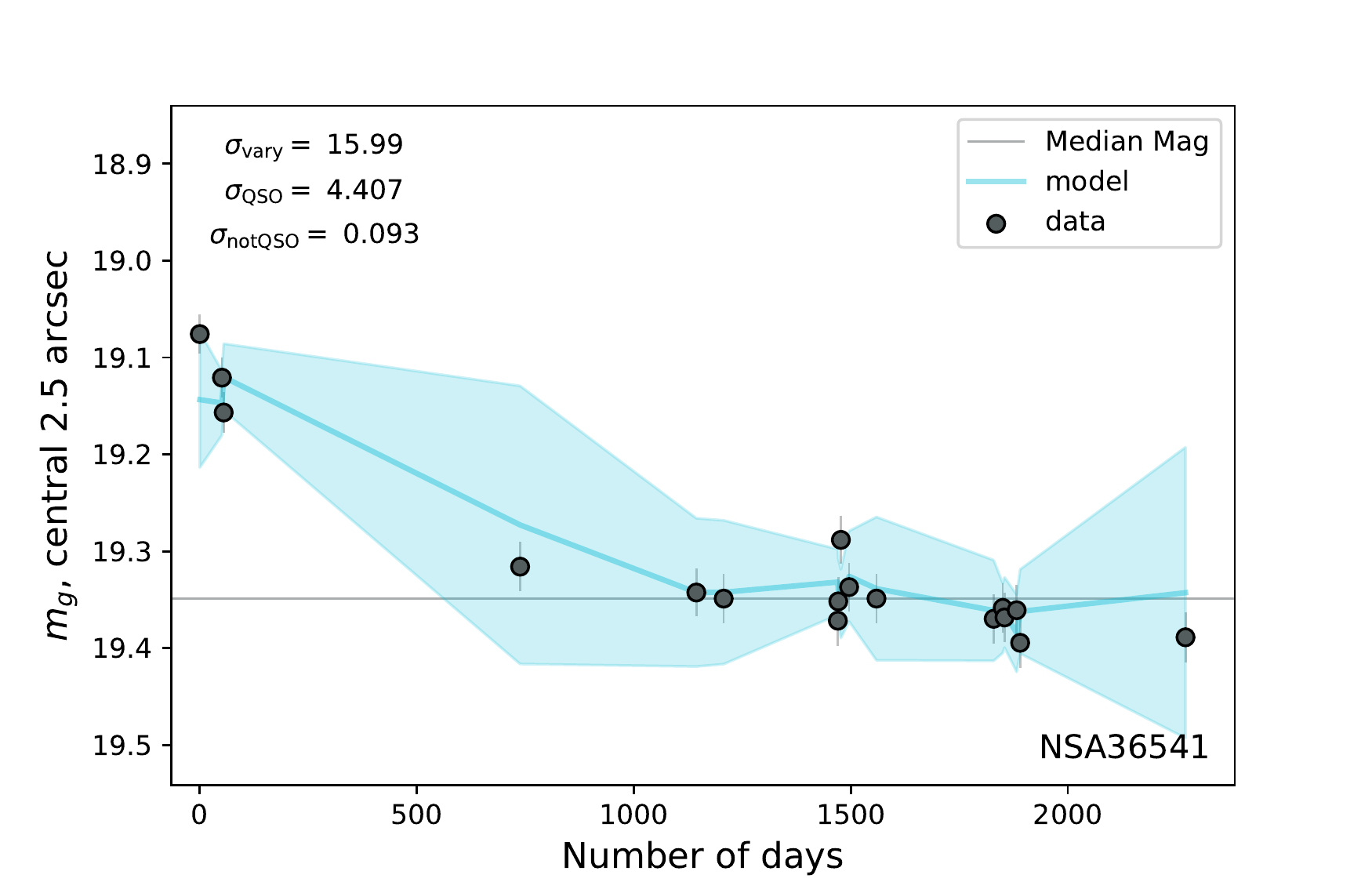}\\
\raisebox{0.5cm}{\includegraphics[width=0.23\textwidth]{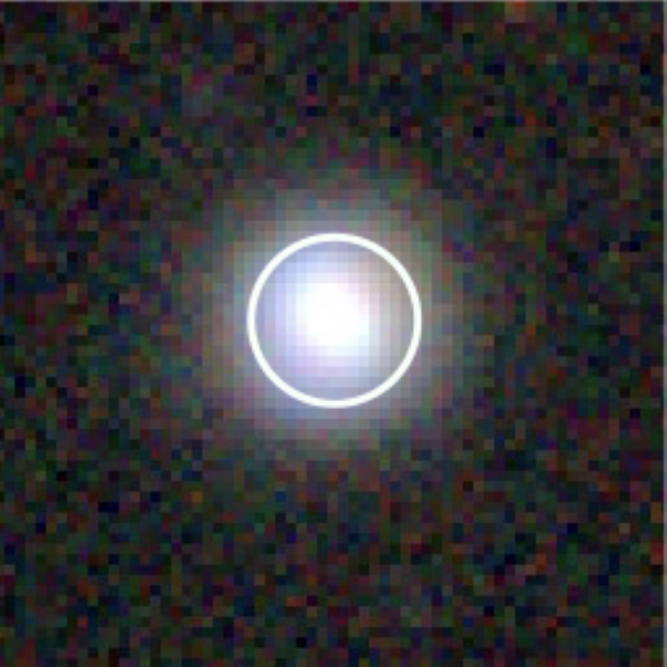}}
\includegraphics[width=0.45\textwidth]{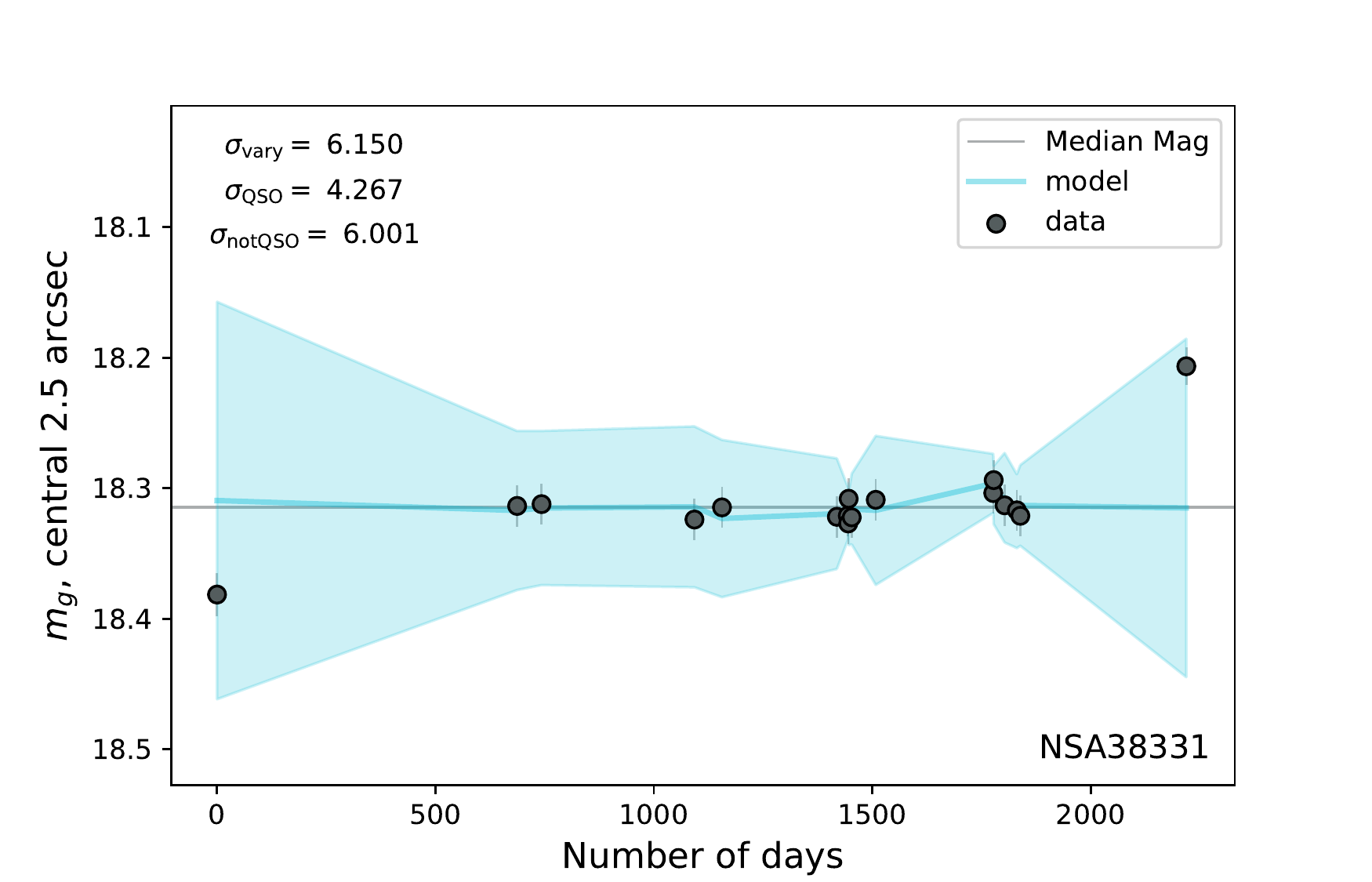}\\
\caption{SDSS \textit{g}-band light curves of low-mass galaxies ($M_{\ast}<10^{10}M_{\odot}$) which meet our AGN variability selection criteria (continued in Figure~\ref{lc_pt5}).  }
\label{lc_pt4}
\end{figure*}

\begin{figure*}
\centering
\raisebox{0.5cm}{\includegraphics[width=0.23\textwidth]{NSA38720_20a_cutout.pdf}}
\includegraphics[width=0.45\textwidth]{ModLC_NSA38720.pdf}\\
\raisebox{0.5cm}{\includegraphics[width=0.23\textwidth]{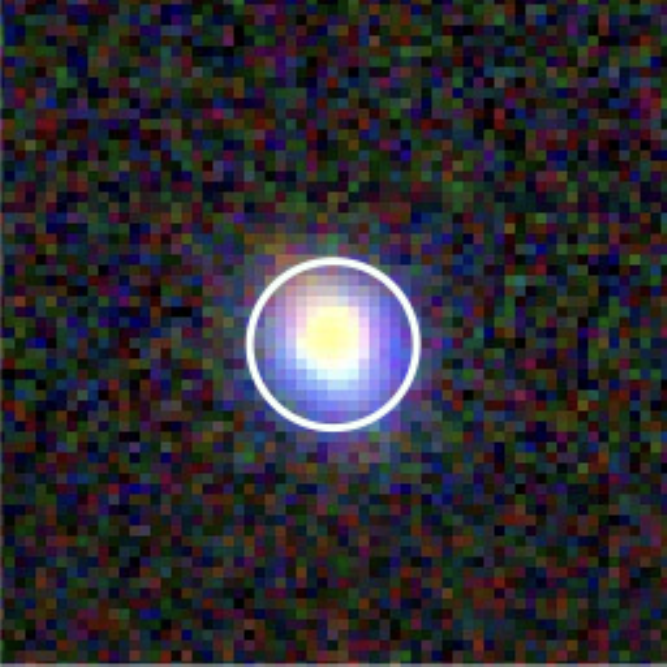}}
\includegraphics[width=0.45\textwidth]{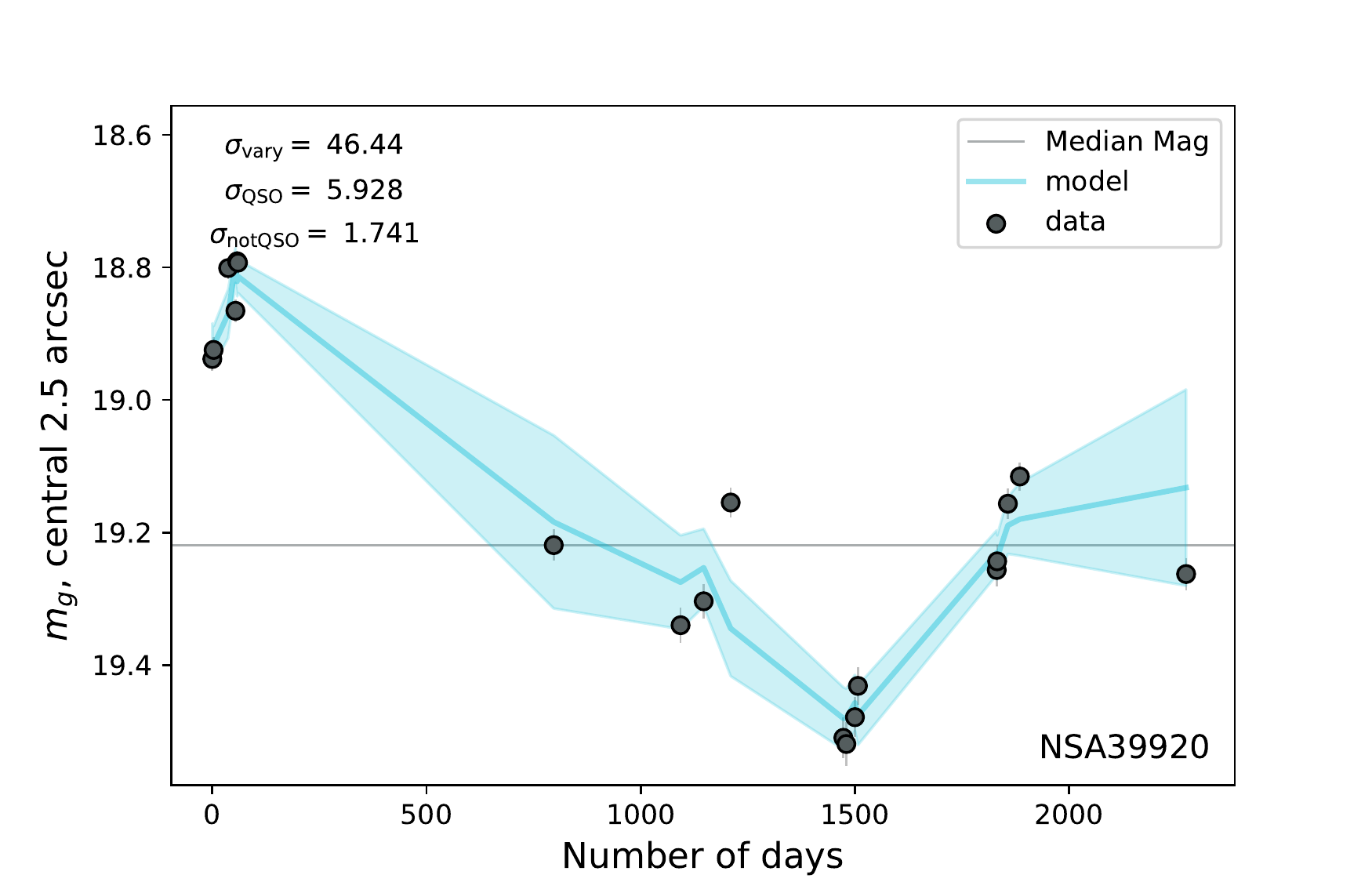}\\
\raisebox{0.5cm}{\includegraphics[width=0.23\textwidth]{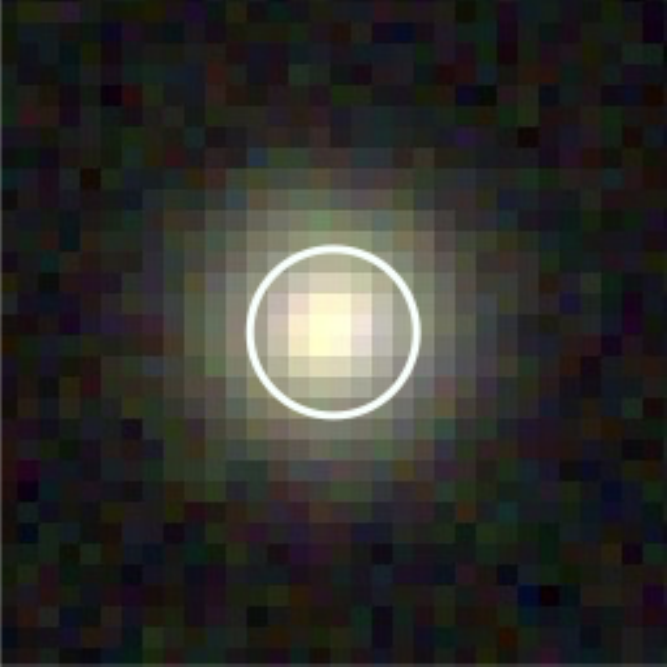}}
\includegraphics[width=0.45\textwidth]{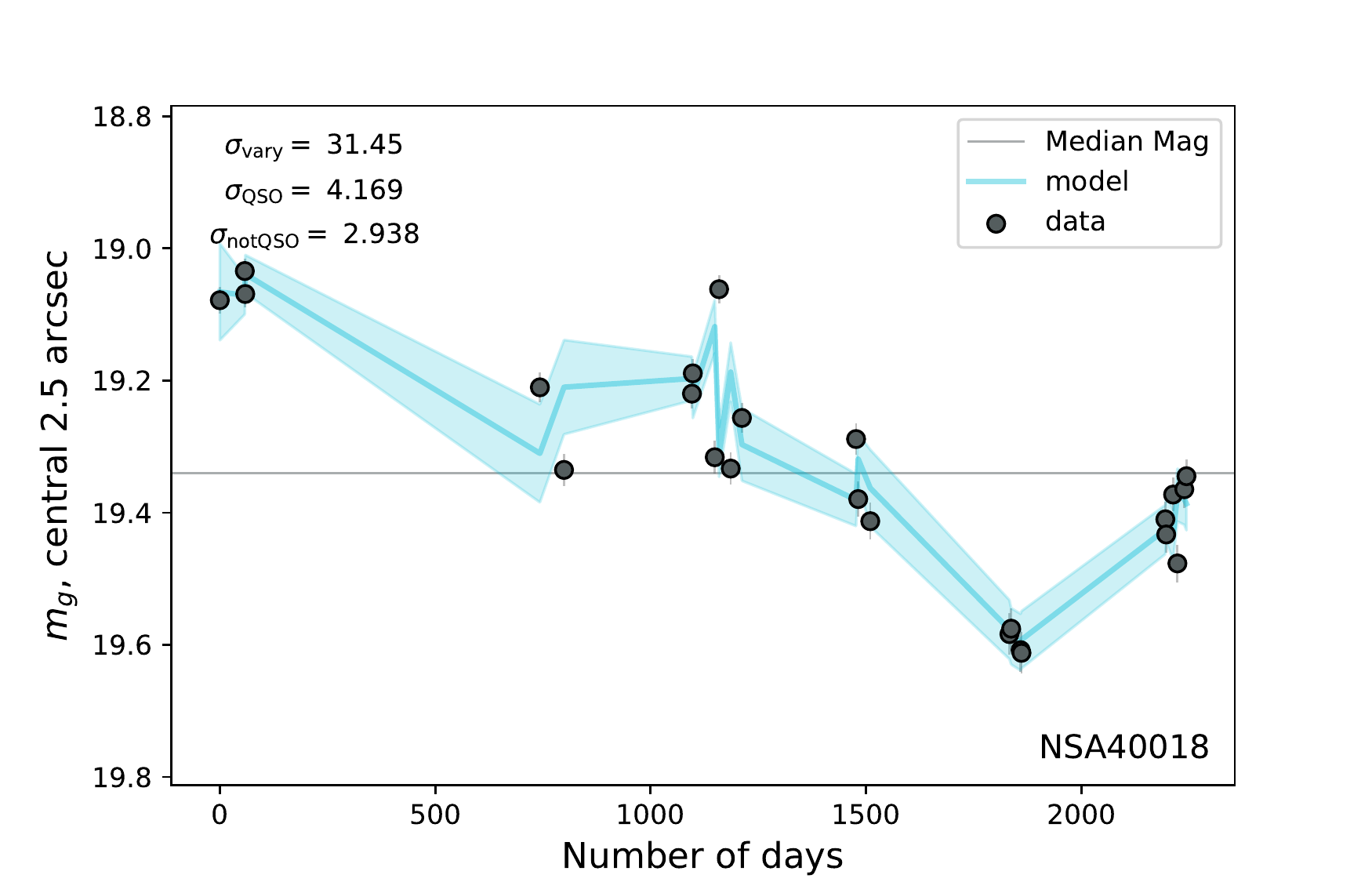}\\
\raisebox{0.5cm}{\includegraphics[width=0.23\textwidth]{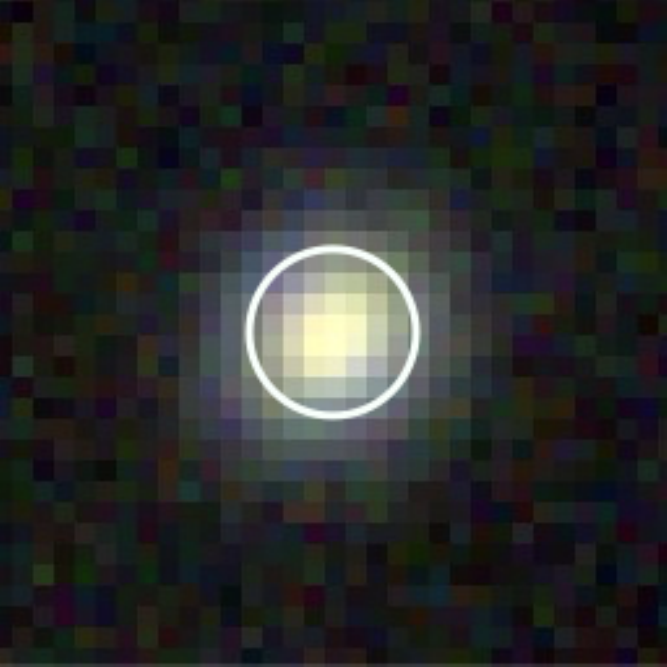}}
\includegraphics[width=0.45\textwidth]{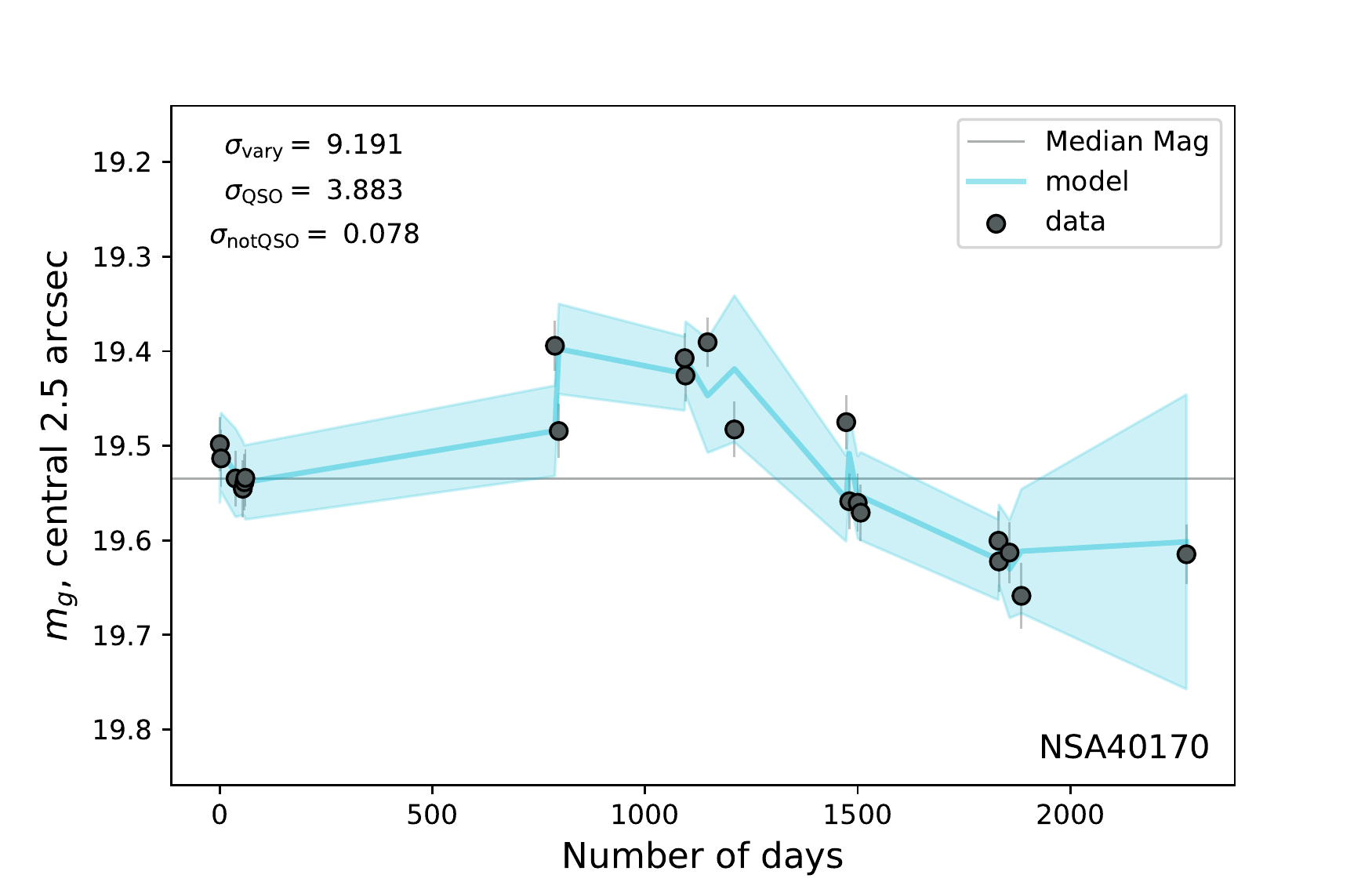}\\
\caption{SDSS \textit{g}-band light curves of low-mass galaxies ($M_{\ast}<10^{10}M_{\odot}$) which meet our AGN variability selection criteria (continued in Figure~\ref{lc_pt6}).  }
\label{lc_pt5}
\end{figure*}

\begin{figure*}
\centering
\raisebox{0.5cm}{\includegraphics[width=0.23\textwidth]{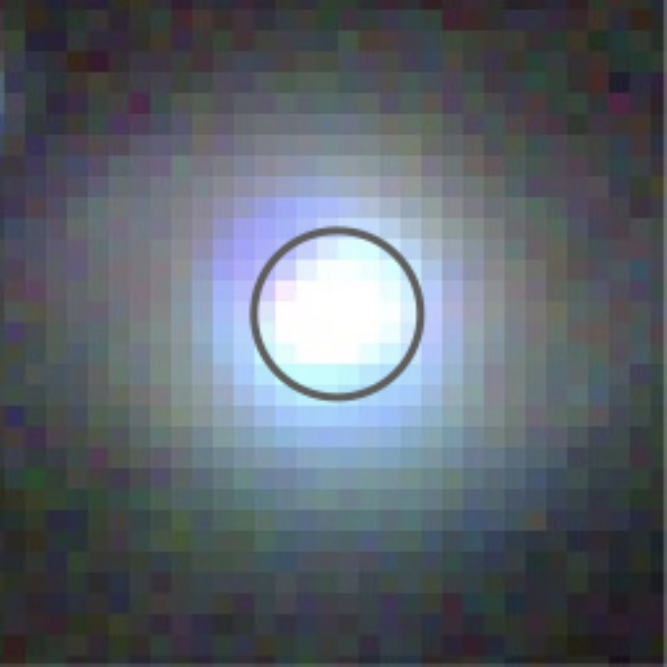}}
\includegraphics[width=0.45\textwidth]{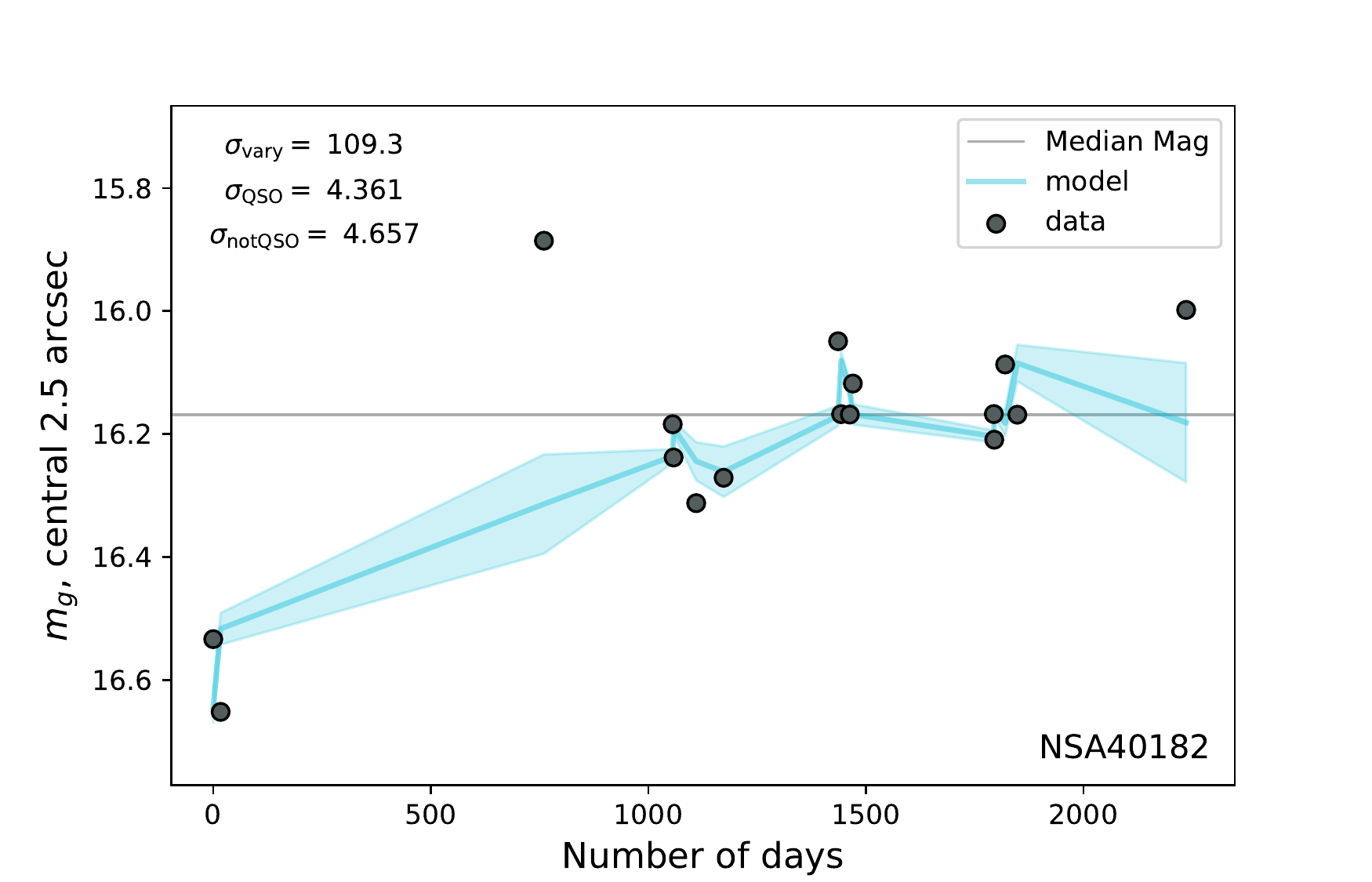}\\
\raisebox{0.5cm}{\includegraphics[width=0.23\textwidth]{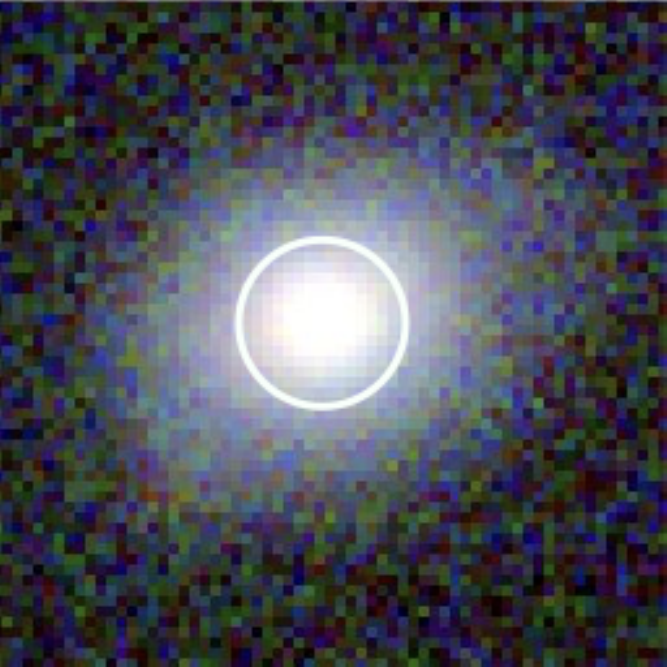}}
\includegraphics[width=0.425\textwidth]{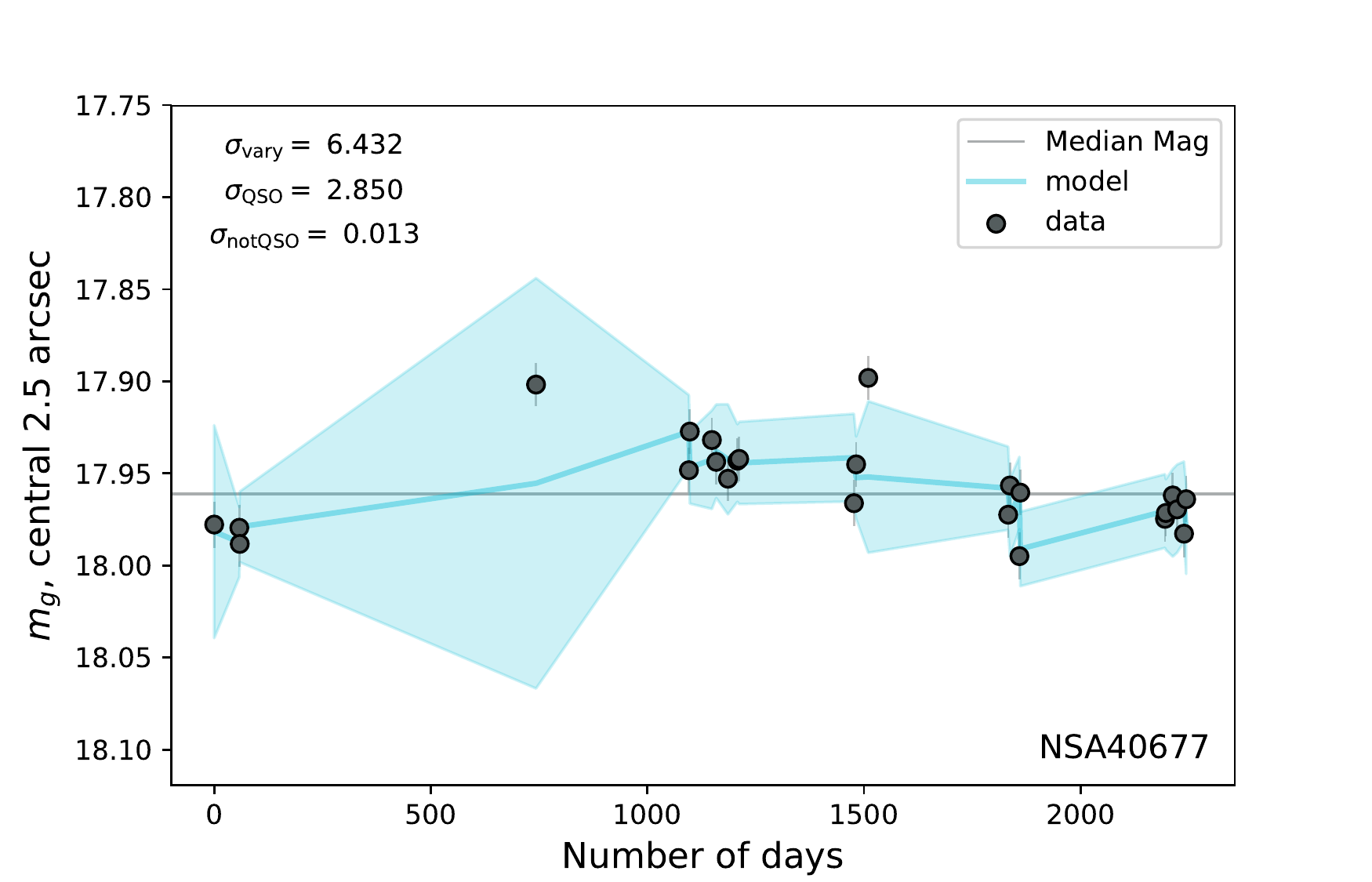}\\
\raisebox{0.5cm}{\includegraphics[width=0.23\textwidth]{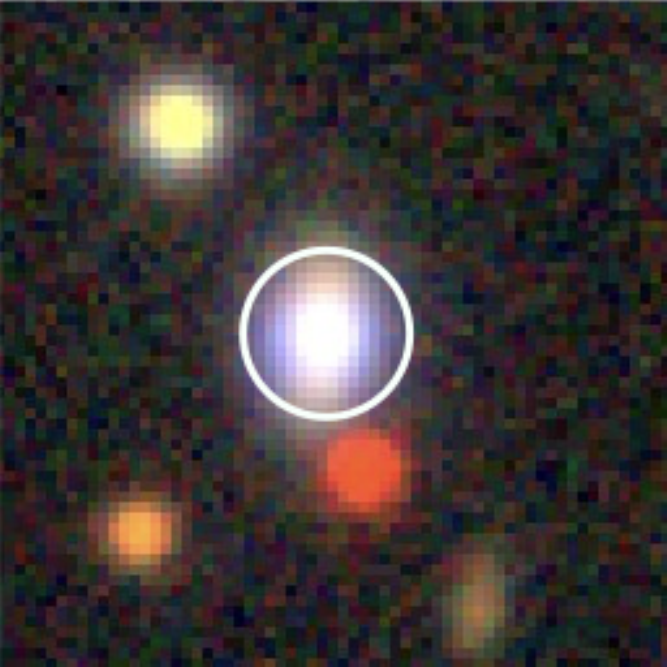}}
\includegraphics[width=0.45\textwidth]{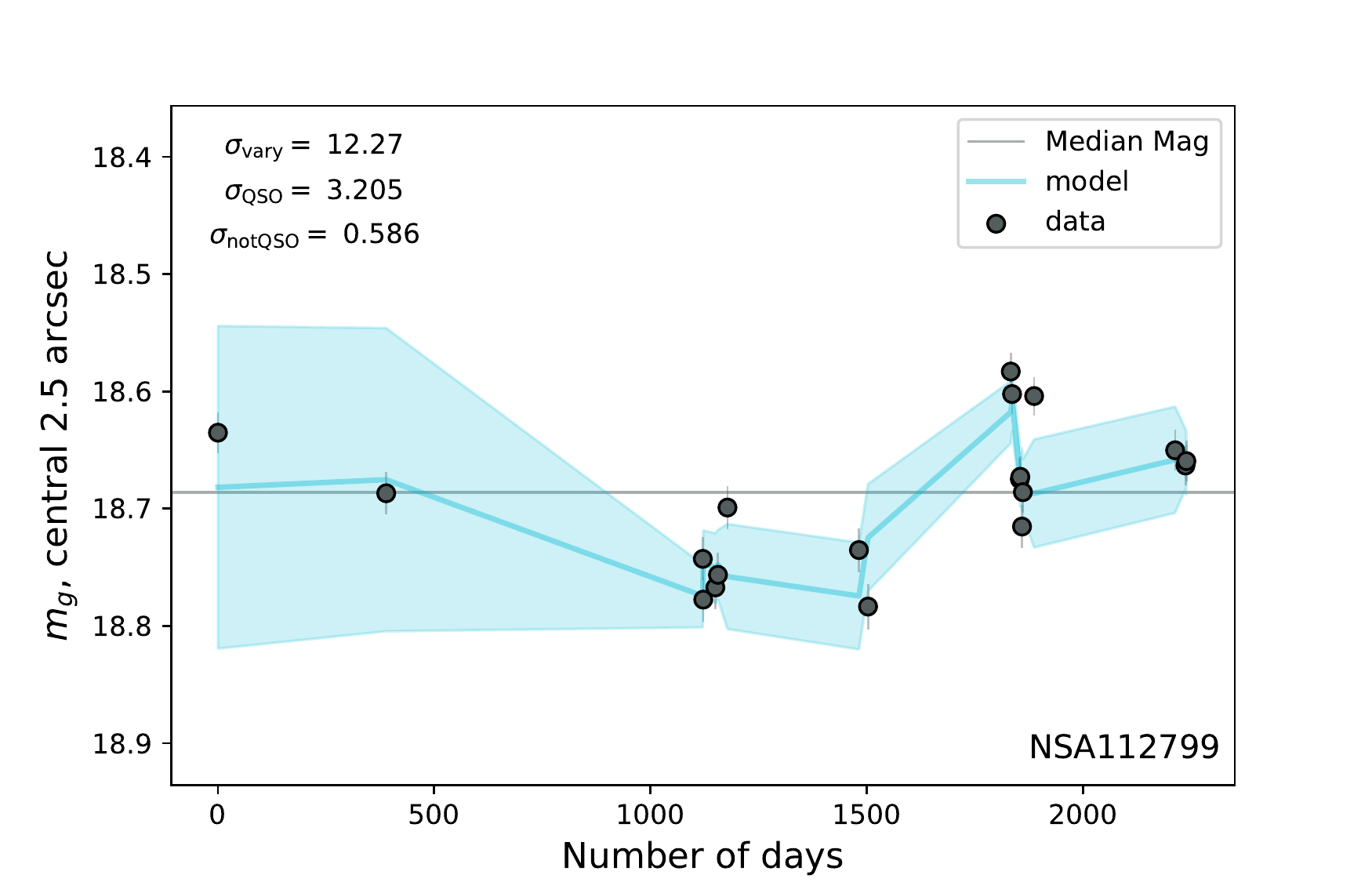}\\
\raisebox{0.5cm}{\includegraphics[width=0.23\textwidth]{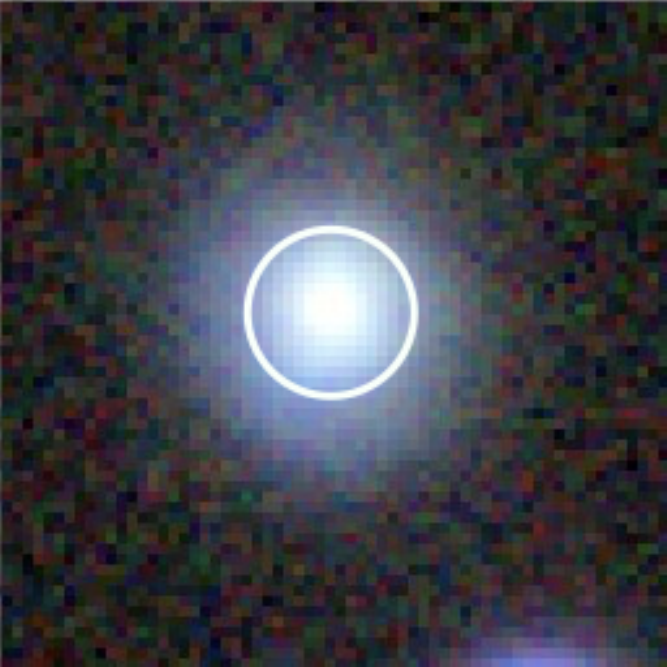}}
\includegraphics[width=0.45\textwidth]{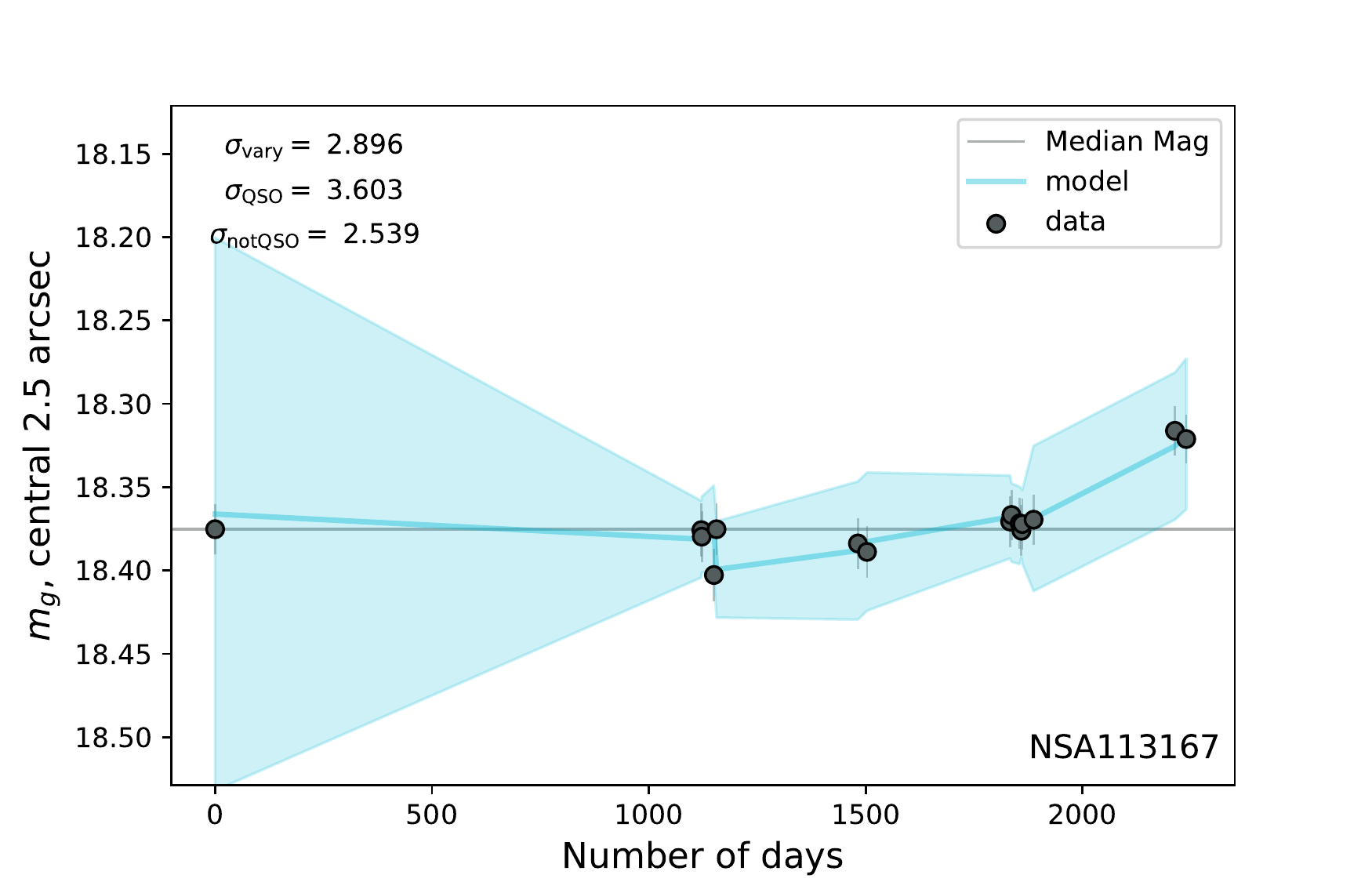}\\
\caption{SDSS \textit{g}-band light curves of low-mass galaxies ($M_{\ast}<10^{10}M_{\odot}$) which meet our AGN variability selection criteria (continued in Figure~\ref{lc_pt7}).  }
\label{lc_pt6}
\end{figure*}

\begin{figure*}
\centering
\raisebox{0.5cm}{\includegraphics[width=0.23\textwidth]{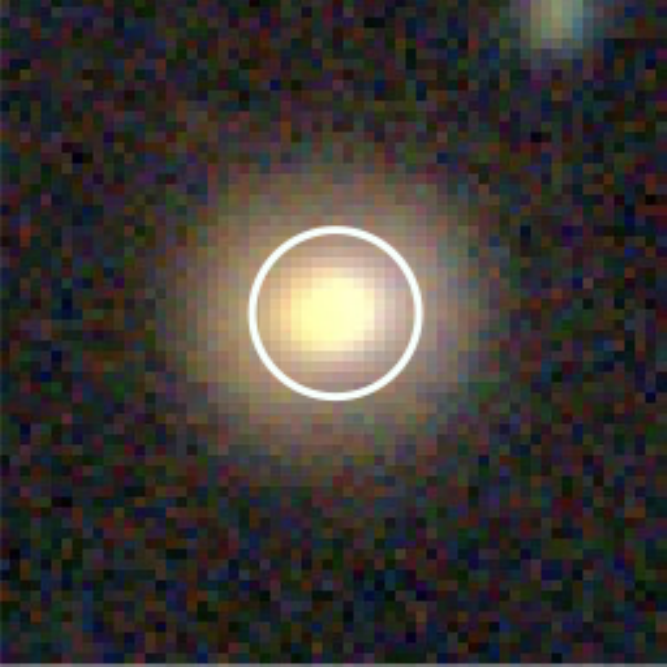}}
\includegraphics[width=0.45\textwidth]{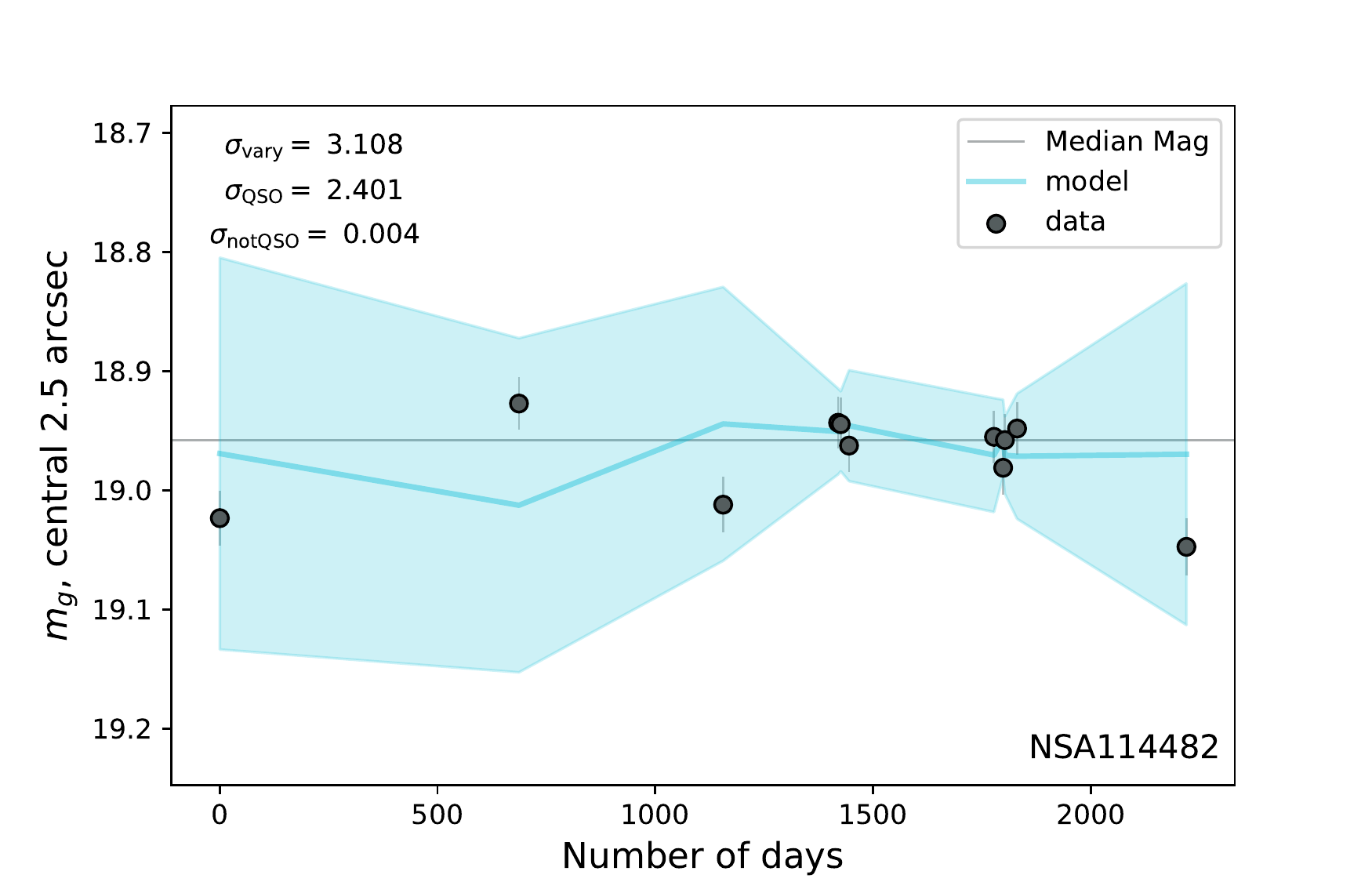}\\
\raisebox{0.5cm}{\includegraphics[width=0.23\textwidth]{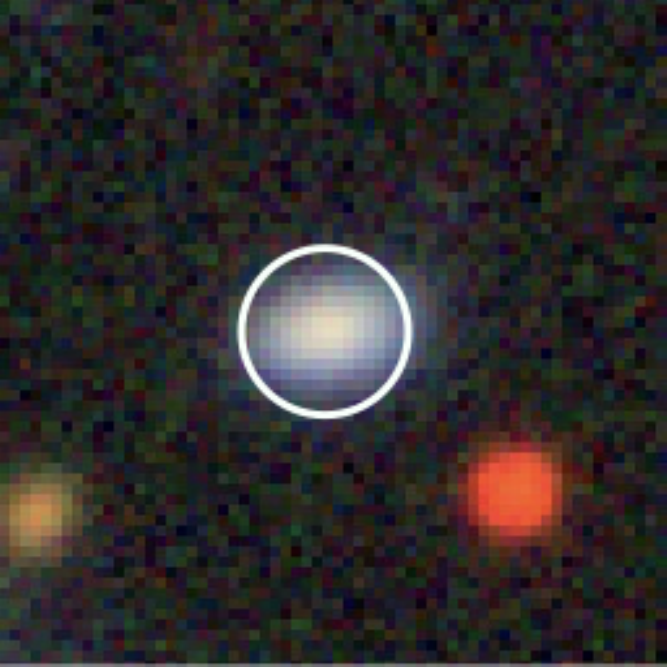}}
\includegraphics[width=0.45\textwidth]{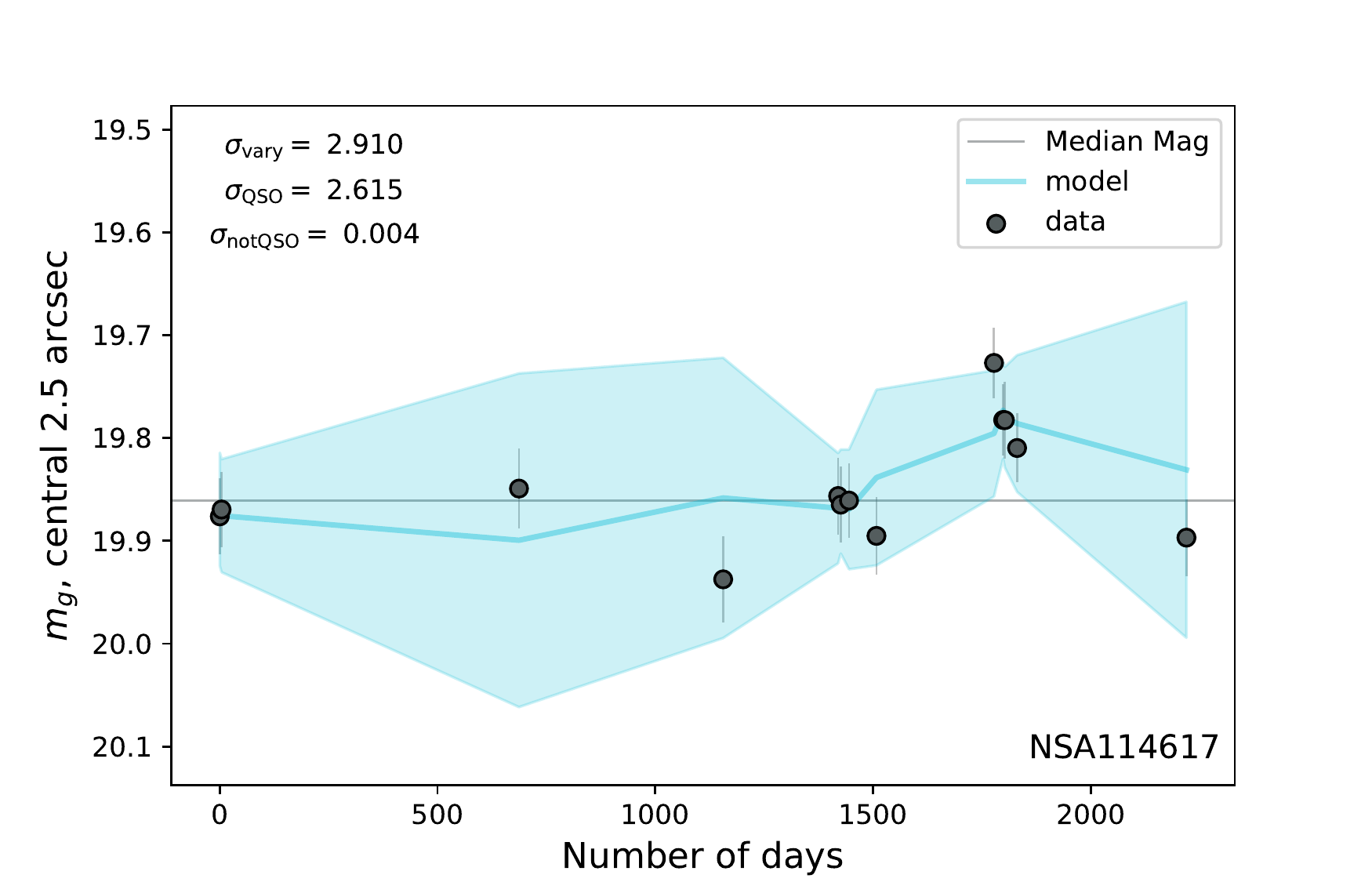}\\
\raisebox{0.5cm}{\includegraphics[width=0.23\textwidth]{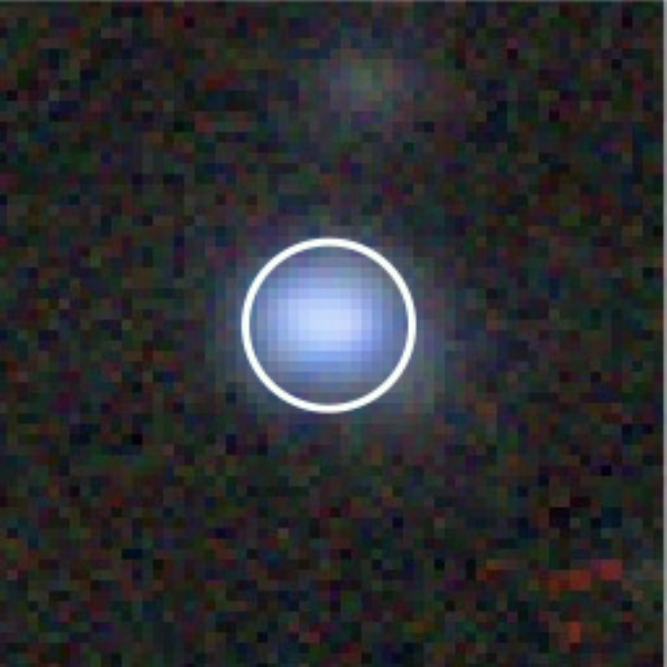}}
\includegraphics[width=0.45\textwidth]{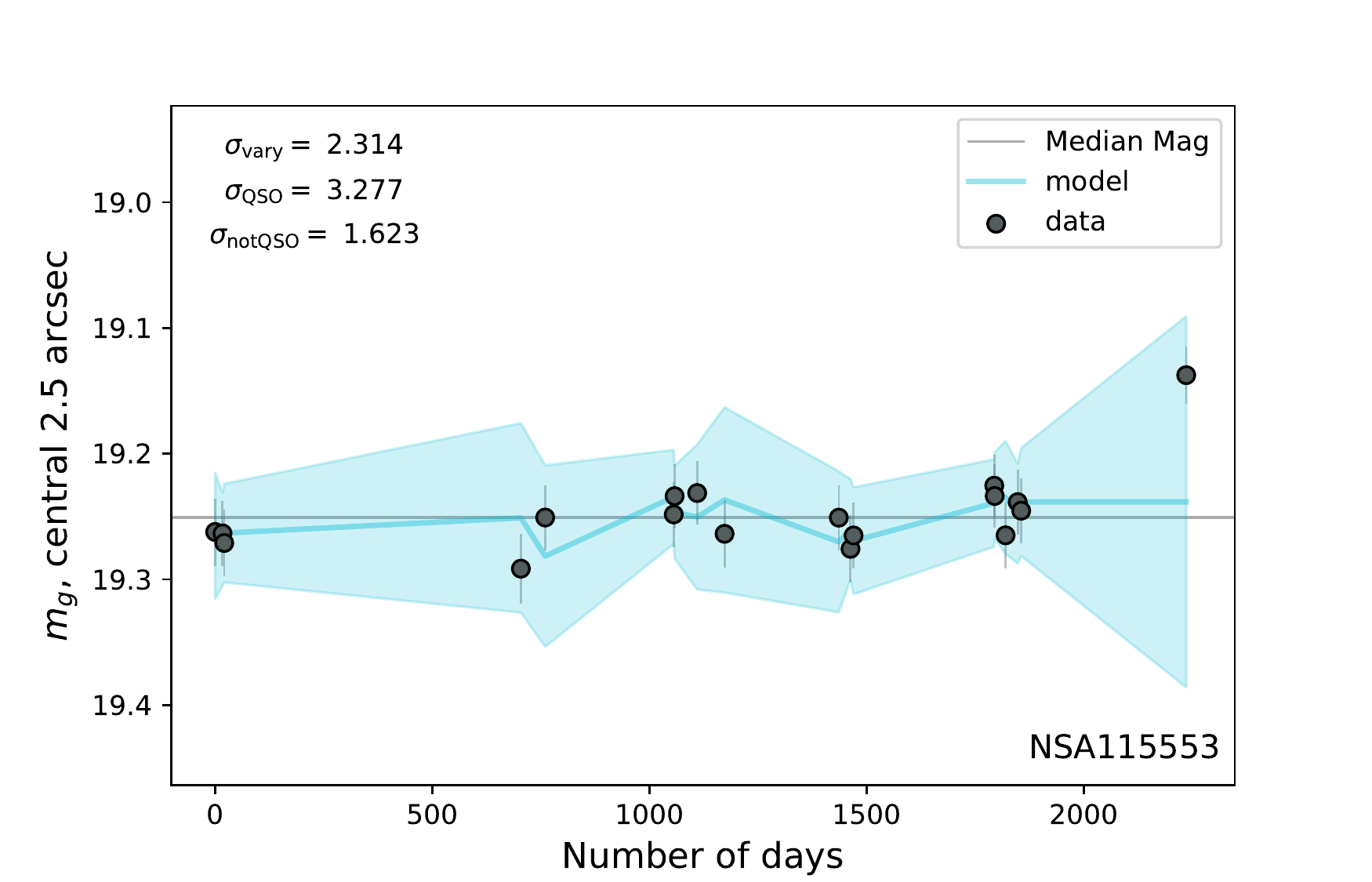}\\
\raisebox{0.5cm}{\includegraphics[width=0.23\textwidth]{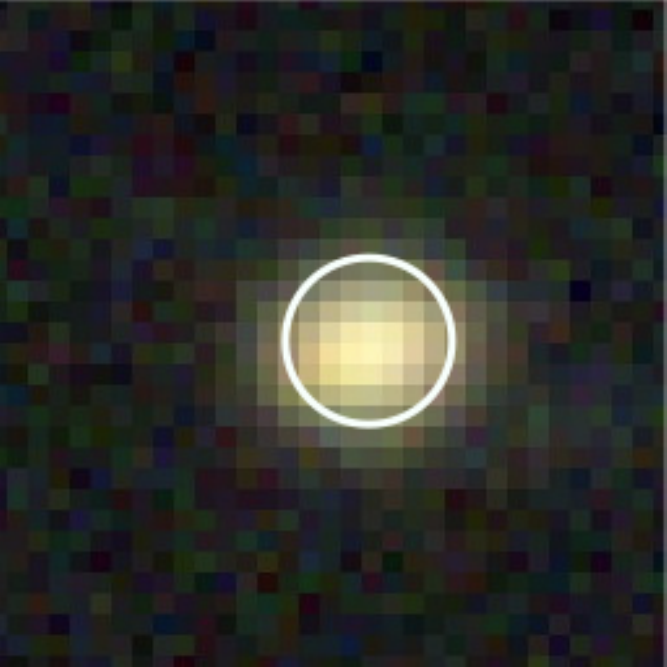}}
\includegraphics[width=0.45\textwidth]{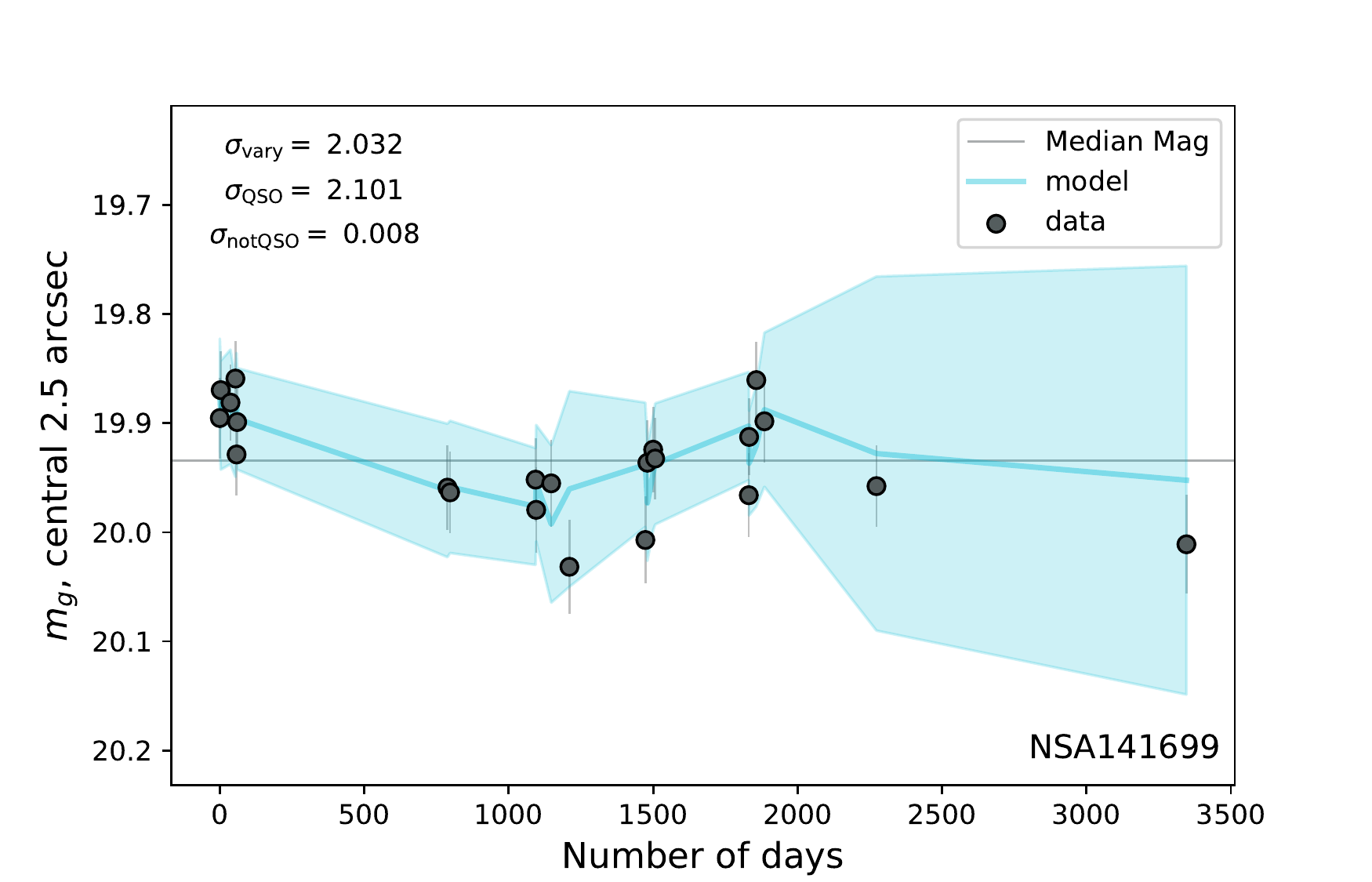}\\
\caption{SDSS \textit{g}-band light curves of low-mass galaxies ($M_{\ast}<10^{10}M_{\odot}$) which meet our AGN variability selection criteria (continued in Figure~\ref{lc_pt8}).  }
\label{lc_pt7}
\end{figure*}

\begin{figure*}
\centering
\raisebox{0.5cm}{\includegraphics[width=0.23\textwidth]{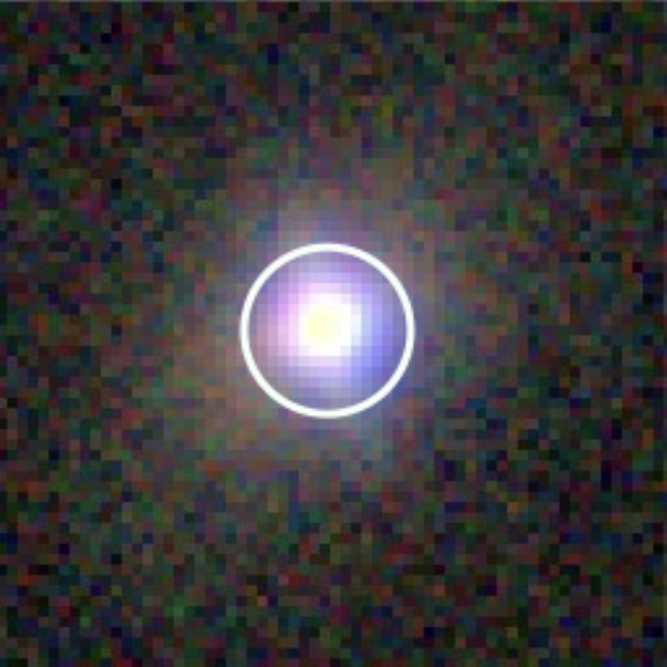}}
\includegraphics[width=0.45\textwidth]{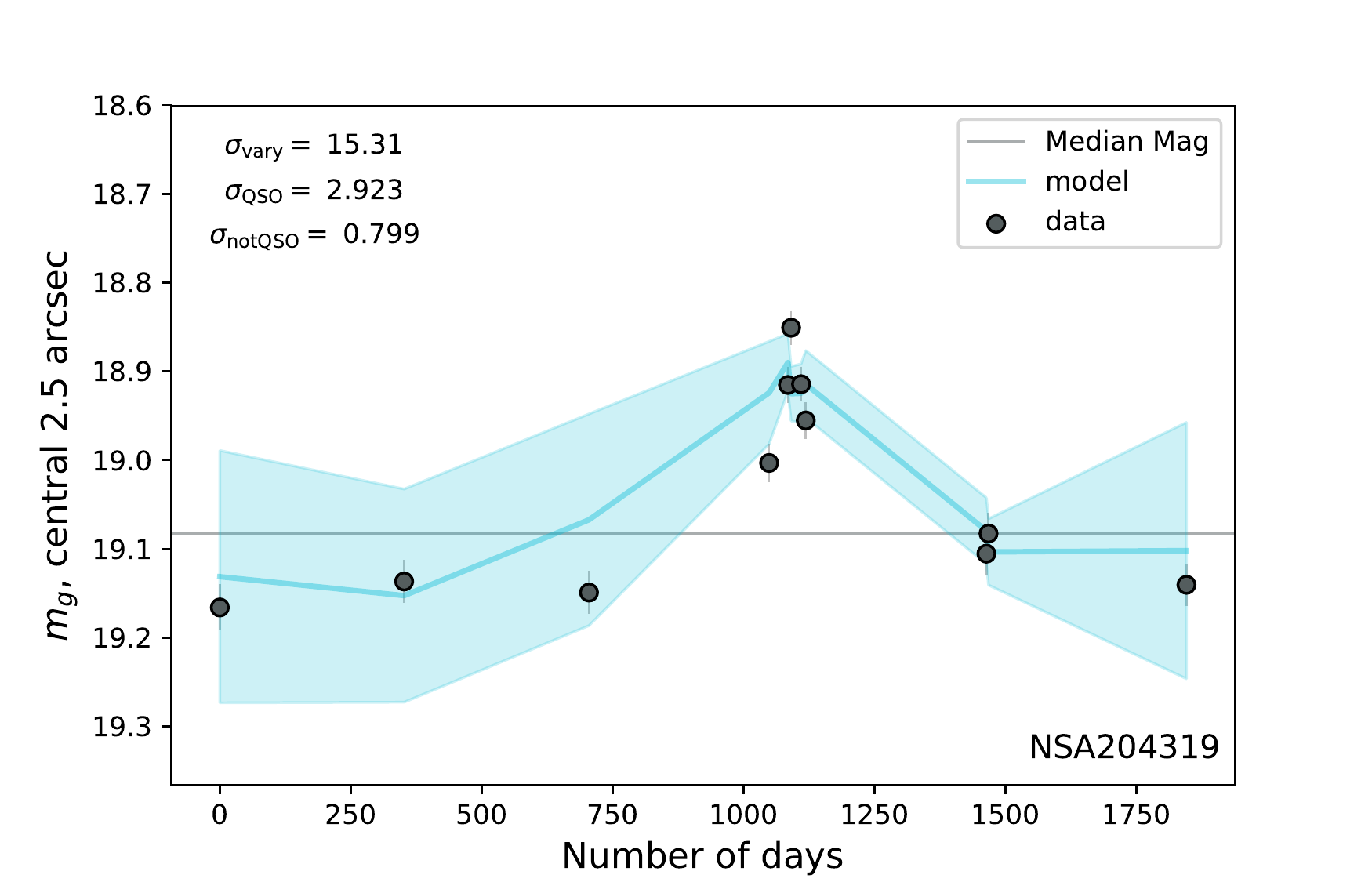}\\
\raisebox{0.5cm}{\includegraphics[width=0.23\textwidth]{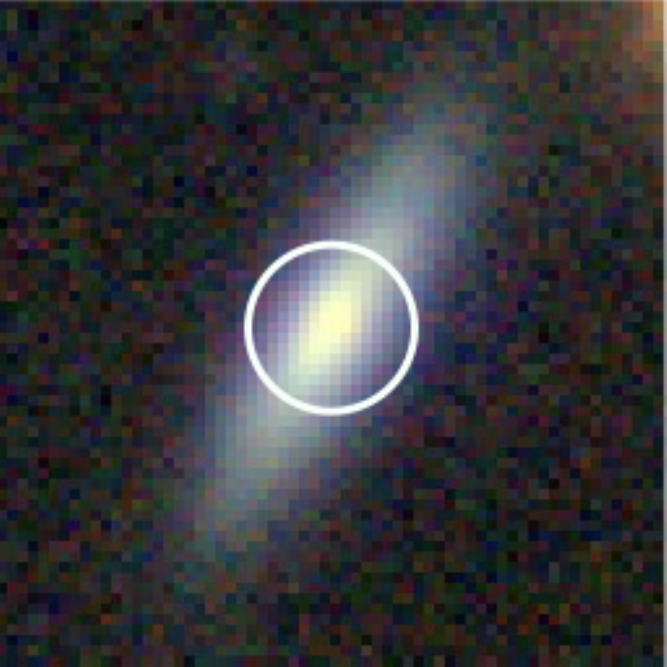}}
\includegraphics[width=0.45\textwidth]{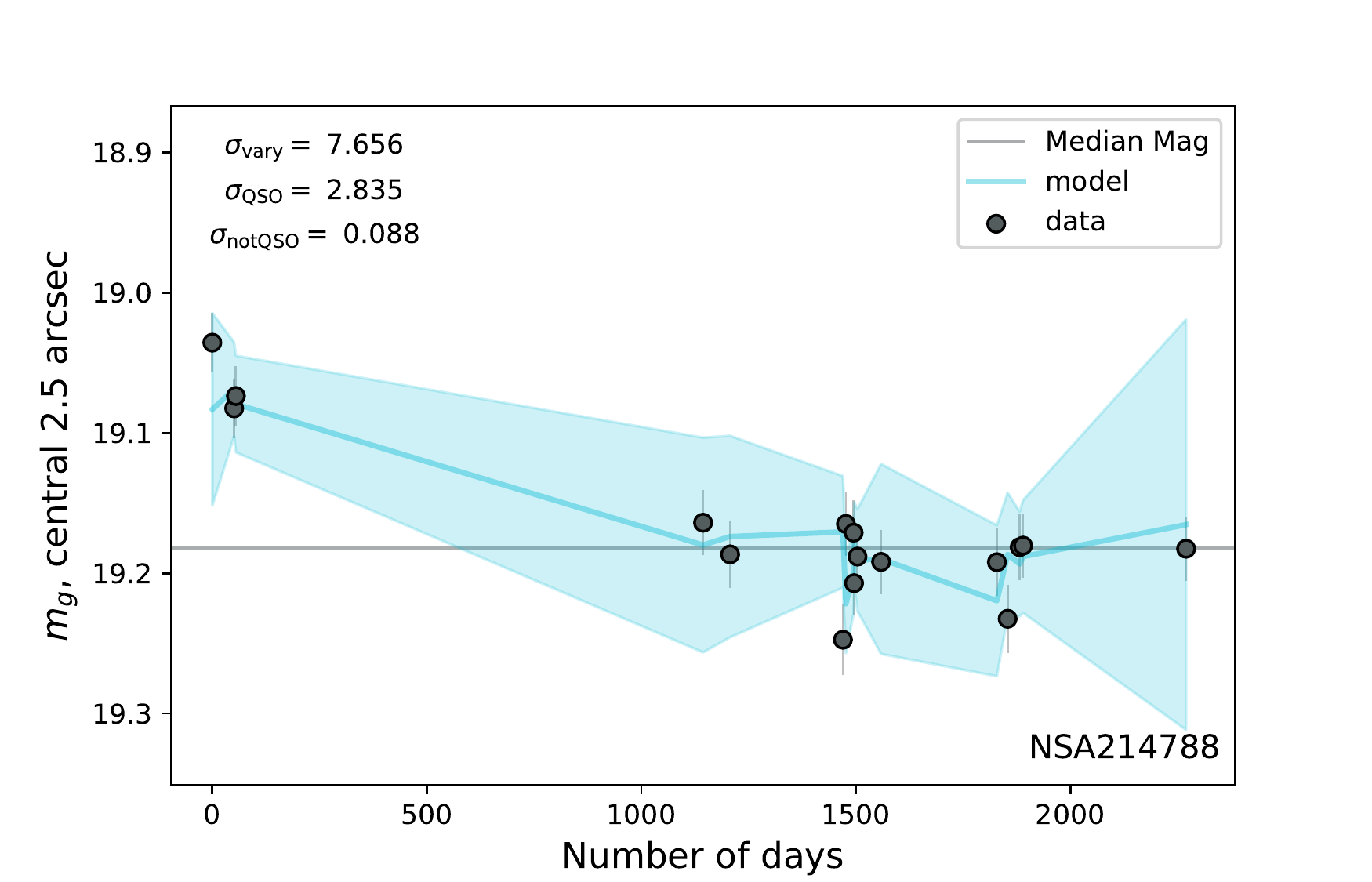}\\
\raisebox{0.5cm}{\includegraphics[width=0.23\textwidth]{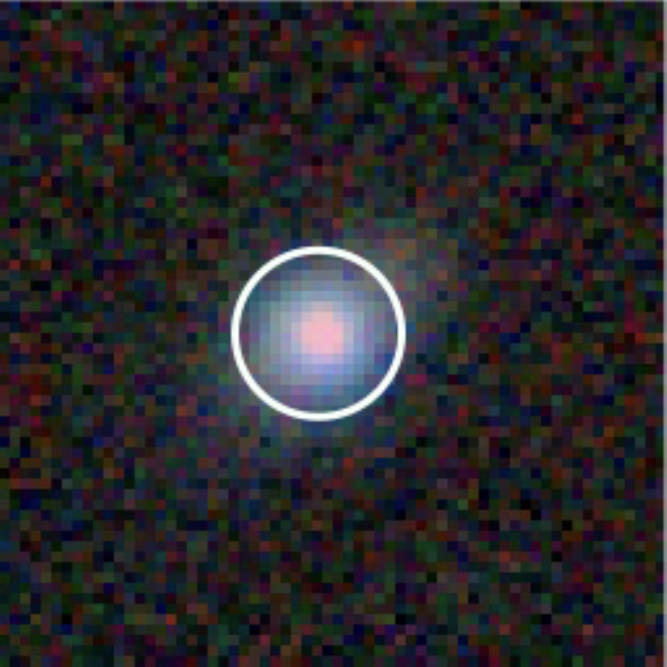}}
\includegraphics[width=0.45\textwidth]{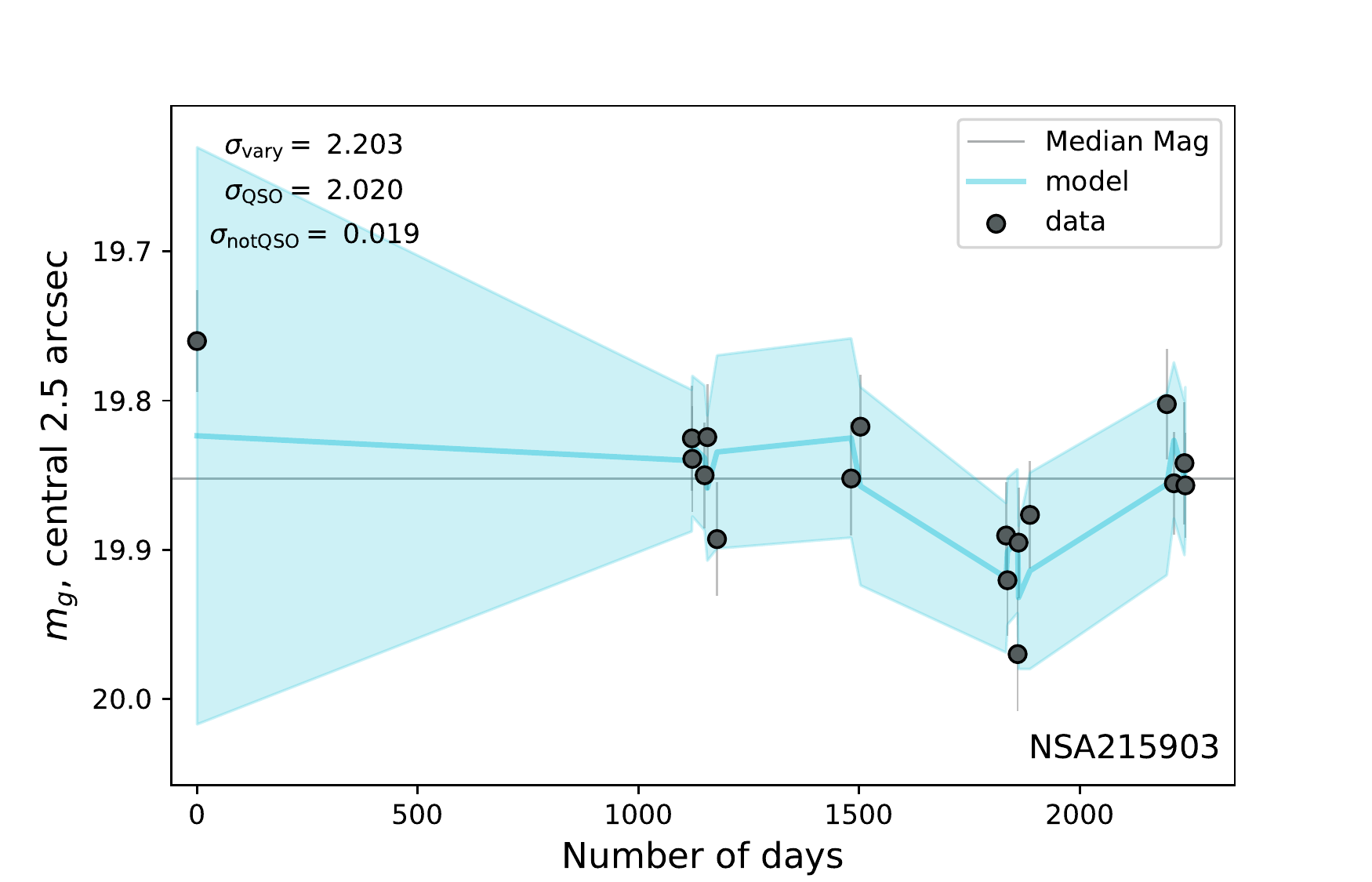}\\
\raisebox{0.5cm}{\includegraphics[width=0.23\textwidth]{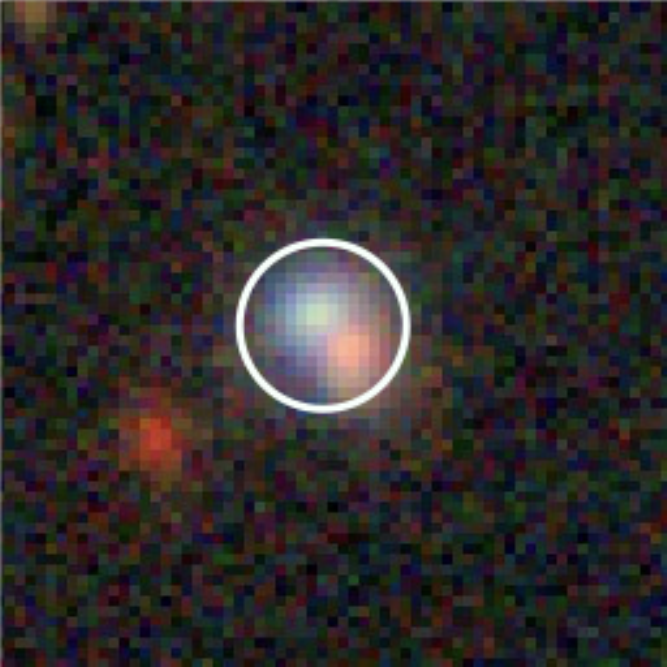}}
\includegraphics[width=0.45\textwidth]{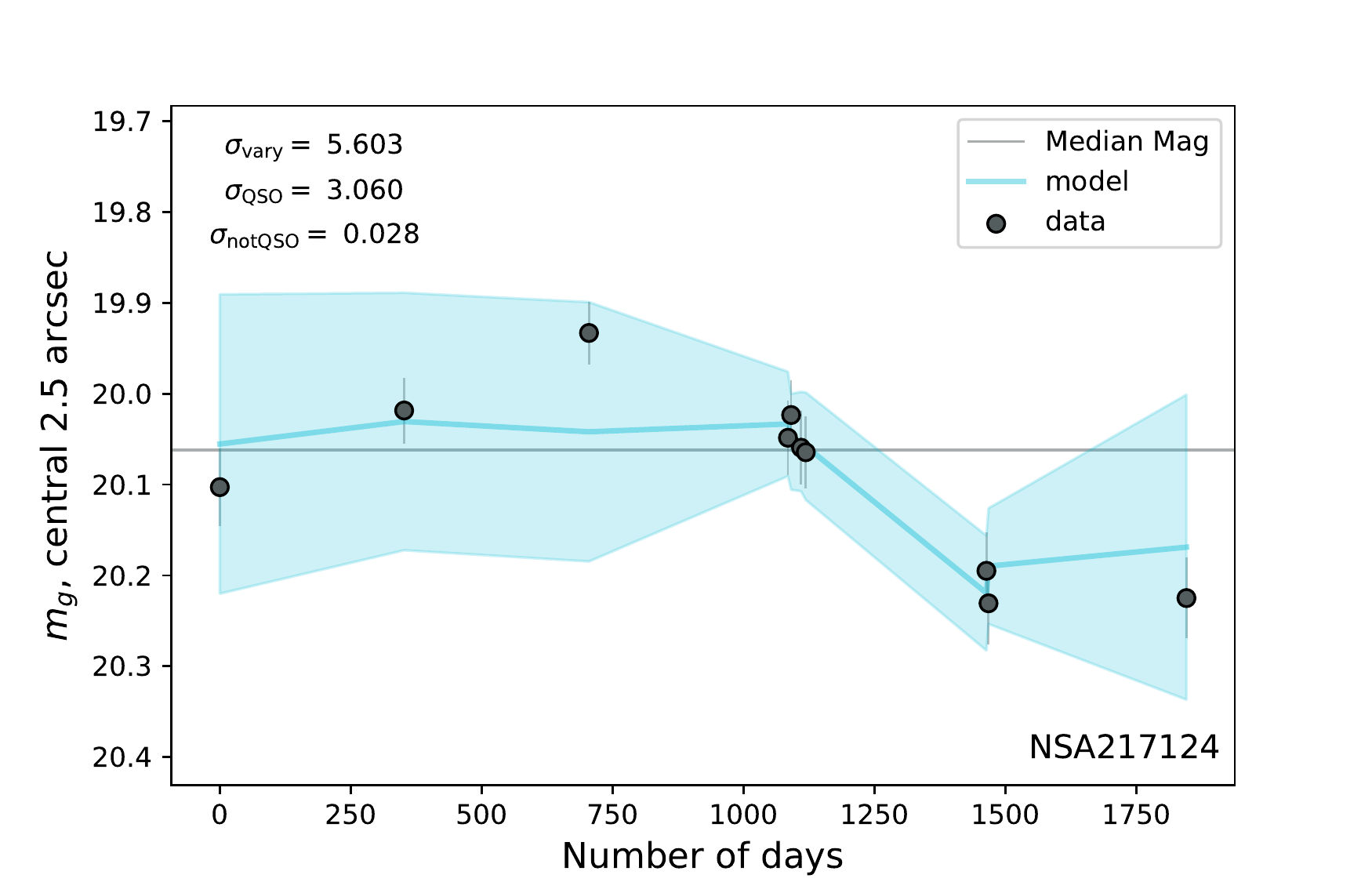}\\
\caption{SDSS \textit{g}-band light curves of low-mass galaxies ($M_{\ast}<10^{10}M_{\odot}$) which meet our AGN variability selection criteria (continued in Figure~\ref{lc_pt9}).  }
\label{lc_pt8}
\end{figure*}

\begin{figure*}
\centering
\raisebox{0.5cm}{\includegraphics[width=0.23\textwidth]{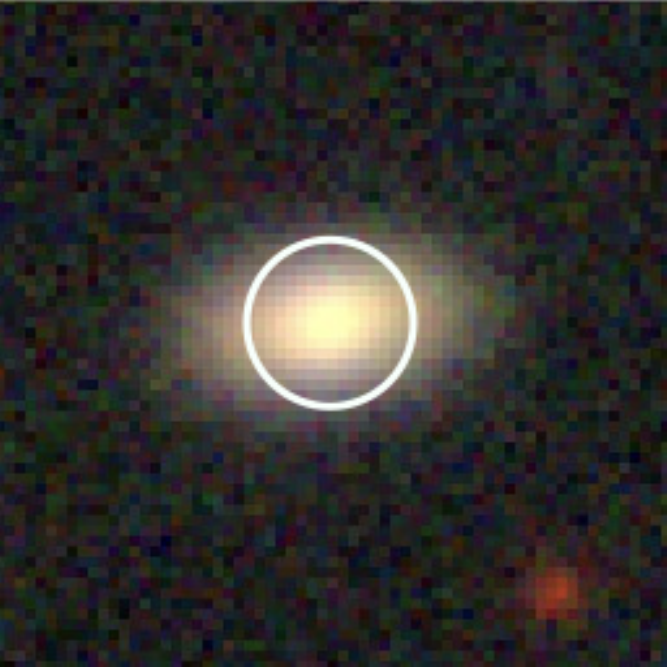}}
\includegraphics[width=0.45\textwidth]{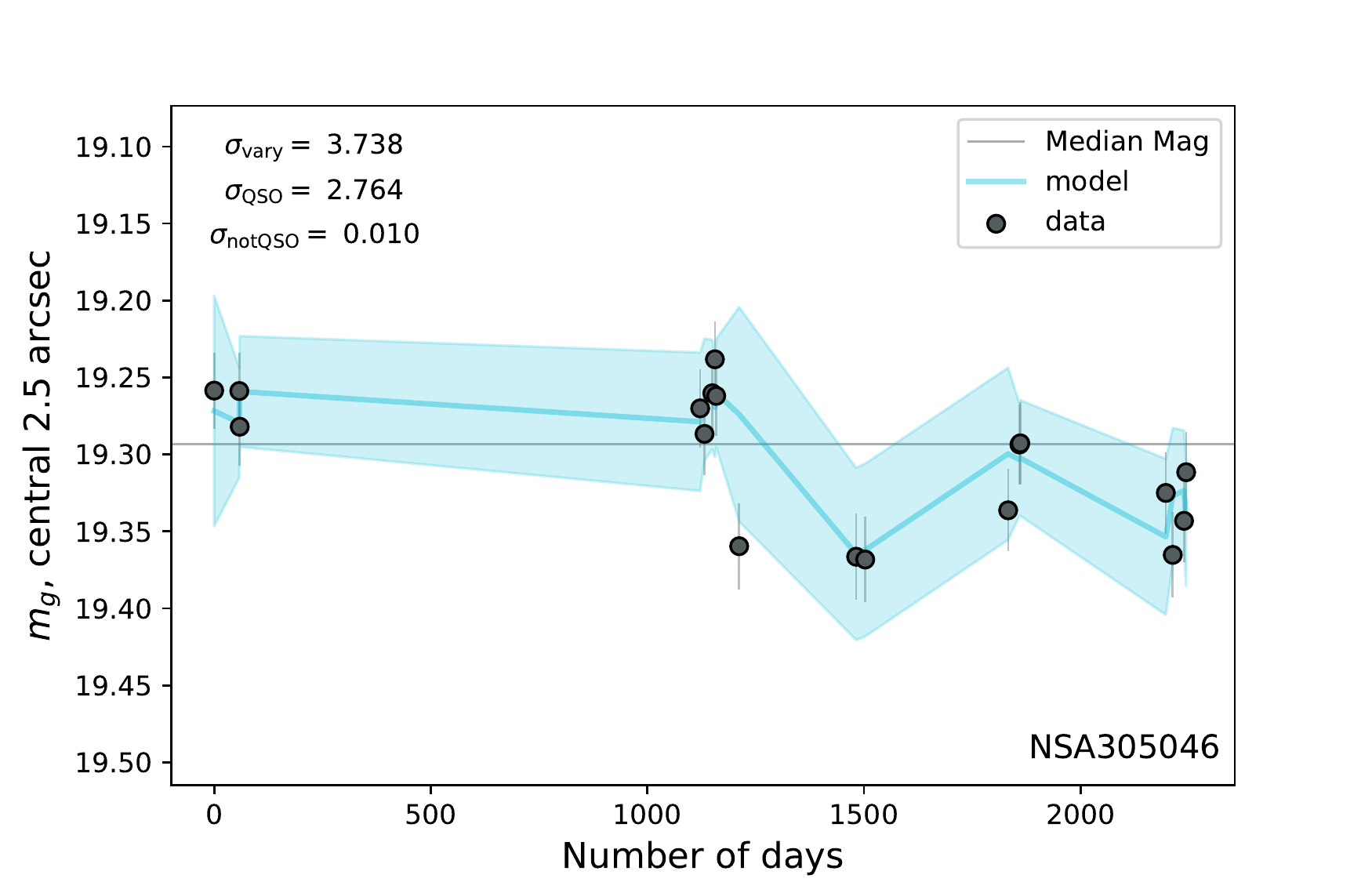}\\
\raisebox{0.5cm}{\includegraphics[width=0.23\textwidth]{NSA305680_20a_cutout.pdf}}
\includegraphics[width=0.45\textwidth]{ModLC_NSA305680.pdf}\\
\raisebox{0.5cm}{\includegraphics[width=0.23\textwidth]{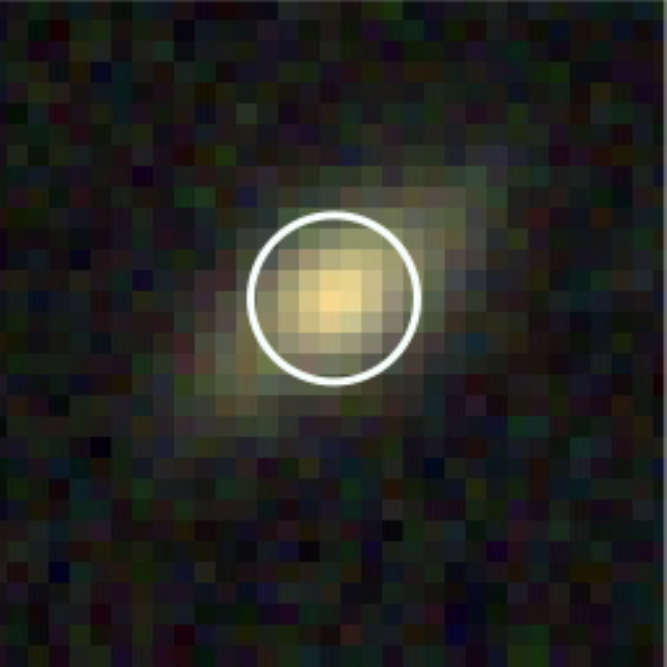}}
\includegraphics[width=0.45\textwidth]{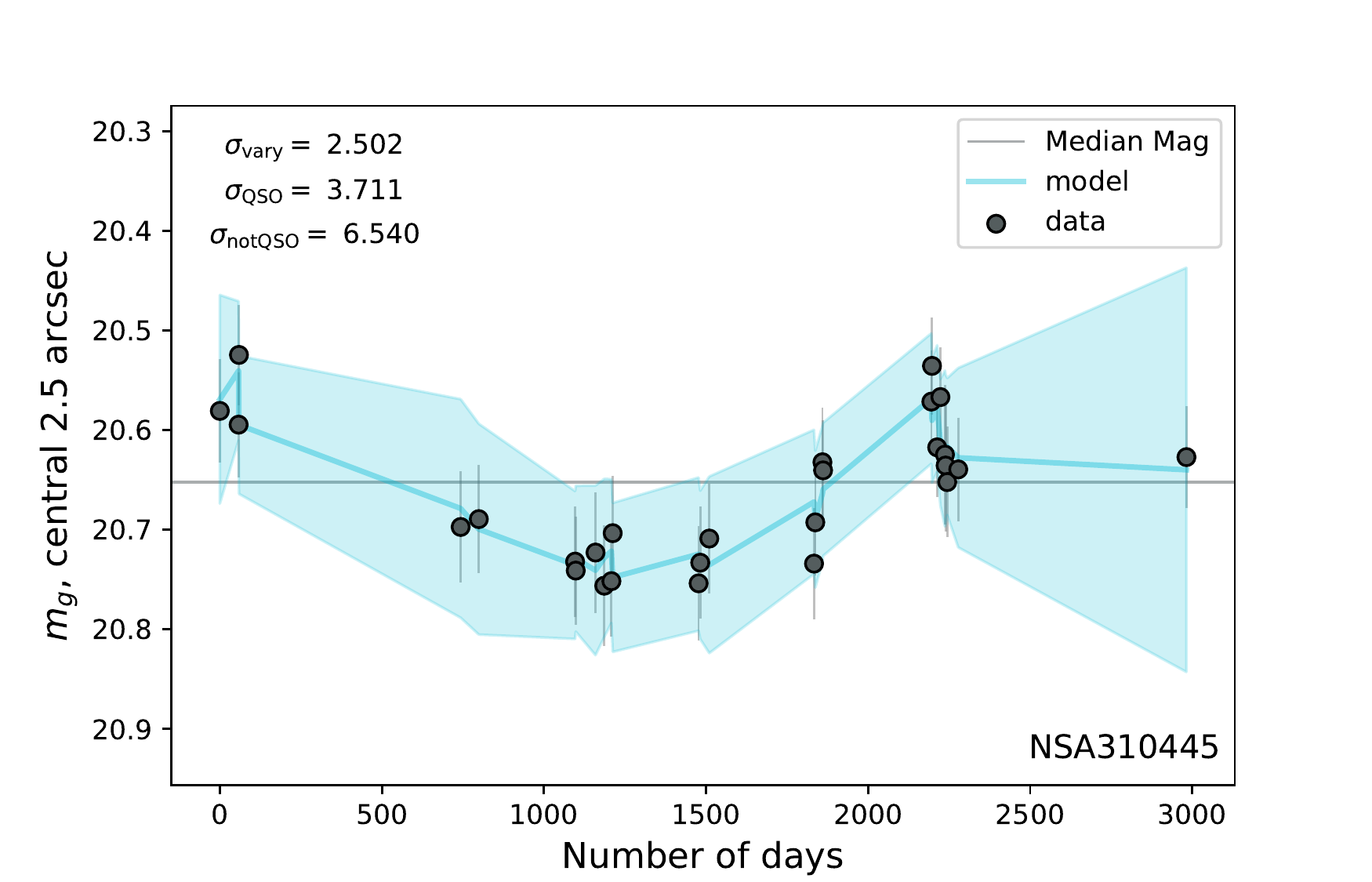}\\
\raisebox{0.5cm}{\includegraphics[width=0.23\textwidth]{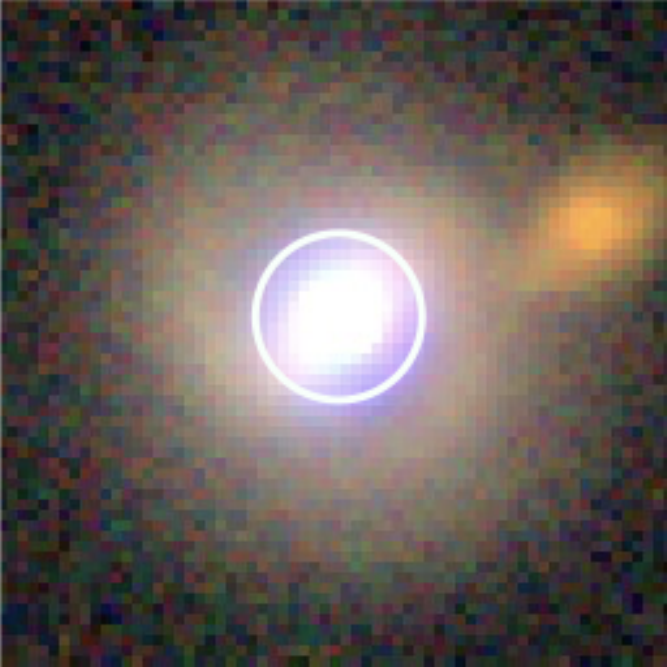}}
\includegraphics[width=0.45\textwidth]{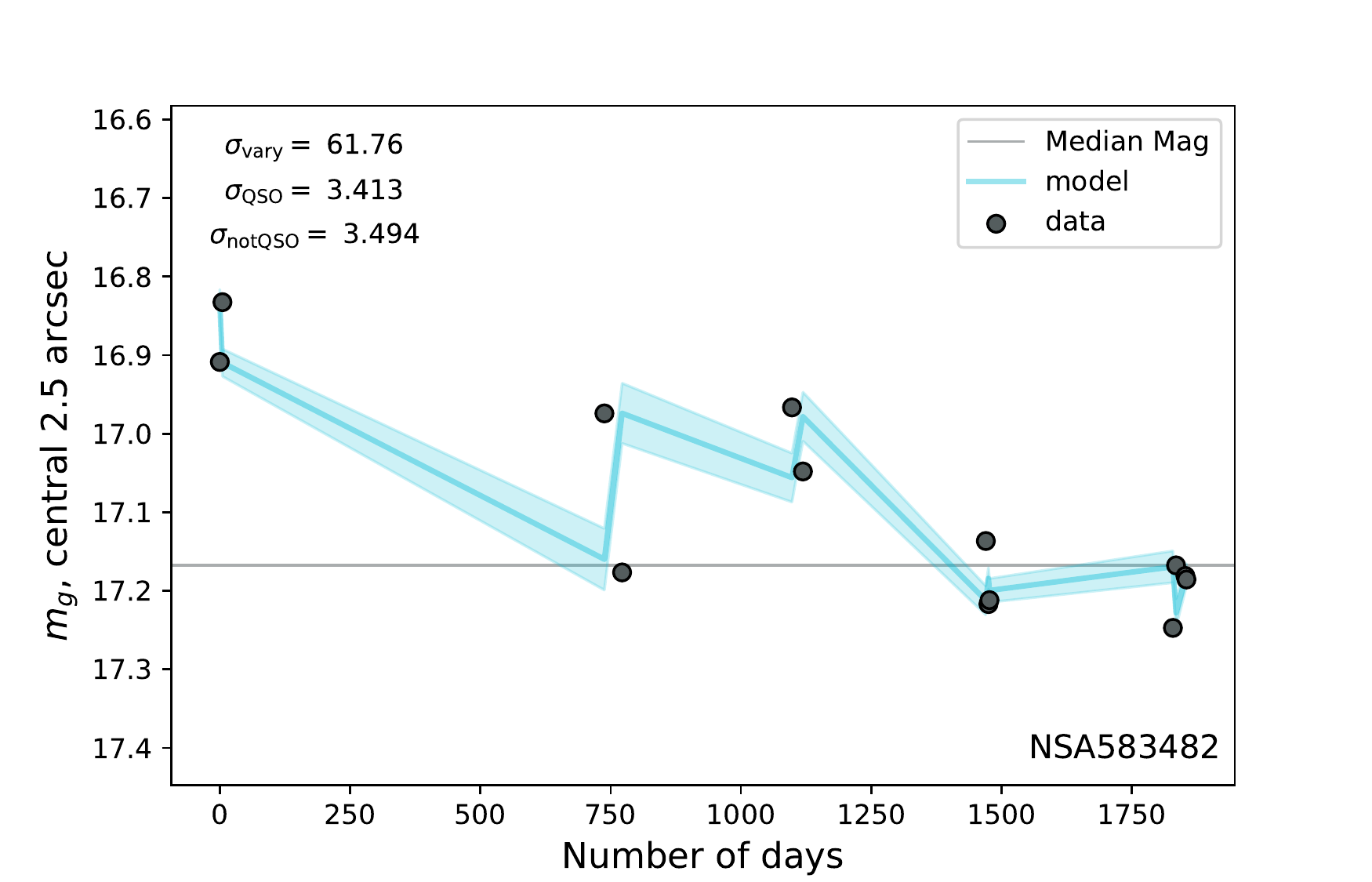}\\
\caption{SDSS \textit{g}-band light curves of low-mass galaxies ($M_{\ast}<10^{10}M_{\odot}$) which meet our AGN variability selection criteria. }
\label{lc_pt9}
\end{figure*}

\end{appendix}

\end{document}